\documentclass[twocolumn]{aastex701}

\usepackage{xspace}
\usepackage{amsmath}
\usepackage{gensymb}
\usepackage{ragged2e}

\newcommand{\lya}{\mbox{Ly$\alpha$}\xspace}
\newcommand{\fesc}{\mbox{$f_{esc}$}\xspace}
\newcommand{\flya}{\mbox{$f_{esc,Ly\alpha}$}\xspace}
\newcommand{\tigm}{\mbox{$T_{\rm IGM}$}\xspace}

\newcommand{\cg}{\mbox{{\tt CIGALE}}\xspace}

\newcommand{\cgLratio}{\mbox{$1.01\pm0.24$}\xspace}

\newcommand{\cgigmcorr}{$100\pm32$}

\newcommand{\objname}{\mbox{MXDFz4.4}\xspace}
\newcommand{\ra}{53.1577218}
\newcommand{\dec}{$-27.7822649$}
\newcommand{\redshift}{{4.442}}
\newcommand{\finalsigma}{\mbox{$5.3\sigma$}\xspace}
\newcommand{\fluxsigma}{\mbox{$5.2\sigma$}\xspace}
\newcommand{\lycflux}{\mbox{$4.2\pm0.8\,\mathrm{nJy}$}\xspace}
\newcommand{\timetoEoR}{\mbox{$\sim0.25\,\mathrm{Gyr}$}\xspace}
\newcommand{\fluxratio}{\mbox{$0.18\pm 0.04$}\xspace} 
\newcommand{\bkgwindow}{66.25}

\newcommand{\rfiftykpc}{$0.63\pm0.04\,\mathrm{kpc}$}

\newcommand{\lyaL}{$1.8\pm0.2\times10^{41}\,\mathrm{erg/s}$}
\newcommand{\finalflux}{$93 \pm 8\times10^{-20}\,\mathrm{erg/s/cm^2}$}
\newcommand{\finalEWobs}{$95\pm11\,\mathrm{\AA}$}
\newcommand{\finalEW}{$17\pm2\,\mathrm{\AA}$}

\newcommand{\fwhm}{$405\pm52\,\mathrm{km/s}$}
\newcommand{\HF}{$0.27^{+0.15}_{-0.18}$}
\newcommand{\asymmetry}{$2.1\pm0.6$}

\newcommand{\lyarfivekpc}{$2.4^{+0.6}_{-0.5}\,\mathrm{kpc}$}

\newcommand{\lyafesc}{\mbox{$2-7\%$}\xspace}

\begin{document}


\title{\objname: A LyC emitter $\mathbf{250\,\mathrm{\textbf{Myr}}}$ after the epoch of reionization and a first test of Ly$\mathbf{\alpha}$ morphology as a tracer of LyC escape at high redshift}

\author[orcid=0009-0007-8470-5946,sname='North America']{Ilias Goovaerts}
\affiliation{Space Telescope Science Institute, 3700 San Martin Drive, Baltimore, MD 21218, USA}
\email[show]{igoovaerts@stsci.edu}  

\author[0000-0002-9946-4731]{Marc Rafelski}
\affiliation{Space Telescope Science Institute, 3700 San Martin Drive, Baltimore, MD 21218, USA}
\affiliation{Department of Physics and Astronomy, Johns Hopkins University, Baltimore, MD 21218, USA}
\email[]{mrafelski@stsci.edu}

\author[orcid=0000-0001-7396-3578]{Alexander Beckett}
\affiliation{Aix Marseille Université, CNRS, CNES, LAM (Laboratoire d’Astrophysique de Marseille), 13388 Marseille, France}
\email[]{alexander.beckett@lam.fr}

\author[0000-0003-0028-4130]{Grecco Oyarz\'un}
\affiliation{Homer L. Dodge Department of Physics and Astronomy, The University of Oklahoma, 440 W. Brooks St., Norman, OK 73019, USA}
\affiliation{Department of Physics and Astronomy, Johns Hopkins University, Baltimore, MD 21218, USA}
\email[]{grecco.oyarzun@ou.edu}

\author[orcid=0009-0000-9676-0538]{Annalisa Citro}
\affiliation{Minnesota Institute for Astrophysics, School of Physics and Astronomy, University of Minnesota, 316 Church Street SE, Minneapolis,
MN 55455, USA}
\email[]{}

\author[0000-0002-0072-0281]{Farhanul Hasan}
\affiliation{Space Telescope Science Institute, 3700 San Martin Drive, Baltimore, MD 21218, USA}
\email{fhasan@stsci.edu}

\author[orcid=0000-0001-5294-8002,sname='North America']{Kalina V.~Nedkova}
\affiliation{IPAC, Mail Code 314-6, California Institute of Technology, 1200 E.~California Blvd, Pasadena, CA 91125, USA}
\email[]{knedkova@caltech.edu}

\author[0000-0003-0145-8964]{Calum Hawcroft}
\affiliation{Space Telescope Science Institute, 3700 San Martin Drive, Baltimore, MD 21218, USA}
\email[]{}

\author[orcid=0000-0002-6610-2048]{Anton M. Koekemoer}
\affiliation{Space Telescope Science Institute, 3700 San Martin Drive, Baltimore, MD 21218, USA}
\email[]{akoekemoer@stsci.edu}

\author[orcid=0000-0002-4917-7873,sname={Revalski}]{Mitchell Revalski}
\affiliation{Space Telescope Science Institute, 3700 San Martin Drive, Baltimore, MD 21218, USA}
\email[]{mrevalski@stsci.edu}

\author[0000-0001-8587-218X]{Matthew J. Hayes}
\affiliation{Stockholm University, Department of Astronomy and Oskar Klein Centre for Cosmoparticle Physics, AlbaNova University Centre, SE-10691, Stockholm, Sweden}
\email{matthew@astro.su.se}

\author[orcid=0000-0002-9136-8876]{Claudia Scarlata}
\affiliation{Minnesota Institute for Astrophysics, School of Physics and Astronomy, University of Minnesota, 316 Church Street SE, Minneapolis,
MN 55455, USA}
\email[]{}

\author[0000-0003-1581-7825]{Ray A. Lucas}
\affiliation{Space Telescope Science Institute, 3700 San Martin Drive, Baltimore, MD 21218, USA}
\email{}

\author[0000-0001-9440-8872]{Norman A. Grogin}
\affiliation{Space Telescope Science Institute, 3700 San Martin Drive, Baltimore, MD 21218, USA}
\email{nagrogin@stsci.edu}

\author[orcid=0000-0002-3746-2853]{David V. Stark}
\affiliation{Space Telescope Science Institute, 3700 San Martin Drive, Baltimore, MD 21218, USA}
\affiliation{Department of Physics and Astronomy, Johns Hopkins University, Baltimore, MD 21218, USA}
\email[]{}

\author[orcid=0000-0001-7044-3809]{Paolo Suin}
\affiliation{Université Paris-Saclay, Université Paris Cité, CEA, CNRS, AIM, 91191 Gif-sur-Yvette, France}
\email[]{}

\author[orcid=0000-0003-3382-5941]{Nor Pirzkal}
\affiliation{Space Telescope Science Institute, 3700 San Martin Drive, Baltimore, MD 21218, USA}
\email[]{}

\author[orcid=0000-0002-7756-4440]{Louis-Gregory Strolger}
\affiliation{Space Telescope Science Institute, 3700 San Martin Drive, Baltimore, MD 21218, USA}
\email[]{}



\begin{abstract}

Assessing the contribution of ionizing sources to cosmic reionization is a central goal of extragalactic astrophysics. Understanding and quantifying ionizing escape remains challenging near the epoch of reionization. We present the highest-redshift Lyman continuum (LyC) emitter detected to date, \objname, at $z=\redshift$ in the MUSE eXtremely Deep Field, observed only \timetoEoR after the end of reionization. A high-confidence \lya line confirms the redshift. LyC flux is detected at \finalsigma in the F435W filter with a flux of \lycflux, corresponding to a flux measurement at \fluxsigma.
After correcting for the intrinsic production of LyC photons and the IGM opacity at $z=4.44$, we derive high escape fractions, \fesc, ranging from $50-100\,\%$. We apply established low-redshift tracers of LyC escape and, for the first time at this redshift, promising \lya morphological tracers such as the halo fraction. SED fitting indicates the presence of a recent burst of star formation; we explore its impact on the production and escape of ionizing photons.
\lya-based tracers of \fesc reveal a complex scenario in which the recent burst strongly influences LyC production and escape, combined with a more evolved stellar population. This interpretation is supported by UV diagnostics, including $\Sigma_{SFR}$ and sSFR. Our results provide cautious support for the \lya halo fraction as a LyC escape tracer at high redshift. Considering the burst-driven enhancement in ionizing photon production and escape, we conclude that stochastic star formation in the early universe likely plays a significant role in the contribution of galaxies to cosmic reionization.

\end{abstract}



\section{Introduction} \label{sect:intro}
The contribution of ionizing sources to the process of cosmic reionization is a central topic of extra-galactic astrophysics and a key goal of the JWST mission. With the first years of JWST data, significant progress has been made in two of three necessary quantities to constrain this contribution: the star formation rate (SFR) density and the production of ionizing photons (per unit SFR). The SFR density during the epoch of reionization (EoR) is determined by the UV luminosity function (LF), which has been robustly constrained with photometry \citep{Donnan2024UVLF,Finkelstein2024UVLF,Asada2025UVLF} and spectroscopy \citep{Harikane2025UVLF_spec} up to $z\sim 13$. Similarly, the production of ionizing photons has been quantified by recent works \citep{Simmonds2023xi_ion_LAEs_z6,Pahl2025xi_ion,Llerena2025_xi_ion,Hayes2025UVspectra_highz}, with excellent statistics, revealing only slight or no evolution with redshift. The third necessary quantity, the escape fraction of ionizing photons, \fesc, remains elusive. 

Ionizing flux cannot be directly observed in the EoR, due to absorption of ionizing photons by the intervening neutral HI gas in the inter-galactic medium (IGM). Therefore, \fesc must first be understood at lower redshifts, and indirect tracers established, which we can use to quantify the escape of ionizing photons within the EoR. Significant progress has been made in recent years by the Low-z Lyman Continuum Survey (LzLCS: \citealt{Flury2022LyC_lowz_survey}, see also \citealt{Chisholm2022LyC_beta,LeReste2025LaCOS}) --- which has studied in detail 35 Lyman continuum emitters (LCEs) at $z\sim0.3$ --- and by other studies \citep[][see \citealt{Jaskot2025LyCreview} for a review]{Izotov2016LyC_compactgals, Verhamme2017LyC_strongLAEs_lowz, Izotov2020lya_lowz_SFGsOIII/OII, Izotov2022Lya_LyC_MgIISFGs, Izotov2024lya_lowz_Zpoor}. 
The most important findings are that (i)~\lya properties, such as equivalent width (EW) and escape fraction (\flya), correlate well with \fesc \citep{Henry2015LyC_Lya_GPs,Verhamme2017LyC_strongLAEs_lowz,Steidel2018KLCS,FLury2022LzLCS_results} and that (ii) the \lya\ line separation from the systemic redshift and the halo fraction show a strong anti-correlation with \fesc \citep{Verhamme2015Lya_to_get_LyC,Izotov2018LyC_OIII/OII,Izotov2020lya_lowz_SFGsOIII/OII,FLury2022LzLCS_results,SaldanaLopez2025HF}. Combined, these results connecting \lya and LyC emission highlight the importance of clear channels in HI column density in allowing the escape of Lyman photons from the interstellar medium of the host galaxy. 

Other galaxy properties that have been found to correlate well with \fesc include SFR surface density, $\Sigma_{\mathrm{SFR}}$ \citep{Izotov2016LyC_compactgals,Vanzella2018Ion3,FLury2022LzLCS_results,Jaskot2024multivariateLyC_lowz}, steep, blue UV slopes \citep[$\beta$,][]{Chisholm2022LyC_beta,Ji2025dust_LyC}, the strength of the [O~III]/[O~II] ratio \citep{Izotov2018LyC_OIII/OII,Izotov2020lya_lowz_SFGsOIII/OII,Jaskot2024multivariateLyC_lowz}, and the presence of strong UV lines such as CIV $\lambda1550$ \citep{Schaerer2022LyC_CIV,Saxena2022LyC_CIV,Izotov2024_CIV_low_metal}. High $\Sigma_{\mathrm{SFR}}$ likely contributes to the creation of clear channels crucial to LyC escape, steep UV slopes are attributed to the role of dust in blocking LyC photons, and High [O~III]/[O~II] ratios in LCEs have been associated with sensitivity to ionization parameter and feedback from massive stars. 

However, no single tracer alone reliably predicts the escape of ionizing photons, underscoring the need for a framework that both links the escape of LyC photons to available tracers and assesses the importance of different variables in tracking the escape of LyC photons, such as in \cite{Maji2022pred_LyC_fromLya,Jaskot2024multivariateLyC_lowz}. We have reached the point that such multivariate predictors can be used on the appreciable sample of low redshift LCEs collated in the literature and on simulated data \citep{Rosdahl2022SPHINX_LyC_EoR,Choustikov2024LyC_tracers}. However, due to a lack of data as comprehensive as the LzLCS for a statistically significant population at $z>2$, we are not yet at this stage at these redshifts.

At $z>2$, which we term ``high redshift'' for the purposes of this work, a number of open questions remain. Recent studies have cast doubt on the ability of \lya properties to reliably trace \fesc at high-$z$ \citep{Rutkowski2017,Kerutt2024LAE_LyC,Citro2025LAEs_noLyC}. For instance, \cite{Citro2025LAEs_noLyC} found galaxies with strong \lya emission but without traces of LyC emission at the expected strength based on low-z estimators. \cite{Kerutt2024LAE_LyC} found a diversion from the well-established relation at low redshift between \fesc and \lya peak separation \citep{Izotov2018LyC_OIII/OII} as well as large scatter in the well-established \lya EW and \fesc relation \citep{Steidel2018KLCS}. 

Therefore, doubts persist over the applicability of low-redshift \fesc tracers during the EoR, which is their primary purpose. As $\mathbf{\sim10\,\mathrm{Gyrs}}$ of evolution have taken place between the EoR and low-redshift measurements, there are numerous factors which may influence the universal validity of \fesc tracers. Examples include the merger fraction, known to increase with redshift \citep{Rodriguez-Gomez2015Illustris_mergers}, the evolution of the star formation efficiency \citep{Madau1996SF_history}, and the UV background \citep{Fan2006UVBackground_EoR}. It is therefore important to search for LCEs as close to the EoR as possible. 

In this work, we describe the detection of an LCE at $z=\redshift$, within \timetoEoR of the EoR, in the MUSE eXtremely Deep Field (MXDF; \citealp{bacon2022musedatareleaseII}), which we henceforth refer to as \objname. 
In \S\ref{sect:MXDF} we describe the MUSE data used to detect the \lya line, which confirms the redshift, as well as the HST and JWST data used to detect the LyC flux and characterize the spectral energy distribution (SED) of the galaxy. \S\ref{sect:candidate} describes \objname and the necessary modeling to derive its properties: SED fitting and modeling of IGM transmission. We evaluate all accessible \fesc tracers which have been established at low redshift, focusing on the important \lya-LyC connection, in \S\ref{sect:tracers}. Finally, we interpret this discovery in relation to the transmission of the IGM and the properties of \objname in \S\ref{sect:discussion}, offering conclusions in \S\ref{sect:conclusions}.

Throughout this paper, we adopt a value for the \textit{Hubble} constant of $H_0=70\,\mathrm{km\,s^{-1}\,Mpc^{-1}}$, and the cosmology used is $\Omega_{\Lambda}=0.7$ and  $\Omega_{\mathrm{m}}=0.3$. All IMFs used are stated throughout the text and magnitudes are given in the AB mag system \citep{OkeGunn1983}. All equivalent widths are given in the rest frame unless otherwise stated. 

\section{Data and Analysis} \label{sect:MXDF}
The MUSE eXtremly Deep Field (MXDF; GTO Program 1101.A-0127, PI R. Bacon, \citealt{bacon2022musedatareleaseII}) is a region of extremely deep observations by the MUSE integral field spectrograph on the VLT, with the deepest area having a total exposure time of 141h. MUSE has a field of view of $1\times1\arcmin$ with a spatial pixel size of $0\farcs2$. The spectra have a wavelength range of $4700-9350~\mathrm{\AA}$ and a spectral pixel size of $1.25~\mathrm{\AA}$. This wavelength coverage enables MUSE to detect \lya emission in the redshift range $2.9<z<6.7$. All MXDF data is publicly available\footnote{\url{https://amused.univ-lyon1.fr}} and viewable using a web interface\footnote{\url{https://amused.univ-lyon1.fr/project/UDF/browse}}. 

The MXDF is situated within the footprint of the Hubble Ultra Deep Field \citep[HUDF;][]{Beckwith2006HUDF}, with additional public HST ACS/WFC and WFC3/UVIS imaging from programs including GOODS \citep{Giavalisco2004GOODS}, CANDELS \citep{Grogin2011,Koekemoer2011CANDELS}, 3D-HST \citep{Brammer2012.3DHST}, UVCANDELS \citep{Wang2025UVCANDELS}, UVUDF \citep{Teplitz2013UVUDF}, and HDUV \citep{Oesch2018HDUV}. The HST filters used in this analysis are WFC3/UVIS F336W and ACS/WFC F435W, F606W, F775W, F814W, and F850LP. The F435W filter, in which \objname's LyC emission is detected, has an average depth of $150\,\mathrm{ks}$. These data have been reprocessed as part of HST archival program PID 16621 (PI: Koekemoer), including calibration improvements from updated darks, biases, flatfields and astrometric alignments, as well as other low-level improvements including improved removal of cosmic rays, satellite trails, and bad pixels.

This field also  benefits from public JWST imaging from the JADES and JEMS programs \citep{Eisenstein2023JADES,Rieke2023JADES_DR1,Williams2023JEMS}\footnote{DOIs: \cite{RiekeJADESdoi,WilliamsJEMSdoi}}, where the JWST/NIRCam filters from these programs that we use are: F090W, F115W, F150W, F182M, F200W, F210M, F277W, F335M, F356W, F410M, F430M and F444W. F460M and F480M are also included in the JADES imaging data release \citep{Rieke2023JADES_DR1}; however, we exclude these filters from our SED fitting, as visual inspection revealed elevated background levels relative to the source signal in these bands, rendering the photometry unreliable and the associated uncertainties likely underestimated.

\subsection{Photometry of \objname} \label{subsect:phot}
\begin{figure}
    \centering
    \includegraphics[width=\linewidth]{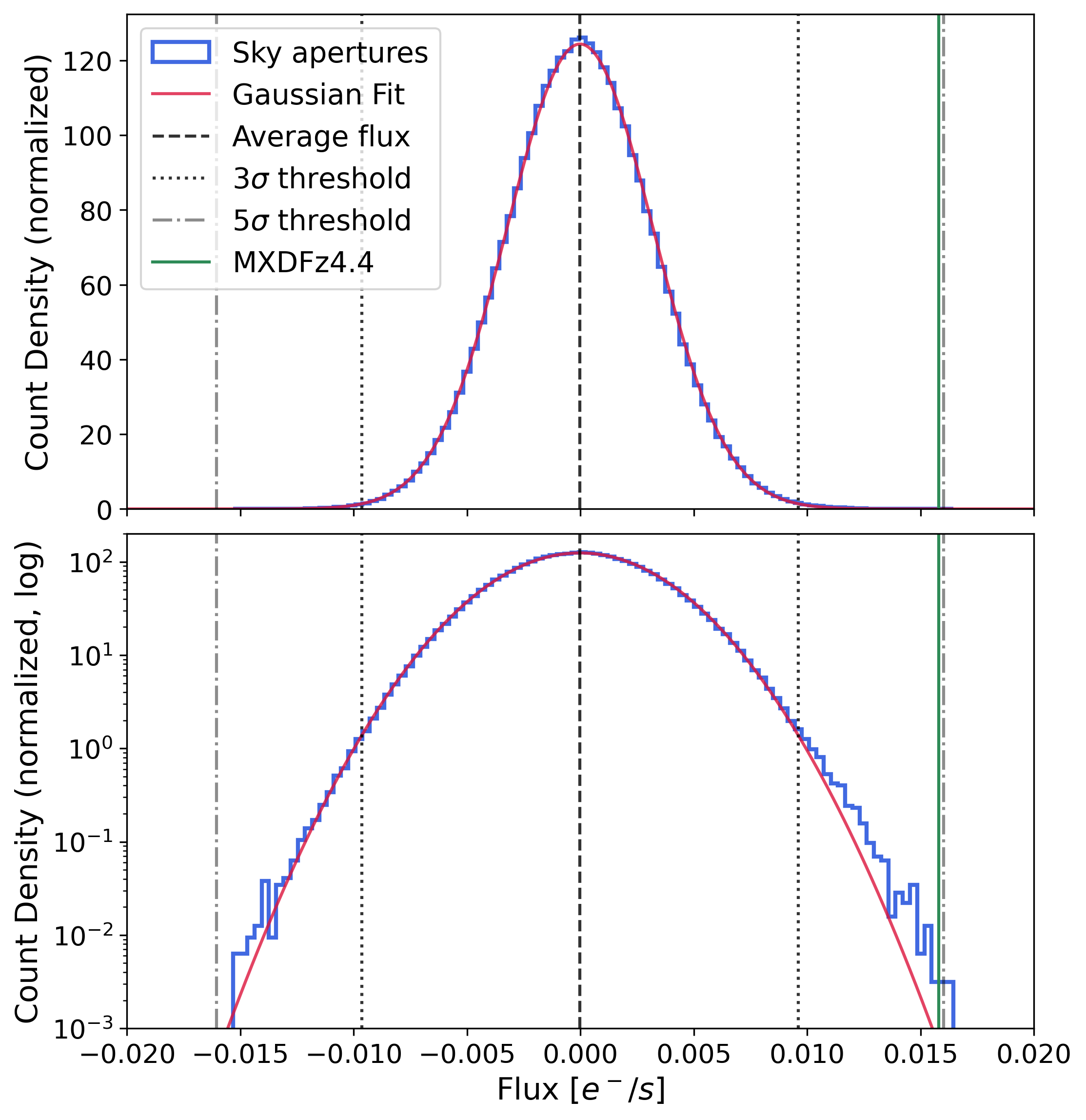}
    \caption{The flux in 1,000,000 empty sky apertures is shown as a histogram along with a Gaussian fit to those fluxes. The detection flux of \objname is shown as the green solid line, coincident with the $5\sigma$ threshold shown in the dot-dashed gray line. The agreement between this empirical test and the formal uncertainty from the RMS map demonstrates the robustness of our LyC detection in F435W. The lower plot has a logarithmic y axis to better visualize the high-flux tail of the distribution.}
    \label{fig:skyap}
\end{figure}
We perform custom photometry on the full-resolution images to maximize the resulting flux SNR of \objname. The segmentation map released by the JADES collaboration blends \objname with a faint neighbor to the southeast, so we create a custom version where these sources are carefully de-blended. We perform Kron photometry \citep{Kron1980photometry} on the full-resolution images and correct these values by 10\%, consistent with curve-of-growth analysis checks we performed. We note that the neighbor is very faint, so this has minimal effects on the photometry published in \cite{Rieke2023JADES_DR1}. 

As the emission in F435W is spatially far more compact than in redder filters, we apply a different method to maximize the SNR in the F435W image, similarly to the method used by \cite{Rafelski2015, Sun2024} for faint UV photometry of the UVUDF and UVCANDELS surveys. Specifically, we calculate an isophotal flux using a segmentation map based on the F606W filter, a neighboring filter where \objname is more compact than the detection image used by JADES, but with high enough SNR to create a reliable segmentation map. This smaller aperture fully captures the flux in the F435W filter and reduces the contribution from sky dominated pixels, thereby increasing the SNR of the flux measurements. The uncertainty on the flux within the aperture is calculated using the F435W RMS map and propagated with the uncertainty on the sky background subtraction, which is done in a wide annulus around \objname, ensuring no flux from any object is taken into account.

We connect the F435W flux to the rest of the photometry by using an aperture correction based on the larger JADES-based isophotal area. This correction is determined by taking the ratio of the isophotal flux measured in the smaller and larger isophotal areas in the F606W image. This correction assumes that the distribution of flux is similar in F606W and F435W, which is confirmed both visually and by comparison with the F435W flux measured in the larger isophotal area. The two flux measurements agree within the smaller isophotal area uncertainties as expected. This F606W isophotal area flux measurement is then used to define the SNR of the detection. The F435W detection significance is \finalsigma.

To verify that correlated noise or large-scale noise patterns do not affect this detection, we calculate the flux in 1,000,000 empty sky apertures of the same size as the aperture used to detect \objname's F435W flux. Apertures are placed at random positions across the image, avoiding regions containing detected sources. A histogram of these fluxes is shown in Fig.~\ref{fig:skyap}, with \objname's detection flux shown with a green solid line coincident with the $5\sigma$ threshold shown in the dot-dashed gray line. The width of this distribution agrees with the formal uncertainty from the RMS map, indicating that correlated noise and large-scale patterns do not significantly affect the LyC flux measurement in F435W.

We correct for the effects of the point spread function (PSF) by applying published encircled energy corrections (similar to PSF matching) to the resultant Kron photometry, which is reasonable given the compactness of our galaxy\footnote{HST: \url{https://www.stsci.edu/hst/instrumentation/acs/data-analysis/aperture-corrections}}\footnote{JWST:~\url{https://jwst-docs.stsci.edu/jwst-near-infrared-camera/nircam-performance/nircam-point-spread-functions\#NIRCamPointSpreadFunctions-Encircledenergiesforresampleddata}}. The alternative standard method of PSF matching the images before measuring the photometry would reduce the SNR of the F435W photometry, as it would smooth the flux over the larger F444W PSF and thus increase our flux uncertainty. We confirm the validity of this method in two ways. First, we check the curves of growth of the photometry including the encircled energy correction and find convergence at large radii. Second, we compare the photometry of the redder HST and JWST photometry with this method to that of the PSF matched photometry from JADES and find good agreement. Therefore, this Kron flux corrected to the total flux (10\%) plus the encircled energy correction is used as our final flux measurement. We propagate the uncertainties from each correction accordingly, together with the uncertainty in aperture flux and background subtraction. The resultant LyC flux measurement in F435W is \lycflux, at \fluxsigma. We subsequently use this photometry for the SED fitting described in \S\ref{subsect:SEDfitting}.

\subsection{Spectral Extraction of \objname from the MXDF datacube}
The MUSE consortium makes their spectra public, the extraction of which follows \cite{bacon2022musedatareleaseII}. We briefly outline this procedure here, specifically that which relates to extraction using the ORIGIN software \citep{Mary2020ORIGIN}, as this is the method used for \objname. The ORIGIN software is optimized to detect faint line emitters in MUSE datacubes. Care is taken to mask any bright continuum sources in a three-stage process. The first two stages involve segmenting and masking out any bright, noisy sources, firstly on the datacube itself, creating a white light continuum image, then on the S/N residual image thereof. Principal Component Analysis is then used iteratively to clean further residuals from each spaxel in the cube. Emission lines are then detected by filtering the datacube with possible spectral line profiles. Each source is then extracted from the original datacube using the algorithm from \cite{Horne1986CCDspectroscopy}. These extracted spectra are publicly available for all line-emitters detected in the MXDF. We take the the spectrum for \objname and apply a custom fitting process which is described in Sect.~\ref{subsubsect:Lya_line}. This process also follows the line-fitting procedure for \lya lines described in \cite{bacon2022musedatareleaseII} but allows us to extract and visualize all necessary \lya properties.


\section{Physical properties of \objname} \label{sect:candidate}

\begin{figure}
    \centering
    \includegraphics[width=\linewidth]{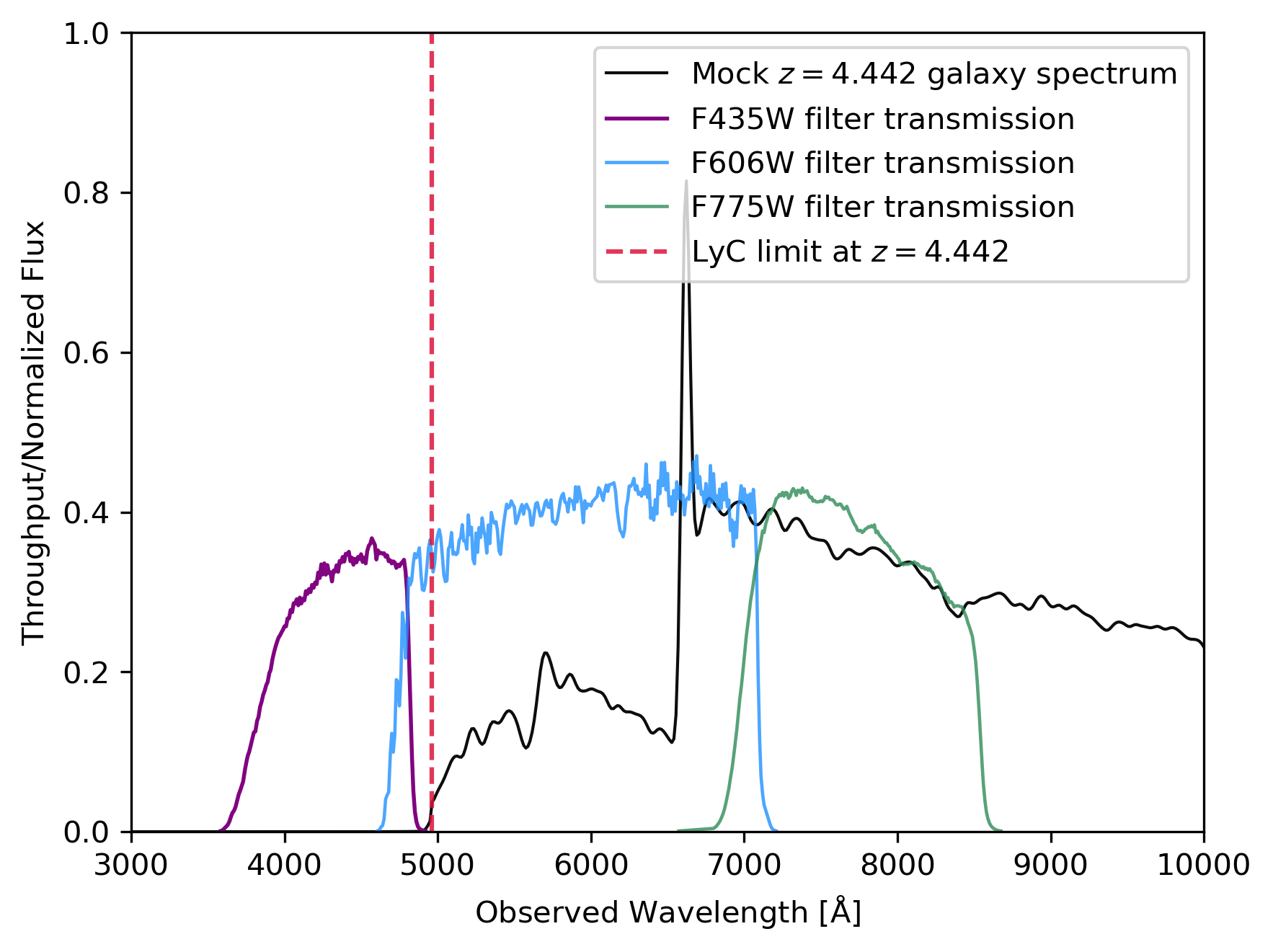}
    \caption{A mock SED of an LAE taken from the JAGUAR mock catalog \citep{Williams2018JAGUAR}. The LAE has the same redshift as \objname and the F435W filter transmission is plotted in purple, along with a dashed red line delineating the Lyman limit at $z=4.442$. The F435W filter probes only LyC emission at this redshift. The F606W and F775W filters are also shown, and used for the bands that see the \lya and rest frame UV emission respectively.}
    \label{fig:Filters}
\end{figure}

\begin{figure*}
    \centering
    \includegraphics[width=0.98\linewidth]{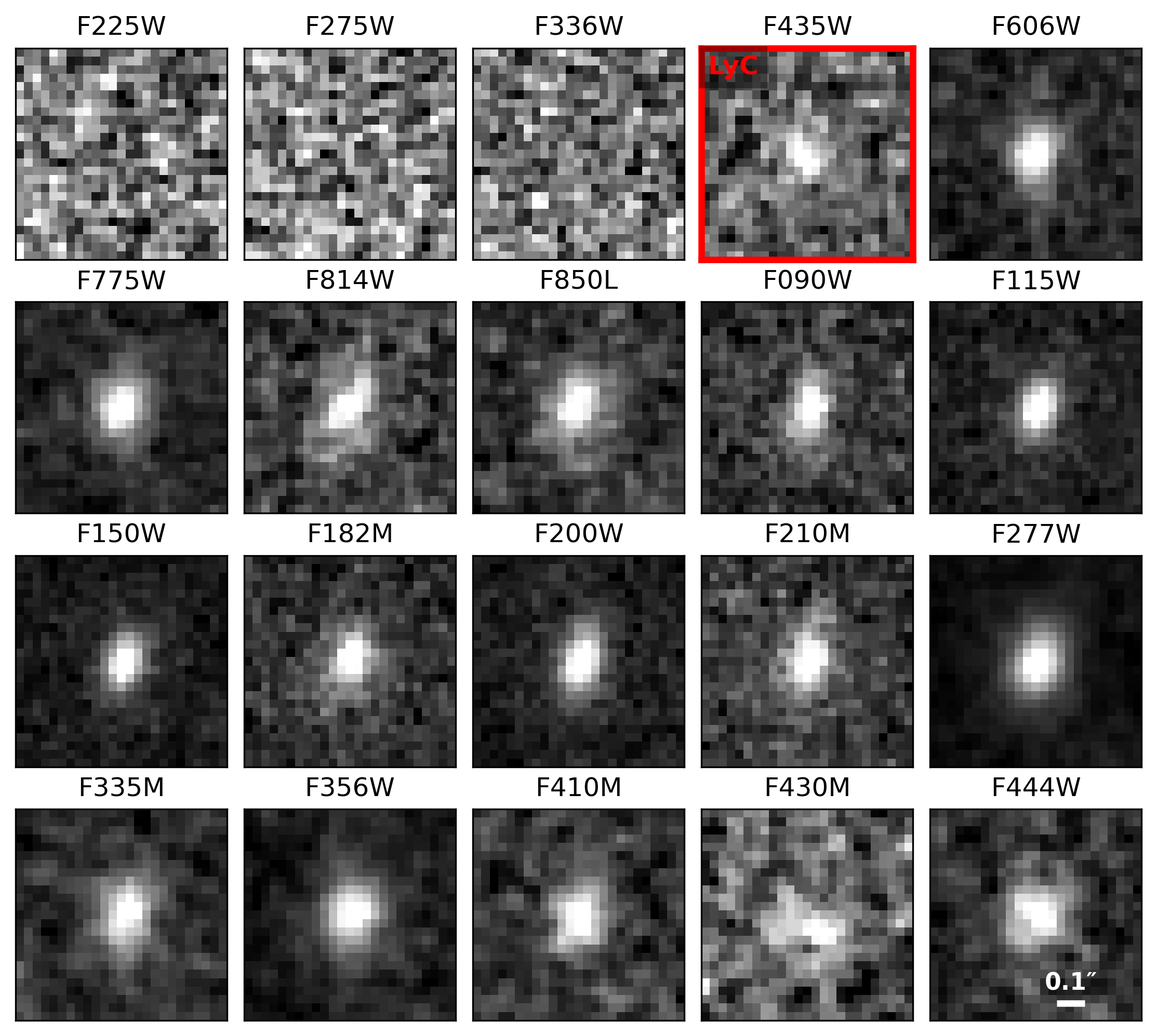}
    \caption{Cutouts of \objname in all filters used for SED fitting, as well as F225W, F275W and F336W to show non-detections. The reduction of the HST data is outlined in Sect.~\ref{sect:MXDF} and the remaining (JWST) filters are from the JADES program \citep{Rieke2023JADES_DR1}. All filters have a pixel scale of $0\farcs03$. The LyC flux in F435W, at \fluxsigma significance, is clearly visible.}
    \label{fig:cutouts}
\end{figure*}

\begin{figure*}
    \centering
    \includegraphics[width=1\linewidth]{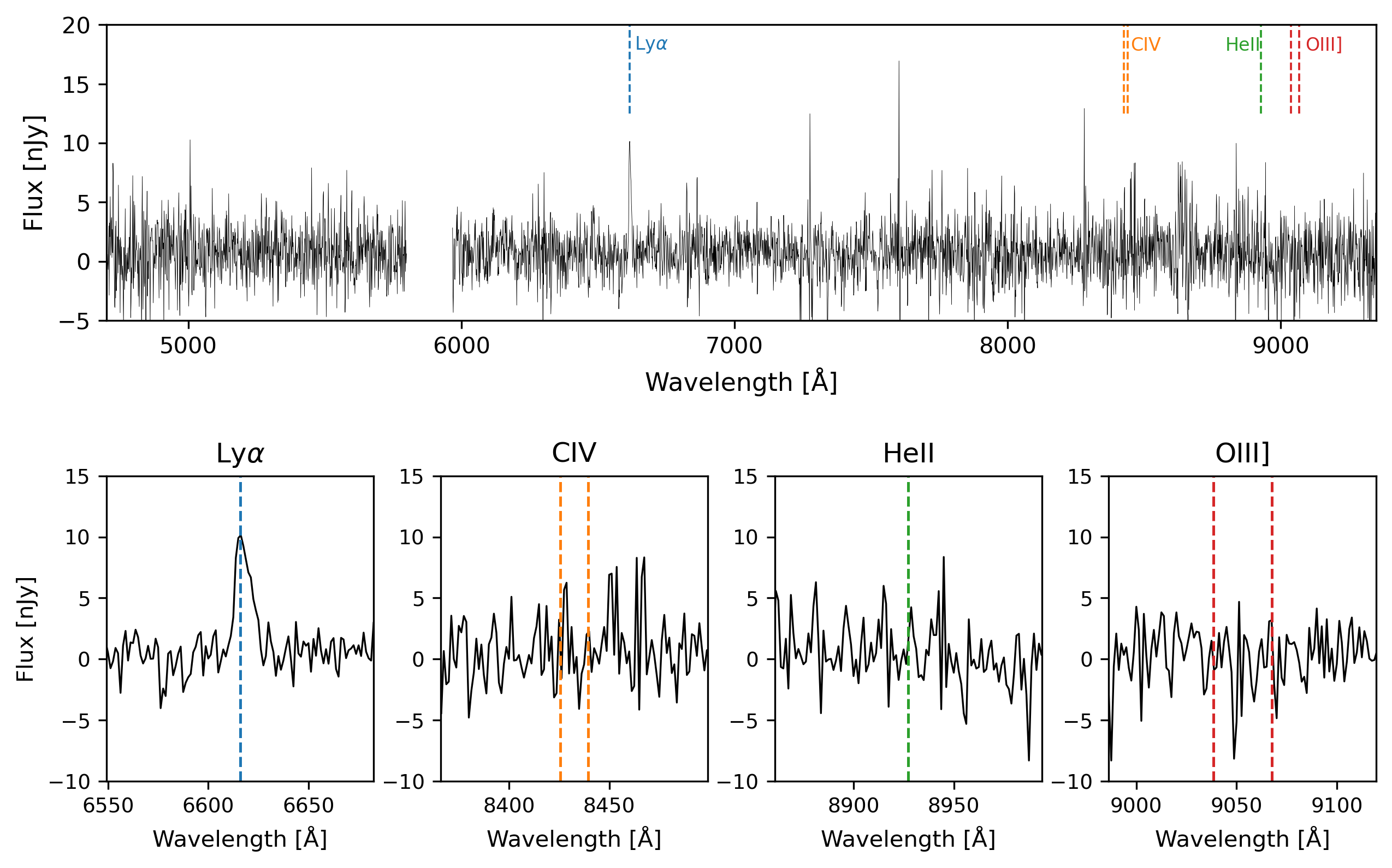}
    \caption{MUSE spectrum of \objname. Cutouts of the UV lines are shown enlarged below the main spectrum. The \lya line at $6615.9\,\mathrm{\AA}$ is evident from its asymmetric line shape. No other emission lines are detected in the MUSE spectrum.}
    \label{fig:MUSEspectrum}
\end{figure*}

\objname is located in the deepest section of the MXDF (exposure time $\sim140\,\mathrm{hrs}$) at an RA and DEC of \ra, \dec, and a redshift (from the \lya emission line) of \redshift. This makes it the highest-redshift directly-observed LCE to date. 

\objname has an observed ratio of ionizing to non-ionizing flux, $\frac{F_{\mathrm{LyC}}}{F_{1500}}$ (in $F_\nu$ units), of \fluxratio, calculated from the F435W to F775W flux ratio. We adopt the nomenclature $F_{\mathrm{LyC}}$ to denote the LyC flux detected in F435W, rather than $F_{900}$, to differentiate from studies where the flux at exactly $900\,\mathrm{\AA}$ is used from spectra. The F435W band covers up to $900\,\mathrm{\AA}$ rest frame, but as it starts below $700\,\mathrm{\AA}$ we term this wavelength range ``LyC''. A diagram of a $z=4.442$ LAE SED taken from the JAGUAR suite of mock SEDs \citep{Williams2018JAGUAR}, the F435W filter transmission, and Lyman limit at $912\,\AA$ are shown in Fig.~\ref{fig:Filters}.

The \fesc values that we derive, depending on the intrinsic ratio of ionizing to non-ionizing flux and the IGM attenuation, are presented in Table~\ref{tab:obj_details} and discussed in \S\ref{subsect:escape_fraction_results}. The properties of \objname derived from SED fitting are detailed in \S\ref{subsect:SEDfitting}, while the detailed analysis of the \lya emission line observed by MUSE is detailed in \S\ref{subsect:Lya-LyC}. The HST and JWST imaging of \objname is shown in Fig.~\ref{fig:cutouts} and the MUSE spectrum is shown in Fig.~\ref{fig:MUSEspectrum}. 

While the spectrum shows only one emission line, \objname has a confident redshift as the line has an asymmetric profile, typical of \lya emission lines. We include a detailed discussion how we rule out other possible redshifts in Appendix~\ref{appendix:interlopers}. Also in this appendix, we calculate the probability of chance alignment with a low-redshift interloper ($0.0076\%$) and present the details of the resolved nature of the flux in F435W, which rules out the possibility of the LyC flux originating from a supernova in a faint, low-redshift host galaxy.

No lines other than \lya are detected in the MUSE spectrum. A redder spectrum of \objname obtained with JWST NIRISS Wide-Field slitless spectroscopy from the NGDEEP program also exists (PID 2079; PIs: S. Finkelstein, C. Papovich, N. Pirzkal, \citealt{Bagley2024NGDEEP}). No lines are detected in the NIRISS spectrum, which is to be expected (see Appendix~\ref{appendix:NGDEEP} for the full details).


\begin{table*}[ht!]
\caption{Escape fraction of \objname and SED-fitting derived properties \label{tab:obj_details}}
\centering

\begin{tabular}{cccccc}
\hline
\hline
$f_{\mathrm{LyC}}/f_{1500}$ & $L_{1500}/L_{\mathrm{LyC}}$  & $T_{\mathrm{IGM}}$ & $f_{\mathrm{esc}}^{\mathrm{rel}}\,[\%]$ & $A_v$ &$f_{esc,abs}^{\mathrm{IGM}}\,[\%]$ \\
\hline
\fluxratio & 3\tablenotemark{a} & 0.18 & $300\%$ & $0.34\tablenotemark{b}-0.68$\tablenotemark{c} & $160 - 220$ \\
 & 1.0-1.5\tablenotemark{d} & 0.18 & $100-150\%$ & $0.34\tablenotemark{b}-0.68$\tablenotemark{c} & $53-109$ \\
 & \cgLratio\tablenotemark{b} & 0.18 &  &  & \cgigmcorr \\
 & 1.0-1.2\tablenotemark{c} & 0.18 &  &  & $100-120$ \\ 
\hline
\end{tabular}

\vspace{0.1cm} 

\begin{tabular}{cccc}
\hline
\hline
$M_{\mathrm{UV}}$ & $\mathrm{SFR_{10}\,[M_{\odot}/yr]}$ & $\mathrm{SFR_{100}\,[M_{\odot}/yr]}$ & $\mathrm{M_{\star}\,[M_{\odot}]}$ \\
\hline
$-18.2\pm 0.1$ & $11.8\pm0.6$\tablenotemark{b} & $2.4\pm0.1$\tablenotemark{b} & $\mathrm{10^{8.3\pm0.2}}$\tablenotemark{b} \\
 & $1.4-3.3$\tablenotemark{c} & $0.29-1.2$\tablenotemark{c} & $\mathrm{10^{7.5-8.0}}$\tablenotemark{c} \\
\hline
\end{tabular}

\tablenotetext{a}{Value from literature.}
\tablenotetext{b}{Value from fitting with BC03 models \citep{Bruzual2003stellar_pops}.}
\tablenotetext{c}{Value from fitting with BPASS models \citep{Eldridge2017}.}
\tablenotetext{d}{Value from pySTARBURST99 models \citep{Hawcroft2025pySTARBURST}.}
\tablecomments{A table of \objname's properties related to escape fraction and properties of the host galaxy derived from SED fitting with \cg (see text). The upper table pertains to escape fraction properties, with the first two rows using luminosity ratios derived from stellar population models, a value commonly assumed in the literature, 3 \citep{Steidel2001LyC_z3,Kerutt2024LAE_LyC} and the values derived from young, metal-poor pySTARBURST99 models. Both these rows use the range of dust attenuation values derived by our SED fitting with \cg. The final escape fractions are given as a range based on the ranges of $f_{\mathrm{esc}}^{\mathrm{rel}}$ and $A_v$. The bottom two rows use the luminosity ratios directly from \cg, which already have dust attenuation included. The luminosity ratio from BPASS is given as a range based on different runs (see \S\ref{subsect:SEDfitting}). All these results use a $T_{IGM}$ value of 0.18 (see text). The lower table contains SED fitting-derived properties of the host galaxy, with the first row using the BC03 models and the second row BPASS. BPASS-derived values are given as a range, based on the values derived in all BPASS runs (see text).}

\end{table*}

\subsection{SED fitting} \label{subsect:SEDfitting}

We use the \cg SED fitting code to derive \objname's properties, fitting the observed photometry with a flexible, non-parametric star formation histories (SFH). In order to predict the ratio of UV to ionizing photons ($L_{1500}/L_{\mathrm{LyC}}$), which we require to calculate the intrinsic escape fraction of \objname, we do not include the IGM-affected bands (F606W and F435W; see \S\ref{subsect:escape_fraction_results}). As mentioned in \S\ref{sect:MXDF}, we also do not fit the F460M and F480M bands. We fit using two different stellar population synthesis models to assess the impact that this choice has on the derived properties.

The \cg SED fitting code \citep{Burgarella2005SED,Boquien2019cigale}, as used for this work, builds composite stellar populations based on flexible SFHs by combining stellar population synthesis models. In our runs, we use two different sets of models: \cite{Bruzual2003stellar_pops} (henceforth BC03) and BPASS v2.2 \cite{Elridge2008BPASS,Eldridge2017}. The BPASS models include the effects of binary populations on stellar evolution. Switching these binary effects on or off, as well as switching between the different extinction laws (MW, LMC, SMC), affects the best fit galaxy properties. Importantly, the affected properties include the intrinsic $L_{1500}/L_{\mathrm{LyC}}$, without a significant change in the reduced chi-squared of the fit. We therefore run CIGALE using the BPASS models in each of these cases (making six separate runs) and utilize the full range of results to report a more realistic value and uncertainty on every measurement.

\begin{figure}
    \centering
    \includegraphics[width=\linewidth]{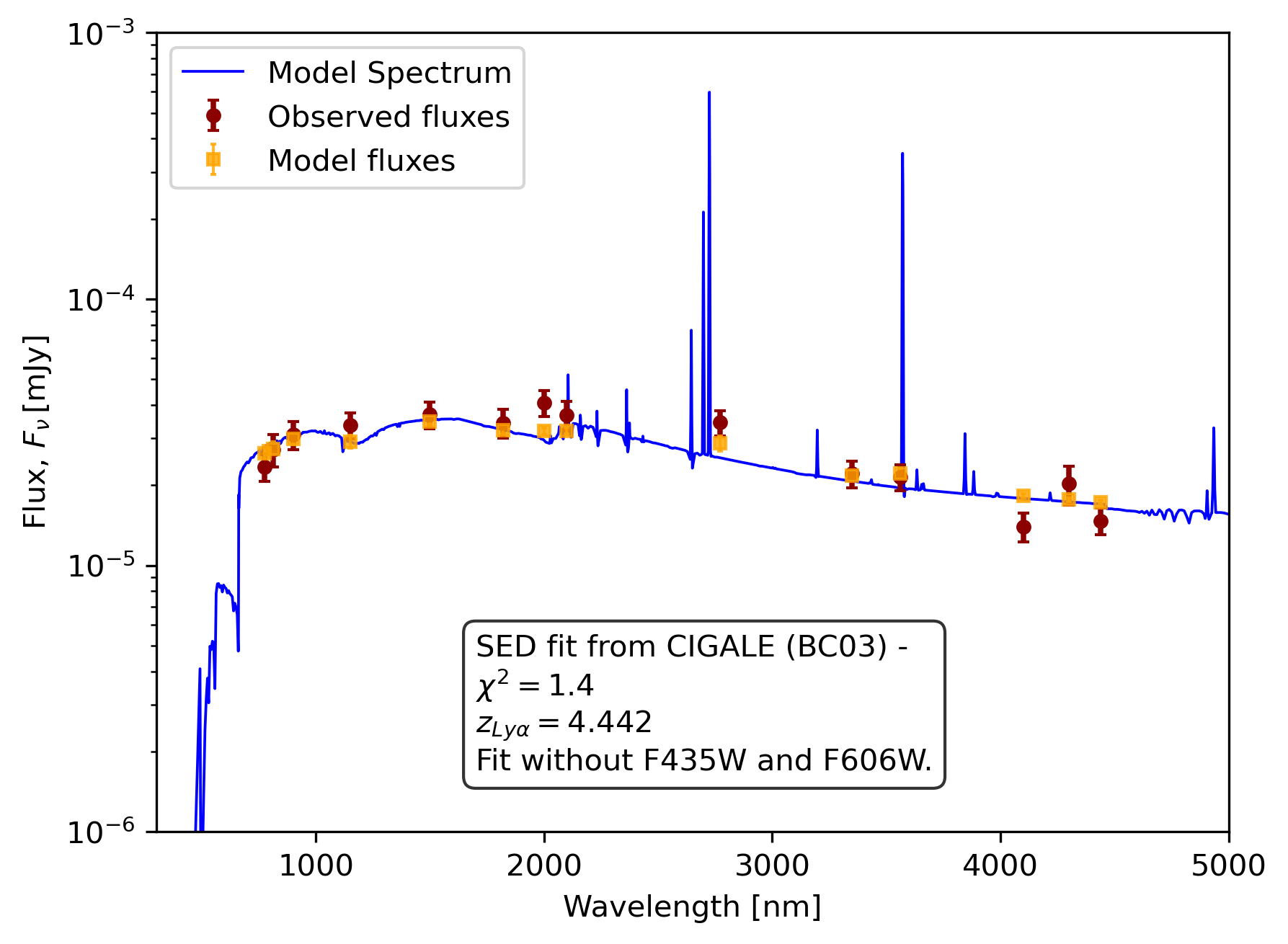}
    
    \includegraphics[width=\linewidth]{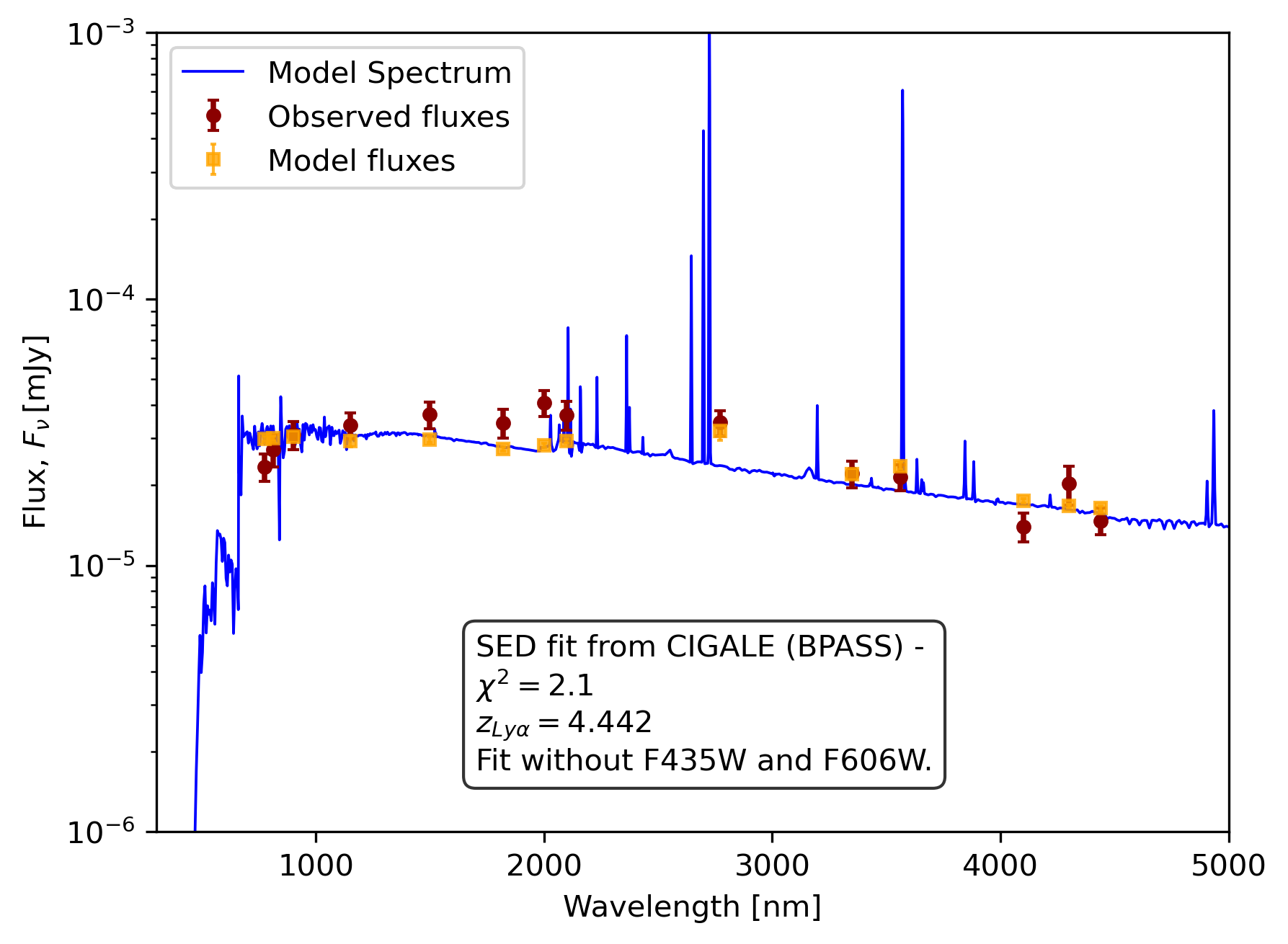}
    
    \caption{The best fit spectra of \objname from \cg using the BC03 and BPASS stellar population models \citep{Bruzual2003stellar_pops,Eldridge2017}, plotted in blue. The observations are overplotted in red and model fluxes in orange. The fit was performed without F435W and F606W in order to predict the IGM affected bands and derive $L_{1500}/L_{\mathrm{LyC}}$. The redshift was fixed to the value of $z$ = 4.442 as derived from the \lya emission line. }
    \label{fig:SEDfit}
\end{figure}

\cg also accounts for gas ionized by massive stars, as well as extinction and re-emission due to dust using a modified \cite{Calzetti2000dust_attenuation} attenuation law. We use a Chabrier IMF \citep{Chabrier2003IMF} and adopt a non-parametric SFH with 10 bins covering the past 600 Myr. The latest bin covers the most recent 5 Myr, with the remaining bins equally-sized in log space. The best-fit is found with an SFH that remains very low until $\approx$~10 Myr before the galaxy is observed, rising into a strong burst in the most recent 5 Myr. The resulting model can be used to predict the intrinsic emission in the LyC regime. By convolving the resulting best fit spectra with the F435W and F775W filter transmissions, we derive luminosity ratios, $L_{1500}/L_{\mathrm{LyC}}$, of \cgLratio and $1.0-1.2$ for the BC03 models and BPASS, respectively. We note that using a `delayed-tau' SFH model returns very similar results, i.e., the best fit is found to have a strong burst $\approx$~5 Myr in the past. The best fits using the BC03 and BPASS models (with the non-parametric SFH) are shown in Fig.~\ref{fig:SEDfit}. All SED fitting-derived results, including derived escape fractions (see also \S\ref{subsect:escape_fraction_results}), are given in Table~\ref{tab:obj_details}. Quantities derived using the BPASS models are given as ranges that encompass the values derived for all runs. 


\begin{figure}
    \centering
    \includegraphics[width=\linewidth]{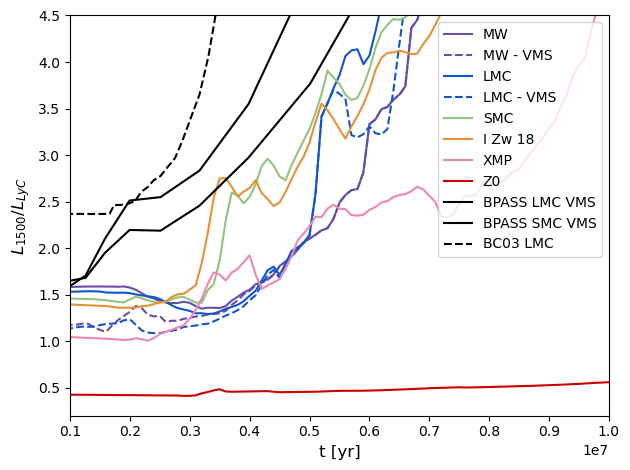}
    \caption{Evolution of $L_{1500}/L_{\mathrm{LyC}}$ over the first $10\,\mathrm{Myr}$ of a starburst, based on the models from pySTARBURST99 models (in color) \citep{Hawcroft2025pySTARBURST} as well as BPASS \citep{Elridge2008BPASS,Eldridge2017} and BC03 \citep{Bruzual2003stellar_pops} (in black). Values have been convolved with the F435W and F775W transmission curves and are in $F_\nu$ such that they are comparable to the observations. Metallcities range from Milky Way metallicity ($Z\sim0.14$) to XMP - extremely metal poor ($Z\sim10^{-5}$), with a zero metallicity line included for comparison. VMS indicates models including very massive stars, up to $300\,\mathrm{M_{\odot}}$.}
    \label{fig:pySTARBURST}
\end{figure}

Due to the likely impact of the recent burst of star formation on \fesc (due to stellar feedback clearing holes in the ISM; \citealt{Trebitsch2017feedback_LyC,Rahner2017stellar_winds_fesc,Carr2025rad+supernovae_LyC}), we additionally consider $L_{1500}/L_{\mathrm{LyC}}$ values taken directly from two different stellar population synthesis models of young stars ($\lesssim10\,\mathrm{Myr}$): BPASS \citep{Elridge2008BPASS} and the recent pySTARBURST99 models \citep{Hawcroft2025pySTARBURST}. Both codes incorporate effects that can be significant for the production of LyC photons. pySTARBURST99 includes the effects of stellar rotation as well as very massive stars (VMS), up to $300\,\mathrm{M_{\odot}}$ (see Fig.~\ref{fig:pySTARBURST}). BPASS takes into account the effect of binary stellar populations. 

\cg may struggle to fit a very young population due to a sparsity of very young star templates, additionally motivating an analysis using these specialized stellar models. Both population synthesis codes utilize a similar framework to make predictions of synthetic observables based on grids of input stellar evolution models and spectral libraries. 
While these codes adopt different stellar libraries, the only significant difference affecting the $L_{1500}/L_{\mathrm{LyC}}$ ratio should be the inclusion of synthetic VMS spectra: while pySTARBURST99 includes them directly, BPASS extrapolates these from a regular OB grid (relevant in the first $\sim3\,\mathrm{Myr}$). In Fig.~\ref{fig:pySTARBURST} we show the evolution of $L_{1500}/L_{\mathrm{LyC}}$ over the first $10\,\mathrm{Myr}$ for a range of metallicities after including the effects of VMS. The $L_{1500}/L_{\mathrm{LyC}}$ from \cg is consistent with the low end of the range shown, which supports the idea of a very recent burst in \objname. We show the escape fractions derived using the luminosity ratios from these models alongside the \cg-derived escape fractions in Table~\ref{tab:obj_details}. We discuss the consequences for \objname's escape fraction in \S\ref{subsect:escape_fraction_results} and \S\ref{sect:discussion}.

\subsection{IGM modeling} \label{sect:IGM}

The IGM is expected to absorb large fractions of ionizing photons at $z\gtrsim3.5$ \citep{Inoue2011IGM_UVspectra,Inoue2014IGM}. Despite this, LCEs have been confirmed at redshifts up to $z\sim4$ \citep{Vanzella2018Ion3,Mestric2025ion3} and candidates exist up to $z\sim4.4$ \citep{Prichard2022LCEs, Kerutt2024LAE_LyC}. The detection of \objname makes it imperative to assess whether it is possible to have an IGM sightline transmissive enough to allow us to observe the ionizing flux, or whether this result calls into question our models of IGM transmission near the EoR.  

We model the IGM using the TAOIST routine described in \cite{Bassett2021IGM_sim}. In brief, mock sightlines are populated with HI absorbers at randomized locations following the redshift-dependent column density distribution function (CDDF) given by \cite{Steidel2018KLCS}. We model individual HI absorbers with column densities between 10$^{12}$ and 10$^{21}$ cm$^{-2}$ which, at wavelengths redder than 912~\AA, produce Voigt profiles with varying Doppler widths (sampled from \citealt{Hui1999lya_profiles}). At bluer wavelengths, the absorption cross-section is proportional to $\lambda^{3}$ \citep{Osterbrock1974}. By integrating the transmission curve through our LyC band, we can calculate the transmission of the IGM (\tigm), i.e., the fraction of flux in F435W that would pass through the IGM along that mock sightline.

\begin{figure}
    \includegraphics[width=0.98\linewidth]{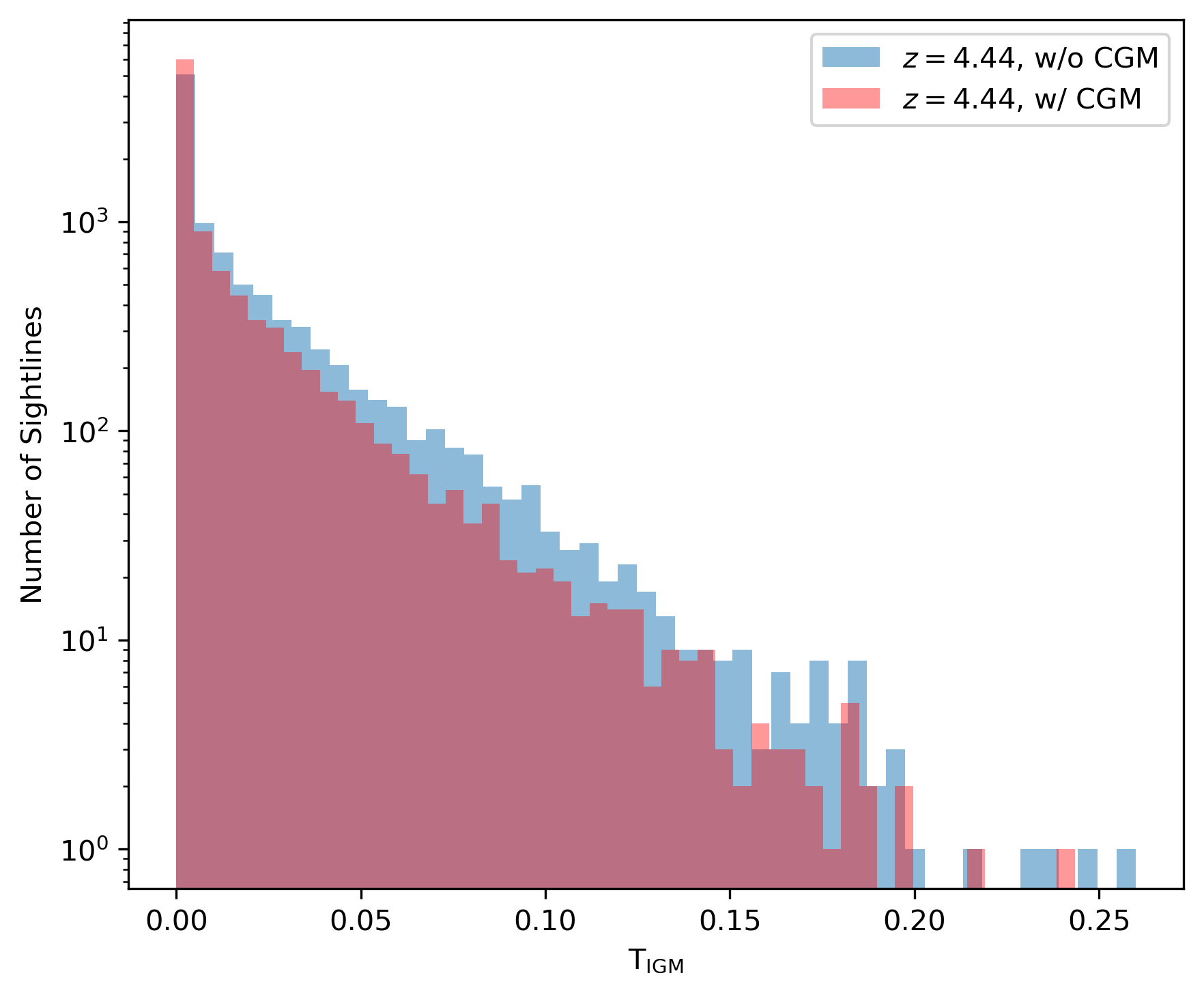}
    \caption{The transmission of the IGM (defined as $e^{\tau_{\mathrm{IGM}}}$) through F435W to the redshift of \objname, based on 10,000 mock sightlines modeled using the TAOIST code as described in S\ref{sect:IGM}. We show the results including and excluding the CGM of \objname.}
    \label{fig:TIGM_dists}
\end{figure}

In order to adequately sample the \tigm distribution, we produce 10,000 mock sightlines. The TAOIST code allows for a different CDDF in the CGM of the galaxy, where the higher gas density leads to more high-column-density absorbers. As the redshift of \objname is based on the Ly$\alpha$ line, it is not clear whether there is Ly$\alpha$ emission at the systemic redshift. We therefore produce two sets of mock sightlines, where one includes absorption from the CGM of \objname and the other assumes a clear channel through the CGM along which LyC photons can escape. The \tigm distribution from both sets of sightlines is shown in Figure \ref{fig:TIGM_dists}. Given that \objname must be in a sightline with higher than average transmission, we select the $3\sigma$ value from our list of modeled sightlines for use in deriving our escape fractions. This value is $T_{\mathrm{IGM}}=0.18$. 


The CDDF used in this modeling was produced using measured HI absorbers at 2.0 $\lesssim$ z $\lesssim$ 2.8 \citep{Rudie2013}, with a power law used to denote the redshift evolution of the CDDF. Extrapolating this redshift evolution to z $>$ 4 may not produce accurate results, as IGM conditions may change close to the EoR. The CDDF also becomes more difficult to measure at low column densities as we consider higher redshifts, as individual absorbers become more difficult to resolve in the Ly$\alpha$ `jungle' \citep{Bielby2020lya_CGM}. This is particularly important as the IGM transmission mostly depends on the prevalence of the lower-column-density absorbers, i.e., the incidence rate of high-column-density Lyman Limit Systems (LLSs) along the line of sight will determine the absorption of LyC photons \citep{Inoue2014IGM}. We also note that this simulation does not include the effect of the large scale structure, which may result in a systematic bias in the estimate of \tigm\ \citep[][]{Scarlata2025}.

An estimate of the \lya transmission in the MXDF, at an impact parameter of $\sim100\,\mathrm{kpc}$ from \objname, already exists \citep{Matthee2024MXDF_lya_transmission}. The authors find a \lya transmission of $0.292\pm0.008$ at $z=4.46$ using the absorption spectrum of a bright LAE at $z=4.77$, making this more transmissive than typical quasar sightlines at this redshift. We take the transmission spectrum of this LAE (MUSE ID 53) and attempt to derive a LyC transmission from it. We obtain $0.05\pm0.03$, a more-than-average transmissive sightline at $z=4.44$. However, the spectral resolution of quasar spectra normally used to study HI gas at these redshifts is a factor of $\sim10$ greater than MUSE \citep{Becker2015quasar_abs_lines,Lofthouse2020MAGG,Bielby2020lya_CGM}. Additionally, the SNR of quasar spectra is generally far higher. We therefore cannot determine the H~I column density with sufficient accuracy to reliably constrain the LyC transmission. Furthermore, the correlation between the sightlines to MUSE ID 53 and \objname may be poor due to the impact of CGM-scale absorbers, which can have a large impact on the LyC transmission (e.g. \citealt{Rudie2013, Inoue2014IGM}). We therefore choose not to trust this derivation of $\mathrm{T_{IGM}}$ and rely on the modeling previously described in this section.


\section{The escape fraction of \objname} \label{subsect:escape_fraction_results}
In this section we start by defining the terms used in the explanation and calculation of the LyC escape fraction. This is followed by the measured and derived quantities for \objname. 

\subsection{Escape fraction definitions}
The observed flux ratio between the photometric bands which measure the flux bluewards of $912\,\mathrm{\AA}$ (LyC flux) and the UV flux is denoted by $f_{\mathrm{LyC}}/f_{\mathrm{UV}}$. This contains no model assumptions and is purely an observed quantity. However, the effect of the absorption by the IGM is not taken into account, neither is the intrinsic ratio between the ionizing and non-ionizing flux. These quantities are captured in the relative escape fraction \citep{Steidel2001LyC_z3,Siana2007LyC_z1};
\begin{equation}\label{eq:fescrel}
    f_{esc,rel} = \left( \frac{f_{\mathrm{LyC}}}{f_{\mathrm{UV}}}\right)_{observed}\left(\frac{L_{\mathrm{UV}}}{L_{\mathrm{LyC}}}\right)_{intrinsic}e^{\tau_{\mathrm{IGM}}}.
\end{equation}

We note that the exponential factor at the end of this expression is equal to $\mathbf{1/T_{\mathrm{IGM}}}$; $\mathrm{T_{IGM}}$ is plotted in Fig.~\ref{fig:TIGM_dists}. Both this and the UV-to-ionizing luminosity ratio are model dependent quantities, the first derived by the TAOIST modeling (\S\ref{sect:IGM}) and the second by our SED fitting (\S\ref{subsect:SEDfitting}).

Typically, a further quantity is defined to account for the effect of absorption by dust in the galaxy, termed the absolute escape fraction;
\begin{equation}\label{eq:fescabs}
    f_{esc,abs}^{\mathrm{IGM}}=f_{esc,rel}\times10^{-0.4A_\mathrm{v}},
\end{equation}

where $A_\mathrm{v}$ is the dust attenuation in the rest frame UV, another model-dependent quantity. We mark this quantity with the superscript IGM to denote that it is corrected for IGM absorption. 

We present two methods of deriving \objname's escape fraction. The first assumes an intrinsic luminosity ratio in Eq.~\ref{eq:fescrel}, which comes from stellar population modeling, and then we correct for dust using Eq.~\ref{eq:fescabs}. We adopt an intrinsic luminosity ratio of 3, common in the literature \citep{Steidel2001LyC_z3,Kerutt2024LAE_LyC}, and a range, $1.0-1.5$, based on our young, metal-poor pySTARBURST99 models (Fig.~\ref{fig:pySTARBURST}).
The second method derives the luminosity ratio directly from the \cg best-fit models (see \S\ref{subsect:SEDfitting} and Table~\ref{tab:obj_details}), which already takes dust attenuation into account, hence yielding $f_{esc,abs}^{\mathrm{IGM}}$ directly. 

\subsection{Escape fraction derivations}

We start by considering luminosity ratios derived by \cg in Eq.~\ref{eq:fescrel}.
For the third term in Eq.~\ref{eq:fescrel}, going forward, we consider that the MXDF is a sightline which has a transmission $3\sigma$ above the mean simulated transmission; $T_\mathrm{{IGM}}\sim0.18$ (see \S\ref{sect:IGM}). We discuss this assumption in \S\ref{sect:discussion}. Using the observed flux ratio of \fluxratio, we calculate  $f_{esc,abs}^{\mathrm{IGM}}$ values of \cgigmcorr\,\% (BC03) and $100-120\,\%$ (BPASS). 

Using a luminosity ratio of 3 and the attenuation derived by our SED fitting, we obtain $f_{esc,rel}$ of $300\,\%$ and a possible $f_{esc,abs}^{\mathrm{IGM}}$ range of $160-220\,\%$. As this results in an escape fraction $\gg100\,\%$, we consider it unlikely that the intrinsic luminosity ratio in \objname is as high as 3. This is further discussed in \S\ref{sect:discussion}. 

The luminosity ratios derived from BPASS and pySTARBURST99 models \citep{Hawcroft2025pySTARBURST} are shown in Fig.~\ref{fig:pySTARBURST}. Given the recent burst in star formation shown by our models and the necessity for \fesc to be $<100\,\%$, we focus on the low $L_{1500}/L_{\mathrm{LyC}}$ end of the range shown. As the $L_{1500}/L_{\mathrm{LyC}}$ value derived from \cg is also low, this plot is useful to visualize the possible nature of the recent burst in \objname. The luminosity ratio range we consider from the young pySTARBURST99 stellar models is $1.0-1.5$, which captures the low-metallicity stellar populations at an age of $\lesssim5\,\mathrm{Myr}$. This gives an $f_{esc,rel}$ range of $100-150\,\%$ and an $f_{esc,abs}^{\mathrm{IGM}}$ range of $53-109\,\%$.

As an escape fraction of $>100\%$ is unphysical, either $L_{1500}/L_{\mathrm{LyC}}$ must be below a certain value or our result invokes a tension with the IGM models described in the previous section. We consider a lower bound of $53\,\%$, as given using the luminosity ratios from the pySTARBURST99 models. While a range of \fesc$=53-100\,\%$ is large in absolute terms, this reflects the uncertainty surrounding the model assumptions inherent to the \fesc calculation at high-redshift, where we do not know the intrinsic ratio of ionizing to UV photons or the exact opacity of the IGM. Moreover, the small number of LCEs in the literature within this range are classed as strong LCEs, making a comparison with \objname useful. We now move on to comparing the properties of \objname to other objects in the literature and \fesc tracers established at low-redshift.

\section{Evaluating \fesc tracers} \label{sect:tracers}



\objname represents a significant step forward in measuring the properties of the LCEs actually responsible for the reionization process, as it is $\sim35$ times closer to the end of the EoR (in terms of time elapsed) than low redshift studies \citep{Flury2022LyC_lowz_survey, Jaskot2025LyCreview} and a factor of $3-4$ less than the time between the end of the EoR and $z\sim3-3.5$, where previous high-redshift \fesc studies based on multiple galaxies have been focused \citep{Steidel2018KLCS,Pahl2021LyCfesc_z3,RiveraThorsen2022LyC_HUDF,Beckett2025PIE}.

First, we focus on the important \lya-LyC connection, taking full advantage of MUSE's spectral resolution and spatial information. In following sections we also connect the SFR surface density, $\Sigma_{SFR}$, and the UV slope with \fesc.  

As mentioned in the previous section, we consider in our analysis an envelope of possible escape fractions: $53-100\%$, ranging from the minimum value (based on $L_{1500}/L_{\mathrm{LyC}}$ ratios) to the maximum possible value, which is implied by our IGM transmission corrections. While this is a large range in absolute terms, there are few LCEs observed in the local universe with \fesc$>50\%$, hence it remains useful to compare \objname's properties with those LCEs of high escape fraction that have been observed. 

\subsection{The \lya-LyC connection} \label{subsect:Lya-LyC}

Thanks to the IFU data of \objname from MUSE, we have full access to both spectral and spatial \lya tracers of \fesc. We start by comparing \objname's \lya properties to low redshift predictions. A summary of all the relevant properties is given in Table~\ref{tab:lya}.


\begin{deluxetable*}{cc}
\tablewidth{0pt}
\tablecaption{\objname\ \lya Properties \label{tab:lya}}
\tablehead{
\colhead{Property} & 
\colhead{Value}
}
\startdata
MUSE ID & 280\\
RA: & \ra \\
DEC & \dec \\
Flux & \finalflux \\
Luminosity & \lyaL \\
EW & \finalEW \\
FWHM & \fwhm \\
Red peak asymmetry & \asymmetry \\
$r_{50,Ly\alpha}$ & \lyarfivekpc\\
\flya & \lyafesc\tablenotemark{a}, $5-55\%$\tablenotemark{b}, $13\pm2\%$\tablenotemark{c} \\
Halo Fraction & \HF
\enddata
\tablenotetext{a}{Using SFR derived from \cg using BC03 models.}
\tablenotetext{b}{Using SFR derived from \cg using BPASS models. Range comes from range of SFRs derived in different runs (see \S\ref{subsect:SEDfitting}).}
\tablenotetext{c}{Derived using UV magnitude observed in F775W filter.}
\tablecomments{
Table~\ref{tab:lya}. A summary of \objname's \lya properties. 
}\vspace{-1em}
\end{deluxetable*}

\subsubsection{\lya emission line properties} \label{subsubsect:Lya_line}

\begin{figure}
    \centering
    \includegraphics[width=\linewidth]{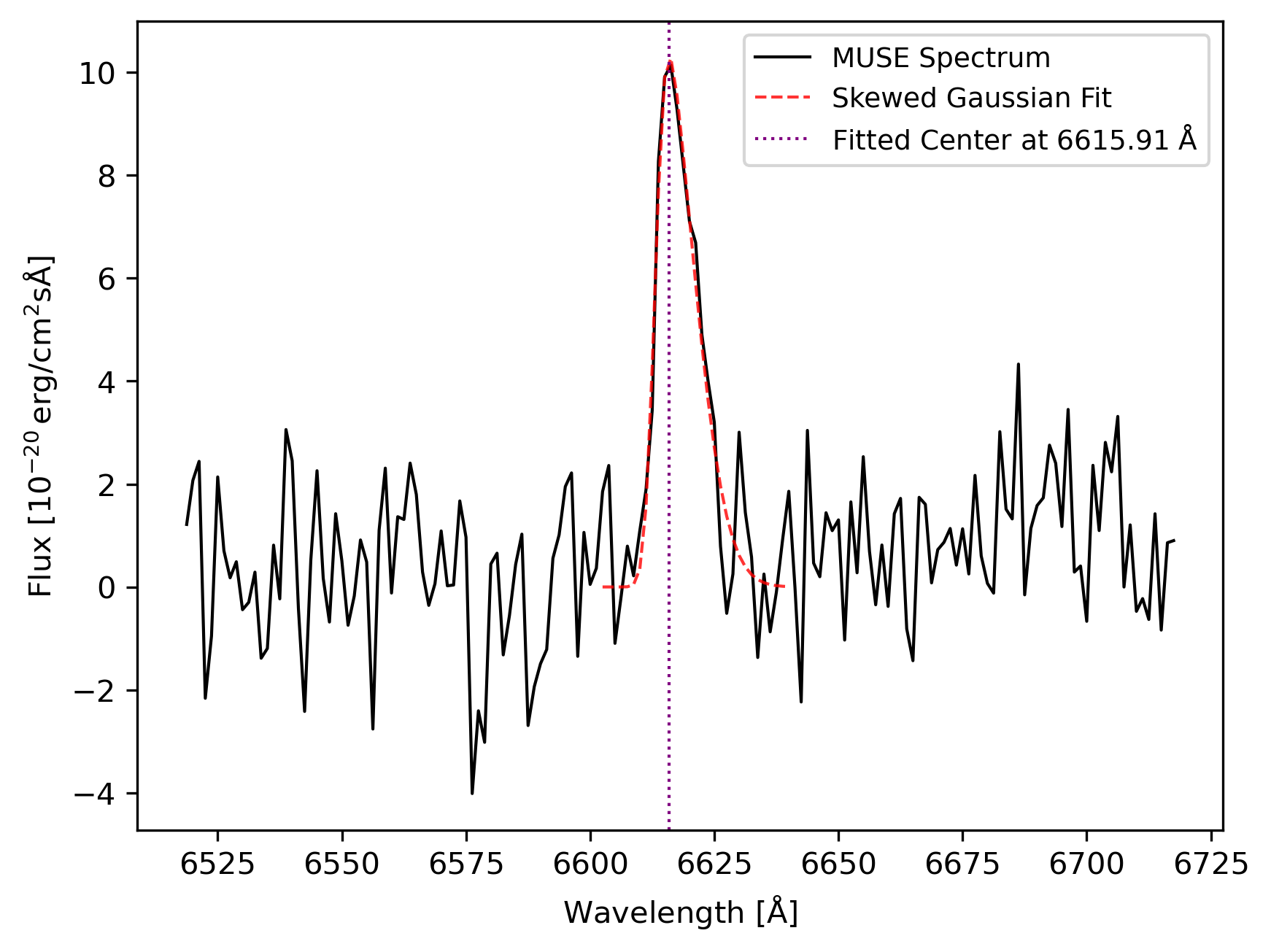}
    \caption{A zoom in on \objname's \lya emission line. The skewed Gaussian fit to the line, described in \S\ref{subsubsect:Lya_line}, is plotted in red, and the fitted peak of the line at $6615.91\,\mathrm{\AA}$ is plotted in purple. The properties of the line derived from the fit are detailed in Table~\ref{tab:lya}.}
    \label{fig:Lya_line_fit}
\end{figure}

The simplest \lya properties we can study are the flux and the equivalent width (EW). 
In order to derive the \lya flux, we perform a custom skewed-Gaussian fit to the line, closely following the procedure in \citep{bacon2022musedatareleaseII}. The skewed Gaussian equation we fit is as follows:
\begin{equation}
F(\lambda) =A \left[ 1 + \operatorname{erf}\!\left( 
\frac{\gamma\,(\lambda - \lambda_0)}{\sqrt{2}\,\sigma} 
\right) \right]
\exp\!\left( -\frac{(\lambda - \lambda_0)^2}{2\sigma^2} \right)
\end{equation}
where $A$ is the amplitude of the Gaussian, $\lambda_0$ is the central wavelength, $\sigma$ is the standard deviation and $\gamma$ is the asymmetry parameter; large positive $\gamma$ in this case leading to a more redward-skewed line. The flux of this line is calculated in the same way as a normal Gaussian function:
\begin{equation}
F_{\mathrm{tot}} \;=\; \int_{-\infty}^{\infty}
F(\lambda)\,d\lambda
\;=\; A\,\sigma\,\sqrt{2\pi}.
\end{equation}
We subtract the background from this flux measured in two representative windows of \bkgwindow\,\AA\ (53 MUSE spectral pixels) around the emission line. We subtract the median pixel flux in these two windows multiplied by the number of pixels underneath the \lya line. The final \lya flux is \finalflux. This is completely consistent with the value from the public MXDF catalog \citep{bacon2022musedatareleaseII}. As there is no continuum detected in the MUSE data, we then calculate the \lya EW by taking the ratio of the \lya flux to the flux in the filter that sees the $1216\,\mathrm{\AA}$ emission, F606W, having first subtracted the \lya flux from this filter. The result is an observed \lya EW of \finalEWobs, which gives a rest frame EW of \finalEW.

\begin{figure}
    \centering
    \includegraphics[width=\linewidth]{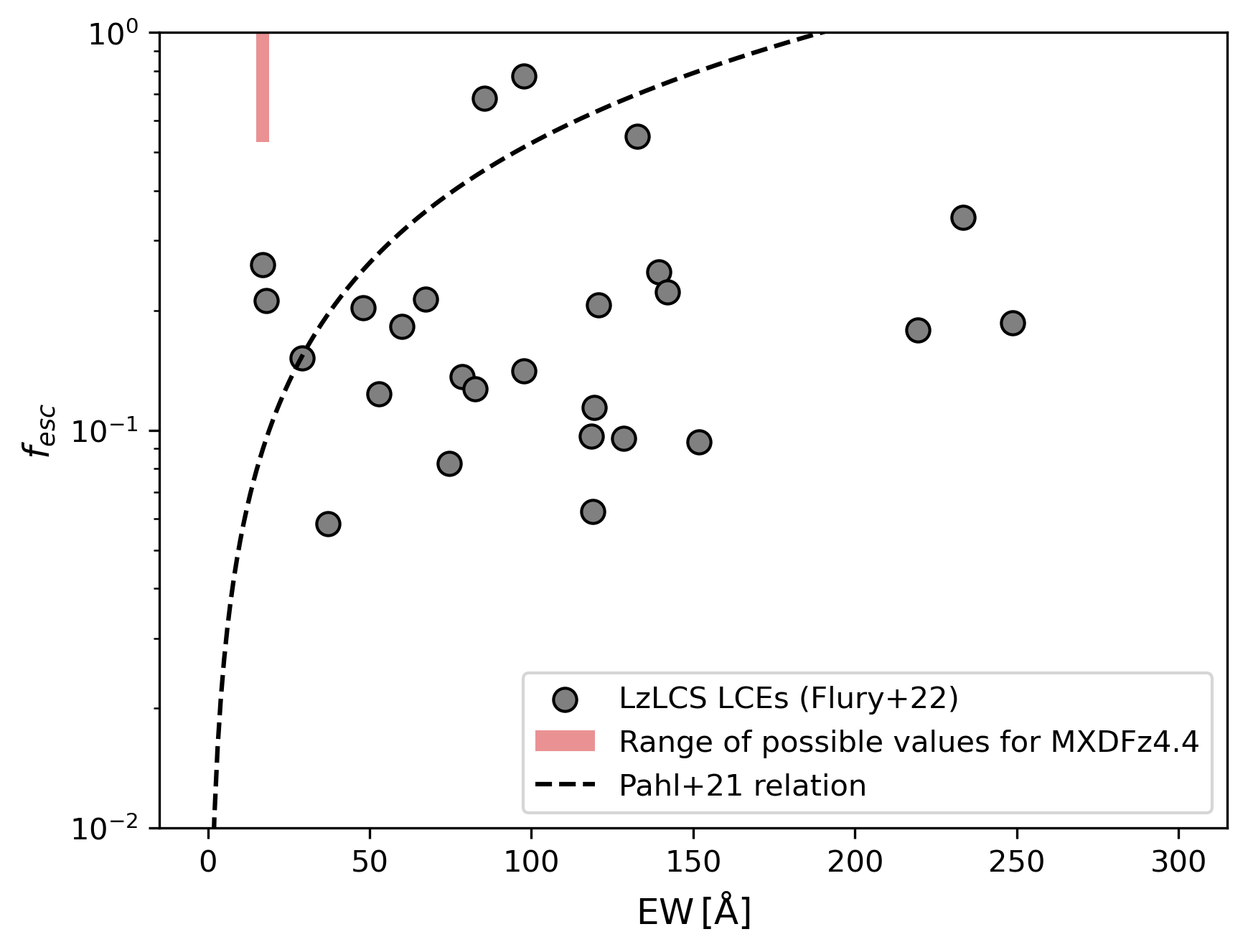}
    \caption{\lya EW vs \fesc, with the range of possible values for \objname highlighted in light red. The darker red area denotes the range derived using the $L_{1500}/L_{\mathrm{LyC}}$ value from \cg. The literature results (grey points) are collected by \cite{FLury2022LzLCS_results} (low-$z$) and the dashed relation is from \cite{Pahl2021LyCfesc_z3} (high-$z$). The red area representing \objname has a width of $\pm1\sigma$ in \lya EW and the full range of possible values for \fesc: $27-100\%$.}
    \label{fig:Lya_EW}
\end{figure}

We compare this \lya EW with literature results in Fig.~\ref{fig:Lya_EW}, where we shade in red the full region that \objname could occupy in terms of \fesc and $\pm1\sigma$ in terms of \lya EW. This region is outside the range of results from low-redshift \citep{FLury2022LzLCS_results}, although there is significant scatter (grey points). However, this region is also in significant disagreement with the relation at high-redshift established by \cite{Pahl2021LyCfesc_z3}. That said, the \lya EW is not expected to be a good tracer at very high \fesc. This stems from the necessary reduction in nebular emission when a high fraction of ionizing photons escape the galaxy rather than powering the nebular emission from the ISM. 

While the \lya escape fraction is generally closely related to the EW \citep{Sobral2019Lya_esc_from_EW,Goovaerts2024Lya_esc_fraction,begley2024lya_LyC_z45}, it should scale with \fesc even to high \fesc values if the \flya is calculated using non-resonant hydrogen lines, such as H$\alpha$. Indeed, a close relation between \fesc and \flya has been found \citep{FLury2022LzLCS_results,Jaskot2024multivariateLyC_lowz,Jaskot2025LyCreview}. However, as we do not have access to these lines for \objname, we resort to comparing \fesc to \flya calculated using our SED fitting and the UV magnitude in the F775W filter. 


We calculate the \flya using the dust-corrected SFRs from our \cg runs. Because the SFR has evolved dramatically over the last $100\,\mathrm{Myr}$ --- and even likely over the last $10\,\mathrm{Myr}$ as our SED fitting suggests ---, we consider both averaged SFR values (see Table~\ref{tab:obj_details}). In the past, \lya escape has been calculated using both UV-derived SFRs and H$\alpha$-derived SFRs. Our results for the $100\,\mathrm{Myr}$-averaged SFRs can be more readily compared with the UV-based method, and our $10\,\mathrm{Myr}$ results with the H$\alpha$ method. We therefore give a range of possible \flya values in Table~\ref{tab:lya} (individual uncertainties are far smaller than the range) for the BC03 models, the BPASS models, and also for a comparison to the value derived using the UV magnitude simply converted to an SFR following the standard prescription from \cite{Kennicutt1998SFR}. We caution that this prescription assumes a constant SFR over longer than $100\,\mathrm{Myr}$, which is very unlikely to be the case for \objname. The \lya escape fractions derived are: $8-38\,\%$ (BC03), $27-300\%$ (BPASS) and $13\pm2\,\%$ (UV magnitude). Because of the numerous assumptions and model dependencies used to obtain these values, it is challenging to meaningfully interpret them alongside low-redshift results. They vary from quite low, given that \objname is a strong LCE, to quite high. Observations of the H$\alpha$ line would be necessary to more reliably determine the \objname's SFR and therefore \flya. However, if it is indeed the case that \objname is a very strong LCE, observing its accordingly reduced H$\alpha$ emission may be challenging. 

The asymmetry of the \lya line is generally thought of as resulting from kinematics in the neutral gas in the ISM and CGM, such as outflows, as well as back-scattering. \cite{Kakiichi2021LyC_Lyaspectra} suggest that galaxies with their ISM in the density-bounded scenario (in which ionizing photons escape) have a lower red peak asymmetry (i.e., a more symmetric line). Thus this quantity may provide a simple way to assess the porosity of the ISM, crucial for allowing LyC escape. This is tentatively borne out in the observational sample of \cite{Izotov2016LyC_compactgals,Izotov2018LyC_OIII/OII,Izotov2018lowz_LyC_highfesc}. We define the red peak asymmetry, $A_f$, similarly to \cite{Kakiichi2021LyC_Lyaspectra};
\begin{equation}\label{eq:asym}
    A_f=\frac{\int^\infty_{\lambda_{red\ peak}}f_\lambda d\lambda}{\int^{\lambda_{red\ peak}}_{\lambda_{valley}}f_\lambda d\lambda},
\end{equation}
where $f_\lambda$ is the \lya flux, $\lambda_{red\ peak}$ is the wavelength at the red peak and  $\lambda_{valley}$ is the wavelength between the red and blue peaks. In this case there is no blue peak, so we take a wavelength $20\mathrm{\AA}$ bluewards of the red peak. The resulting $A_f$ is robust to this choice of lower integration limit. We calculate $A_f$ to be \asymmetry, with an uncertainty calculated by MC sampling Eq.~\ref{eq:asym} within the $1\sigma$ limits of our skewed Gaussian fit to the \lya line. This is a low red peak asymmetry according to \cite{Kakiichi2021LyC_Lyaspectra}, consistent with the galaxies they find to be strong LCEs, with mostly density-bounded ISM conditions.


The FWHM of the \lya line is, however, in light contradiction with expectations. \cite{Verhamme2015Lya_to_get_LyC} showed that LAEs with narrow peaks, close to the systemic redshift, are good candidates to be LCEs. However, \objname has a relatively large FWHM: \fwhm~(corrected for the MUSE line spread function). Larger FWHM values generally indicate higher neutral hydrogen column densities. However, \cite{Kerutt2024LAE_LyC} also find reasonably high FWHM values for 12 LCE candidates at $z>3$, some of which are consistent with \objname. 

These slightly conflicting results for the spectral properties of \objname's \lya emission stress the need to also take into account its spatial properties. We do so in the following sections.

\subsubsection{\lya halo fraction} \label{subsubsect:halo_fraction}
\begin{figure}
    \centering
    \includegraphics[width=\linewidth]{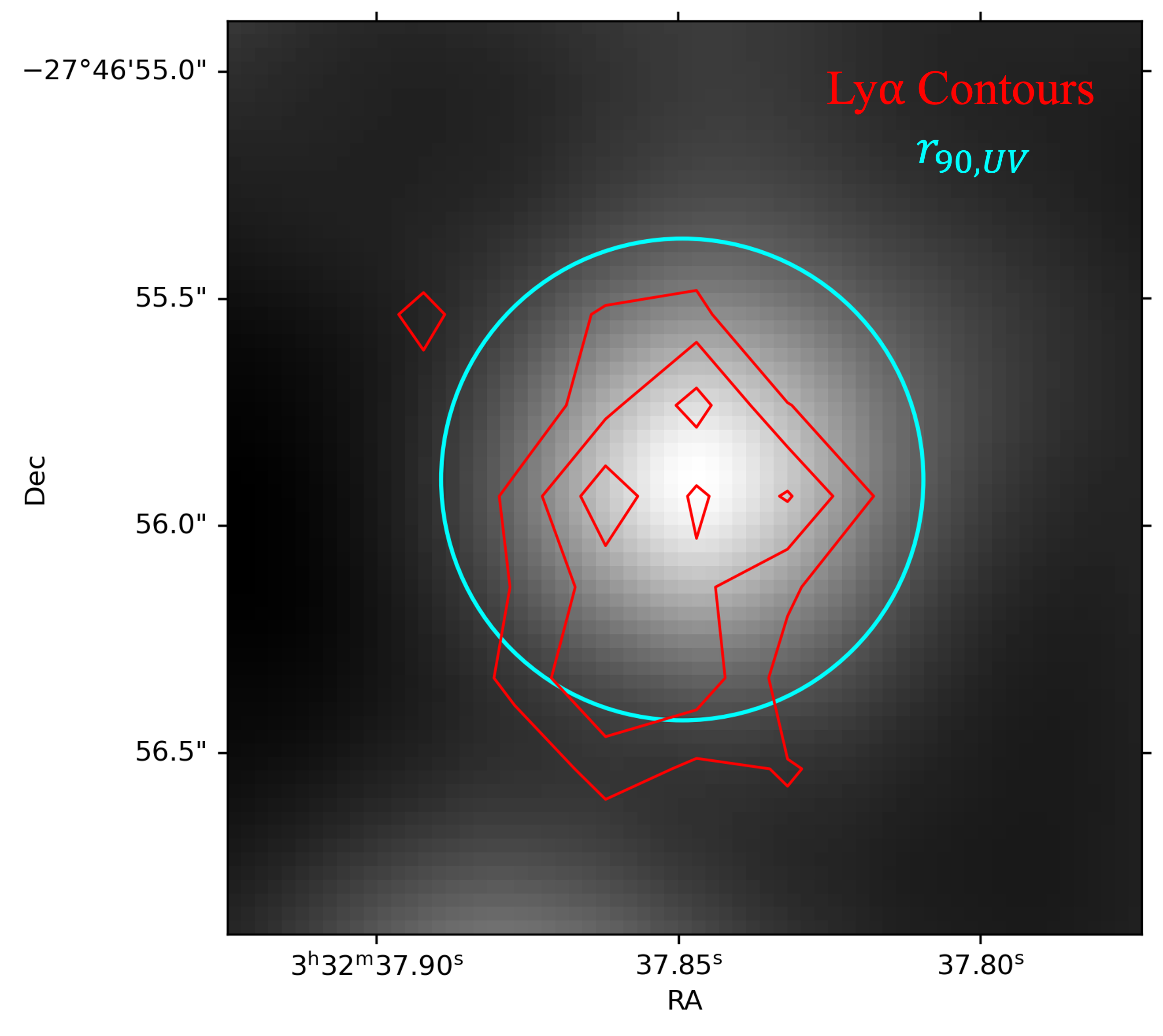}
    \caption{$2,\ 3,\ 4\ \mathrm{and}\ 5\sigma$ contours of \lya (red) shown over the F775W image which sees the rest frame UV emission, PSF matched to MUSE. The cyan circle denotes $r_{90,UV}$.}
    \label{fig:lya_contours}
\end{figure}
The \lya halo fraction, HF, is currently a very promising tracer of \fesc at low redshift \citep{Puschnig2017LyC_ISM_HF,SaldanaLopez2025HF}. Broadly, it encapsulates the amount of \lya emission that escapes the host galaxy in an extended halo, rather than co-spatially with the UV emission. A low halo fraction indicates the presence of clear channels in the ISM and CGM of the galaxy through which \lya photons escape with minimal scattering events, facilitating LyC photon escape. \cite{SaldanaLopez2025HF} fit a relation between the HF and \fesc: $\mathrm{log_{10}(\fesc)=(-2.32\pm0.41)\times HF - (0.38\pm0.25)}$, at low redshift from the Lyman Alpha and Continuum Origins Survey (LaCOS, \citealt{LeReste2025LaCOS}). We study this quantity, and its relation to \fesc, for the first time at high redshift. 

In order to compute the HF for our data, where the \lya comes from MUSE and the UV continuum comes from HST's F775W filter, we broadly follow the method laid out by \cite{Wisotzki2016lya_halos} and close to that employed by \cite{SaldanaLopez2025HF} to facilitate comparison. The key differences between our available data and that of \cite{SaldanaLopez2025HF} are that \lya is detected in MUSE rather than HST and that no continuum is detected in \lya. Thus, we must combine the two datasets to calculate the HF. 

Firstly, a narrow band (NB) image must be created around \objname\ from
the MUSE datacube. We create the NB image for our HF calculation by summing the datacube in a $20\,\mathrm{\AA}$ window around the wavelength of the \lya emission, which contains all the \lya flux. 
However, this NB image still has a much larger PSF than the HST image, so morphological comparisons are not yet possible. We therefore convolve the HST image to the MUSE PSF, using the Moffat profile described in \cite{bacon2022musedatareleaseII} for the MXDF. Fig.~\ref{fig:lya_contours} shows the contours of \lya emission overplotted on the UV emission from F775W, PSF-matched to MUSE. The $r_{90,UV}$ is also shown for reference.

We then cut out a $3\times3\arcsec$ window around \objname and subtract the local background from each, masking \objname and nearby sources in the process. Nearby sources remain masked and radial profiles are calculated for both \lya and UV. 
These radial profiles are then compared after normalizing the profile to the core of the \lya profile. The extended \lya emission over the UV profile is then considered the \lya halo, and the fraction of the total emission that this constitutes is considered the HF. 

Using this method we derive a HF of \HF. The significant uncertainty on this value stems from the uncertainty on the \lya profile. As the integrated SNR of the \lya emission is only $\simeq11\sigma$, the outskirts of the profile have low SNR. Another issue affects the center of the profile, caused by the size of the MUSE pixels, $0.2\arcsec$. In the inner few tenths of arcseconds of the \lya profile, only a small number of pixels are considered, which raises the uncertainty. Both of these effects naturally stem from the MUSE data, and cannot be mitigated save in a source with much higher SNR (ideally $\geq30\sigma$ integrated). 

\begin{figure}
    \centering
    \includegraphics[width=\linewidth]{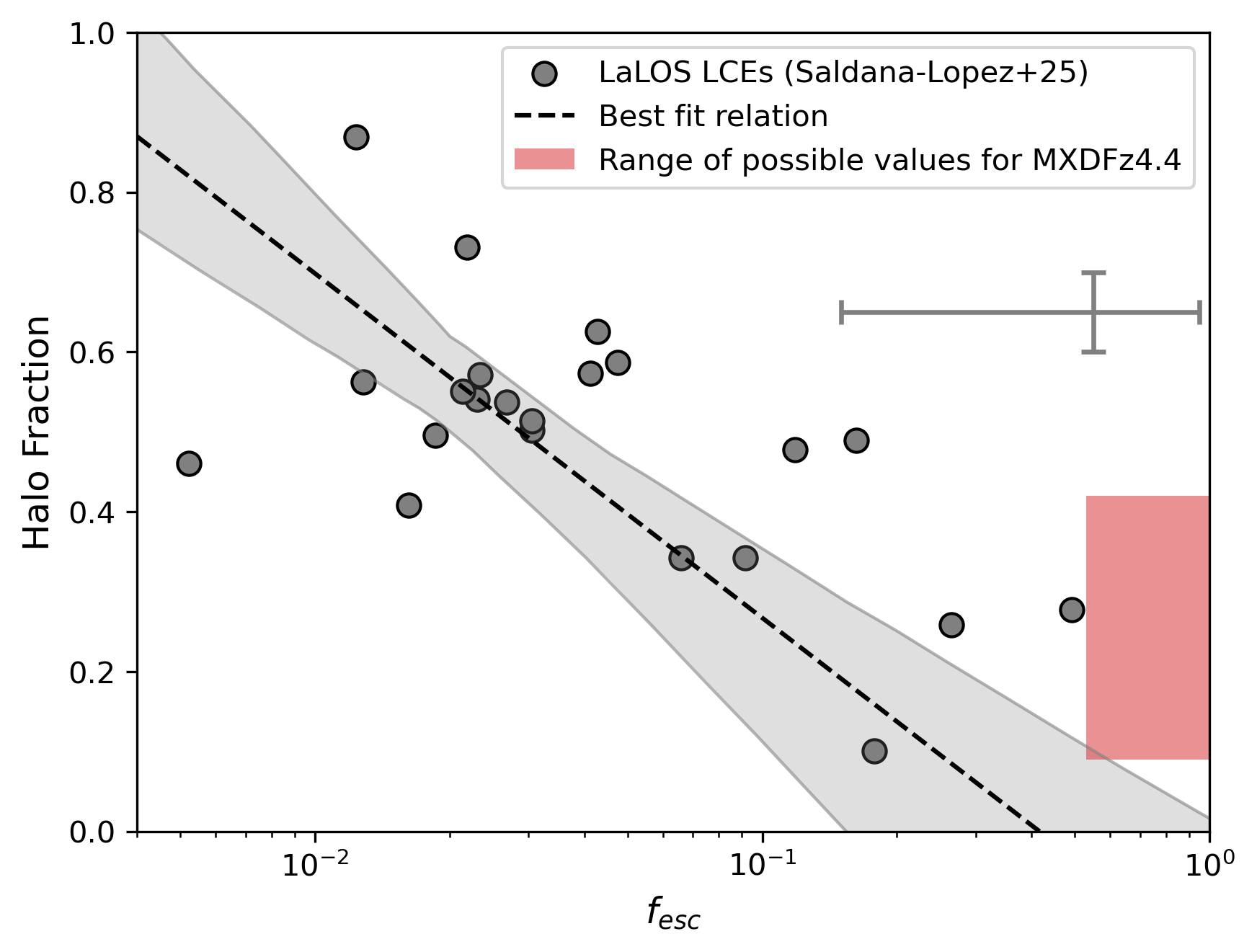}
    \caption{The \lya Halo Fraction (HF) vs \fesc. Grey points, representative error bars, and best-fit relation are from the LaCOS sample in \cite{SaldanaLopez2025HF}. The range of possible values for \objname is highlighted as in Fig.~\ref{fig:Lya_EW}.}
    \label{fig:HF}
\end{figure}

We plot \objname's HF on Fig.~\ref{fig:HF}, together with the results from the LaCOS survey \citep{SaldanaLopez2025HF}. As before, we plot the $\pm1\sigma$ area for the HF and the full range of possible escape fractions. This range is mostly outside the $1\sigma$ uncertainty on the relation established by \cite{SaldanaLopez2025HF}; \objname has a higher fraction of \lya flux emitted in the halo than expected, or equivalently, a higher \fesc than expected based on its HF. We note that the next two strongest LCEs also lie above the relation and its uncertainty. A lack of very high-\fesc LCEs mean it is challenging to truly test this relation in the regime \objname occupies. \\
On the other hand, we show that it is possible to constrain (albeit with large uncertainties) very small halo fractions with MUSE and HST data, lending confidence, from a methodological standpoint, to the possibility of the application of the HF relation across the full range of \fesc at high redshift. We do expect large uncertainties to remain for MUSE LAEs of lower \lya SNR ($\ll30$).


Going forward, increased statistics (most likely coming from $z\sim3$) will greatly benefit the assessment of whether the HF is a reliable tracer of LyC escape at high-redshift.

\subsubsection{\lya $r50$}\label{subsubsect:lya_r50}
Additional \lya morphological parameters have been suggested as \fesc tracers. The compactness of the \lya emission can be expressed more simply than the HF, using $r_{50,Ly\alpha}$. This has also been shown to be a potentially valuable tracer of \fesc \citep{SaldanaLopez2025HF}, and one which does not rely on a detection of the continuum. More compact galaxies, therefore with lower $r_{50,Ly\alpha}$, exhibit higher \fesc values. The suggested interpretation broadly follows the logic used to explain the correlation between the concentration of the UV starlight and \fesc (see the following Section). Additionally, as with the interpretation of the HF, \lya escaping mostly centrally indicates clear channels through the ISM towards the observer, which also elevate \fesc. 

We derive $r_{50,Ly\alpha}$ for \objname via forward modeling of the same local background-subtracted NB image as for the HF. The intrinsic surface brightness distribution is modeled as a Sersic profile and convolved with the MUSE PSF (see \S\ref{sect:MXDF}). Parameters are inferred using MCMC sampling with the {\tt python} package {\tt emcee} \citep{Foreman-Mackey2013,ForemanMackey2019emcee}.

\begin{figure}
    \centering
    \includegraphics[width=\linewidth]{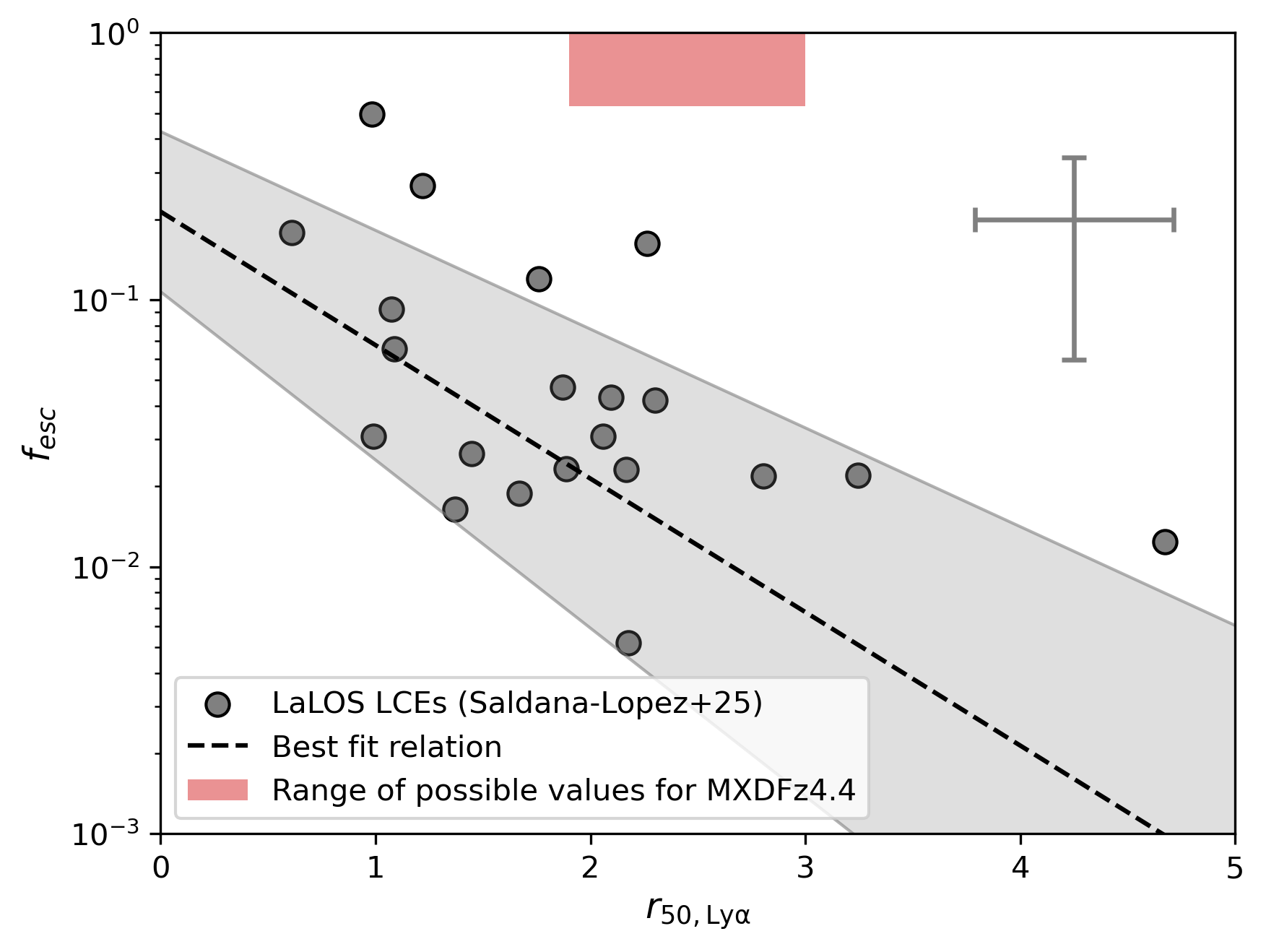}
    \caption{\objname's $r_{50,Ly\alpha}$, plotted against its \fesc, compared to the sample of \cite{SaldanaLopez2025HF} in grey. Colours denote the same as in Figs.~\ref{fig:Lya_EW} and \ref{fig:HF}.}
    \label{fig:Lyar50}
\end{figure}

We find a MUSE PSF-adjusted $r_{50,Ly\alpha}$ value of \lyarfivekpc, which we plot as the red shaded area in Fig.~\ref{fig:Lyar50} together with the results from \cite{SaldanaLopez2025HF}, where once again this area encompasses the full range of possible \fesc values and $\pm1\sigma$ in $r_{50,Ly\alpha}$. We find less agreement here than in the HF, with the region demarcating \objname significantly outside the grey uncertainty region from \cite{SaldanaLopez2025HF}. However, we note that there are other LAEs of high \fesc that reside above this region, showing more extended \lya emission than would be expected given their escape fractions.

Following a similar logic to $r_{50,Ly\alpha}$, the offset of the \lya and UV centroids has also been suggested as a tracer of \fesc \citep{SaldanaLopez2025HF} through the $\frac{\Delta_{Ly\alpha-UV}}{r_{90,UV}}$ statistic. However, the value we measure is consistent with zero after accounting for uncertainties introduced by a potential, small WCS offset between F775W and MUSE. The offset matches expectations from comparing images of extremely different pixel sizes (10-20\% of the larger pixel size, in this case $0\farcs02-0\farcs04$) and the centroiding uncertainty itself, particularly for \objname's clearly non-Gaussian profile. We therefore accept that it will always be challenging to extract significant conclusions from this statistic when combining MUSE and HST/JWST data. 


\subsection{Star formation rate surface density and specific star formation rate} \label{subsect:SFRD}
A high SFR surface density, $\Sigma_{SFR}$, has been highlighted as a galaxy property which can correlate with \fesc \citep{Izotov2016LyC_compactgals,Naidu2020UVbright_high_fesc,Gazagnes2020LyaLyCescape,Jaskot2024multivariateLyC_lowz}. The explanation centers on stellar feedback creating holes in the ISM more efficiently when the stellar environment is more concentrated (higher $\Sigma_{SFR}$).

We define $\Sigma_{SFR}$ within \objname's UV half-light radius using the typical equation for $\Sigma_{SFR}$ \citep{Shibuya2019morphologies_lya,Naidu2020UVbright_high_fesc}: $\Sigma_{SFR} = \frac{\mathrm{SFR}/2}{\pi r_{50}}$. The half light radius of \objname is $r_{50} =\ $\rfiftykpc, as derived in \cite{Nedkova2024ApJ} using a multi-wavelength version of GALFIT \citep{Peng2002Galfit, Peng2010Galfit} that fits Sérsic profiles in all available filters simultaneously \citep{Haussler2013MNRAS}. For \objname, the 30mas HST mosaics in nine filters were used to constrain the half-light radius: F435W, F606W, F775W, F814W, F850LP, F105W, F125W, F140W, and F160W. While the re-processing of the ACS/WFC imaging as part of the HST archival program ID 16621 provide significant improvements to the image quality, these improvements are most significant in F435W and less so in F775W, the latter of which we use to measure the half-light radius. We therefore opt to use the results from \cite{Nedkova2024ApJ} that are constrained over multiple wavelengths, noting that we also find good agreement using the {\tt photutils} package on the re-processed F775W image, lending confidence to this result. 

We derive a number of different SFRs for \objname (see Table~\ref{tab:obj_details}) based on SED fitting and different timescales ($10-100\,\mathrm{Myr}$). The low end of these ranges gives $\Sigma_{SFR}$ values less than $1\,\mathrm{M_{\odot}/yr/kpc^2}$, which appears far too low for a strong LCE such as \objname. The higher end of these ranges ($\gtrsim10\,\mathrm{M_{\odot}/yr}$) gives moderate $\Sigma_{SFR}$ values around $5-6\,\mathrm{M_{\odot}/yr/kpc^2}$. This is still not high considering the very high \fesc \objname likely exhibits. However, when looking in more detail at the SFH plots produced by \cg, the peaks of the recent burst (which is present in the SFH regardless of stellar population model choice) are around $9\,\mathrm{M_{\odot}/yr}$ (BPASS) and $40\,\mathrm{M_{\odot}/yr}$ (BC03).  An SFR of $40\,\mathrm{M_{\odot}/yr}$ gives a $\Sigma_{SFR}$ value of $17\,\mathrm{M_{\odot}/yr/kpc^2}$, which is more consistent with expectations for a strong LCE \citep{Jaskot2024multivariateLyC_lowz}. However, such peak SFRs are not typically used for this calculation. The fact that \objname only has a $\Sigma_{SFR}$ in line with low-redshift predictions when the peak of the recent starburst is considered is likely further evidence of the impact of this burst on the escape of LyC photons. 



Although there is less of a correlation found between \fesc and sSFR (specific star-formation rate; SFR divided by stellar mass), we note that most of the strong LzLCS LCEs have sSFR values above $10^{-8}\,\mathrm{(yr^{-1})}$, between $10^{-8}\,\mathrm{(yr^{-1})}$ and $10^{-7}\,\mathrm{(yr^{-1})}$ \citep{FLury2022LzLCS_results}. High \fesc in high sSFR galaxies is also found in simulations \citep{Menon2025Burst_rad_outflows_LyC}. Using the same SFR ranges, and our \cg and BPASS-derived stellar mass of $10^{8.3}\,\mathrm{M_{\odot}}$, we derive sSFR values between $1.2\times10^{-8}\,\mathrm{(yr^{-1})}$ and $6\times10^{-8}\,\mathrm{(yr^{-1})}$, consistent with the LCEs of very high \fesc in \cite{FLury2022LzLCS_results}.


\subsection{UV slope} \label{subsect:UVslope}
The UV slope, $\beta$, has been suggested as a promising tracer of \fesc at low redshift \citep{Chisholm2022LyC_beta}. It is particularly attractive as it is easy to measure at any redshift (albeit subject to uncertainties, e.g. \citealt{Austin2025UVslopes}). The UV slope is sensitive to the dust content of the galaxy, as dust absorbs these UV photons and re-emits them in the infrared. The UV slope is additionally sensitive to the age and metallicity of the stellar population, with younger, metal-poor populations having bluer (more negative) slopes. However, nebular continuum emission from hot gas in the ISM created by such populations serves to redden the UV slope. It is argued that $\beta$ is mainly sensitive to dust, implying that slope steepness should correlate with \fesc. This is indeed found at low redshift \citep{Chisholm2022LyC_beta}. 

We fit the UV slope of \objname and find a redder slope than expected based on the relation found by \cite{Chisholm2022LyC_beta}, both in the best-fit spectra from \cg and in the photometry. Avoiding the $2175\,\mathrm{\AA}$ feature in the spectrum by calculating the slope in windows of $1250-1750\,\mathrm{\AA}$ and $2400-2600\,\mathrm{\AA}$, we obtain a value using \cg and the BC03 models of $\mathbf{-1.6\pm0.5}$. This is similar to the value derived using the observed photometry: $\beta=-1.4^{+0.3}_{-0.4}$. \cite{Chisholm2022LyC_beta} find that galaxies with very high \fesc, such as \objname (\fesc$\gtrsim20\%$), mostly have $\beta<-2.6$. This suggests the presence of greater dust content than would be expected for a strong LCE, although the older, more chemically evolved component of the composite stellar population found by our SED fitting could also play a role in reddening the UV slope. We note that using the best-fit spectrum from \cg and BPASS gives a bluer slope ($\mathbf{-1.9\pm0.2}$), but that these fits fit the observed photometry slightly worse in the wavelength range where the UV slope is calculated (Fig.~\ref{fig:SEDfit}). 

We also caution that while the UV slope does depend on the dust attenuation along the line of sight --- the same line of sight as the escape of LyC photons ---, it is far less impacted by morphology and the potential presence of clear channels in the ISM. The same applies to the escape of \lya, as recently highlighted by \cite{Markov2025dustyLAE_lensed,Ejdetjarn2026Haro11LyC_lya}. Different components of galaxies can contain differing amounts of dust and neutral hydrogen in their ISM, allowing LyC escape from only some parts of the galaxy. Hence, the integrated light of a galaxy may show the presence of dust, despite the existence of dust-clear channels through which LyC escapes \citep{Ji2025dust_LyC}. 

\section{Discussion} \label{sect:discussion}
In this section we collect the measured properties of \objname, as well as those that are dependent on models, discuss their validity, and  infer the likely nature of \objname.

Due to its very high redshift, and therefore the high opacity of the IGM, we conclude that \objname's intrinsic escape fraction is very high, likely $\geq50\%$. Given that an escape fraction of $>100\%$ is unphysical, we are left with two avenues to explain the detection of \objname, based on Eq.~\ref{eq:fescrel}. Either the intrinsic luminosity ratio of non-ionizing to ionizing photons (the second term) must be low or the transmission of the IGM must be low (the third term). 

We consider the two cases from Table~\ref{tab:obj_details}; using $L_{1500}/L_{\mathrm{LyC}}$ from stellar models (pySTARBURST and BPASS) and using $L_{1500}/L_{\mathrm{LyC}}$ directly from \cg (which includes dust). In the first case, (using the dust attenuation we derive with \cg in Eq.~\ref{eq:fescabs}), $L_{1500}/L_{\mathrm{LyC}}$ must be $\leq1.88$ ($A_v=0.68$) or $\leq1.38$ ($A_v=0.34$). In either case $L_{1500}/L_{\mathrm{LyC}}$ must be close to unity. This brings \objname's observed \fesc into consistency with models of IGM transmission. 
For the \cg-derived case $L_{1500}/L_{\mathrm{LyC}}$ must also be very near unity to avoid $f_{esc,abs}^{\mathrm{IGM}}>100\,\%$. As we can see from Fig.~\ref{fig:pySTARBURST}, $L_{1500}/L_{\mathrm{LyC}}$ ratios of near unity are achieved only by very young and/or very metal-poor populations. 

Even with such extreme stellar populations, only a small fraction of sightlines are transmissive enough to allow the \fesc that we observe from \objname, according to our modeling (\S\ref{sect:IGM} and Fig.~\ref{fig:TIGM_dists}). Using the stellar modeling approach (row 2 of Table~\ref{tab:obj_details}), a $T_{\mathrm{IGM}}$ value of $>10.4$ is necessary to ensure $f_{esc,abs}^{\mathrm{IGM}}<100\,\%$ even with the best case dust attenuation. This is satisfied by just over $2\%$ of sightlines. Using the \cg approach (rows 3 and 4 of Table~\ref{tab:obj_details}), this reduces to $0.3\%$ of sightlines ($T_{\mathrm{IGM}}<0.18$). This can be interpreted as a ``$3\sigma$-unlikely sightline''. The previous case, $T_{\mathrm{IGM}}>10.4$, is equivalent to a little over $2\sigma$ unlikely.
 
It is possible that the IGM actually allows a greater transmission of ionizing photons than the TAOIST models predict. Although the HI CDDF used by TAOIST is redshift dependent, it is primarily calibrated using observations from the Keck Baryonic Structure Survey (KBSS, \citealt{Rudie2013}) at 2 $\lesssim$ z $\lesssim$ 3, and may therefore be inaccurate close to the EoR. TAOIST and other Monte-Carlo based IGM codes also generally do not account for HI clustering, which can be due to large scale structure \citep{Scarlata2025}. This could also increase the transmission of the most transparent sightlines and reduce the possible tension \citep{Kakiichi2018AGN_gals_reion}.  

However, given that we do observe ionizing escape from \objname, and that a young and/or potentially very metal poor stellar population can account for this, we now discuss its physical nature by combining all the \fesc tracers that we have analyzed and the physical conditions we have derived. We caution that comparisons to other LCE samples may be limited by the lack of sufficient LCEs of very high \fesc in the literature.
 
The \lya EW is in significant disagreement both with results found at low redshift and at high redshift \citep{FLury2022LzLCS_results,Pahl2021LyCfesc_z3}. However, this may be expected for an LCE of very high \fesc, as \objname is likely to be. The strength of nebular lines is expected to reduce if significant amounts of ionizing photons escape, as these then no longer go to producing recombination lines like \lya. 

For this reason, \flya is expected to correlate with \fesc up to very high values, however only if it is calculated using non-resonant hydrogen recombination lines, such as H$\alpha$. This is not the case for \objname and given the difficulty in deriving precise SFRs using SED fitting, we are left with a large range of \flya values, derived in a number of ways. High values, as would be expected, are included, but unfortunately our results are not very constraining and therefore are difficult to compare to others in the literature. 
 
Considering the \lya asymmetry and HF, these agree fairly well in that a low HF indicates a direct path for LyC photons to escape the galaxy towards the observer, and a symmetric line suggests a density-bounded scenario which also allows for high levels of escape. 

The high FWHM, however, is less in line with expectations from simulations and from low redshift observations, as this can indicate high neutral hydrogen column densities, which would impede LyC escape. These slightly contradictory \lya properties may be explained by the effect of the recent burst combined with an older, more evolved population (which likely exists given the redder-than-expected UV slope). \lya photons can take time to escape from the galaxy (e.g. \citealt{Mas-Hesse2003lya_escape}). At stellar ages younger than $\sim2.5\,\mathrm{Myr}$, \lya struggles to escape from the molecular clouds surrounding stars \citep{Hayes2007Lya_escape_starburst}. Then, feedback from massive stars, winds, and the first supernovae, such as we expect in \objname's recent burst, progressively clear the clouds and facilitates both \lya and LyC escape \citep{Rahner2017stellar_winds_fesc,Trebitsch2017feedback_LyC,Carr2025rad+supernovae_LyC,Komarova2025winds_LCEs}. If the burst in \objname is at this stage, rapid development of the emergent \lya emission would be expected, with some component remaining from the older stellar population and its ISM. 

If \objname's \lya line is close to the systemic redshift, which has been shown to correlate well with \fesc \citep{Verhamme2015Lya_to_get_LyC, Verhamme2017LyC_strongLAEs_lowz, FLury2022LzLCS_results}, this may alleviate the unexpectedly high FWHM. It would mean that significant fraction of the \lya photons are escaping at the systemic redshift rather than undergoing an extensive radiative transfer process through surrounding neutral hydrogen. This would help to explain the high fraction of LyC photons that also escape. Unfortunately, we lack the additional lines in \objname's spectrum to verify this.

Among all the \lya-based tracers, we note that one of the most promising at low redshift, the HF, qualitatively agrees with predictions, which moderately increases the confidence in the possibility of using it at high redshift. Further confidence would necessitate testing the HF method for much larger samples, especially in the lower escape fraction regime where the HF relation from \cite{SaldanaLopez2025HF} is more constrained. Samples of LCEs in the \fesc range of \objname are likely to remain small for the foreseeable future, although will remain desirable to better constrain the high-\fesc regime.

With these conclusions based on \objname's \lya properties in mind, we suggest that the \objname's recent star formation and the nature of its burst are the driving force behind its high observed escape fraction and highlight the importance of $\Sigma_{SFR}$ and sSFR as tracers. It may be for this reason that the \lya HF still acts as a good tracer of \fesc, while for example the \lya $r_{50}$ is less in line with low-redshift findings \citep{SaldanaLopez2025HF}; the HF is the most sensitive of the \lya diagnostics to the holes created in the ISM for \lya and LyC to escape. Still, the $\Sigma_{SFR}$ is lower than expected for all but the peak of the recent star formation, which likely means that the nature of the stellar population also plays a significant role, i.e.~a young, metal-poor stellar population producing many ionizing photons per UV photon. To illustrate the importance of this, we refer the reader to the earlier discussion about the necessity of $L_{1500}/L_{\mathrm{LyC}}$ being near unity (note that this keeps all other parameters, as well as the actual escape of LyC photons, the same). 
Even the upper bounds on the possible $L_{1500}/L_{\mathrm{LyC}}$ ratios we calculate are low compared to the average galaxy population, and are surpassed, according to Fig.~\ref{fig:pySTARBURST}, for all but the most metal poor stars in $6-7\,\mathrm{Myr}$.

Therefore, it is likely that the very young and potentially metal-poor nature of the stellar population in \objname's recent burst is contributing strongly to its observation as an LCE. This is supported by the recent local Universe work of \cite{Ejdetjarn2026Haro11LyC_lya}, where LyC is boosted due to recent star formation and is not necessarily co-spatial with \lya, which the authors attribute to LyC and \lya\ originating in different locations (i.e., stars versus the ISM). A post-starburst boost in the \fesc of a galaxy has been well studied, both in simulations \citep{Gnedin2008LyCfesc_sims,Yajima2009LyCfesc_supernovae_sim,Kimm2014LyC_supernovae_sim,Trebitsch2017feedback_LyC,Rosdahl2022SPHINX_LyC_EoR, Menon2025Burst_rad_outflows_LyC} and in observations \citep{Flury2025LyC_ISMgeom_feedback,Komarova2025winds_LCEs,Carr2025rad+supernovae_LyC}, with these studies also commenting on the importance of young, metal-poor stars in this process.

The burstiness of early galaxies within the EoR has recently been invoked to explain the excess of UV-bright galaxies observed at very high redshifts \citep{Sun2023burstiness_highz,Shen2023UVvariability_highz,Munoz2026burstiness}. These galaxies have already been suggested as efficient ionizers \citep{Endsley2023UVfaintgals_reion,Simmonds2024bursty_gals_ionizing}, mostly due to their enhanced ionizing photon production efficiency and a potential higher \fesc \citep{Endsley2024SF_ionizing_gals_z6-9, Hayes2025UVspectra_highz}. Our interpretation of \objname's properties supports this, and it also suggests that galaxies with successive strong bursts of star formation may also experience episodes of extremely high LyC production and \fesc. We caution that this is dependent on metallicity, since more enriched populations would produce fewer ionizing photons. This process is cumulative, with successive bursts producing more enriched populations and fewer ionizing photons. However, elevated $\Sigma_{SFR}$ and sSFR during the burst may still facilitate high levels of escape. Such bursty galaxies are therefore strong candidates to emit a large share of the ionizing photons needed to reionize the IGM at early times. 

If high-redshift galaxies have a duty cycle closer to one (i.e. they spend close to 100\% of their time in the star-forming phase), it might provide a mechanism for EoR galaxies to emit enough ionizing photons into the IGM to create the reionization history that we observe, a similar argument to that presented in \cite{Naidu2020UVbright_high_fesc} for $\Sigma_{SFR}$. For this reionization history, the necessary escape fraction on a population level has generally been placed around $5\%$ \citep{Finkelstein2019reionization_fesc}, which is far greater than what we observe in the local Universe. The increased burstiness of high redshift galaxies might therefore help alleviate this issue. 

\section{Summary and conclusions} \label{sect:conclusions}
We present the detection of the highest-redshift Lyman continuum emitter to date, \objname, at $z=\redshift$. This redshift is confirmed by a high-confidence \lya line in extremely deep, 140h MUSE spectroscopy. 
We apply all available tracers established by studies at low redshift and conclude the following about the nature of \objname:


\begin{itemize}
    \item Our SED fitting with \cg and two different stellar population models (Fig.~\ref{fig:SEDfit})
    suggests a very recent and vigorous burst of star formation, with numerous stellar models predicting a very low $L_{1500}/L_{\mathrm{LyC}}$ ratio at these young ages. This would mean that \objname's observed escape fraction is not in tension with models of IGM transmission at $z=4.44$.
    \item \objname's \lya EW is low (Fig.~\ref{fig:Lya_EW}) in comparison to both low and high-redshift studies from the literature. However, this may be explained by \objname's very high escape fraction. When \fesc is very high, nebular emission, including \lya, is expected to decrease as ionizing photons are escaping rather than being absorbed and re-emitted as \lya. 

    \item \objname's \lya escape fraction, \flya, is challenging to constrain without the H$\alpha$ emission line. We use SED fitting and \objname's UV magnitude instead, finding a wide range of \flya values, ranging from low to $>50\,\%$.
    
    \item The \lya shape and FWHM are in slight conflict, with the line being symmetric (as expected for a strong LCE) but with a large FWHM (unexpected given simulations). This may be alleviated by \objname's \lya line being close to the systemic redshift, shown to be important in facilitating LyC escape. However, we lack the other emission lines to assess this. 
    
    \item The \lya half-light radius, $r_{50,Ly\alpha}$, is also larger than expected for high-\fesc galaxies, lying outside of the predicted range based on low-redshift results. Both this and the \lya line FWHM could be attributed to rapid evolution of the \objname's \lya emission due to its recent burst, some contribution from an older stellar population, and more chemically evolved parts of the galaxy. 
    
    \item We tested the promising halo fraction tracer established by \cite{SaldanaLopez2025HF} for the first time at high redshift and find qualitative agreement with low-redshift results, although \objname occupies a sparsely populated area of the parameter space. Nevertheless, our results support the use of this promising tracer for high-redshift \fesc studies and demonstrate how this can be achieved with MUSE data. These results also reinforce the importance of geometry and orientation/viewing angle in the observation of escaping LyC photons. Applying the HF method to more high-redshift LCEs is necessary to truly ascertain its validity towards the EoR and therefore the potential importance of the presence clear channels in the ISM and CGM which this tracer probes. In particular, more high-\fesc LAEs are needed to assess whether \objname is a typical or an outlier case.
    
    \item Among \objname's star formation properties, sSFR is elevated and $\Sigma_{SFR}$ is moderate, rising to very elevated when we consider the peak SFR of the recent burst. Both of these point to the importance of the recent burst of star formation in clearing the ISM and facilitating LyC escape. 

    \item We find a redder UV slope, $\beta$, than expected: $\sim-1.4$ from photometry and $\sim-1.5--1.9$ from the best-fit \cg models. This indicates the presence of some dust in the galaxy and means it is likely we are observing the LyC escaping thanks to a specific ISM geometry, also likely due to feedback from the recent burst.

\end{itemize}

Based on these results, we suggest that galaxies in the early Universe, which have been shown to undergo multiple bursts of vigorous star formation, may also undergo successive periods of extremely high ionizing photon production and escape, although the production is likely also dependent on metallicity. While more work is needed to confirm this, the bursty nature of high-redshift galaxies may provide an explanation for the discrepancy between the low levels of \fesc observed among the general galaxy population at low redshift and the number needed to reionize the IGM. 

While the detection and characterization of \objname shows MUSE to be an extremely useful instrument for high-redshift LCE investigation, we warn that the HF method will not work with all MUSE-detected LAEs, as a significant fraction (10-30\%, depending on the relative depths of the MUSE data and photometry) are undetected in the continuum \citep{bacon2022musedatareleaseII,Goovaerts2023LAEs_LBGs}. Still, the $r_{50,Ly\alpha}$ and all the line shape diagnostics would remain available, making MUSE an efficient instrument to quantify \fesc in LAEs up to $z\sim6.5$, well into the end of the EoR.



\begin{acknowledgments}

Based on observations with the NASA/ESA Hubble Space Telescope
obtained at the Space Telescope Science Institute, which is operated by the Association of Universities for Research in Astronomy, Incorporated, under NASA contract NAS5-26555. Support for Program number 16621 provided through a grant from the STScI under NASA contract NAS5-26555. MJH is supported by the Swedish Research Council (Vetenskapsr{\aa}det) and is Fellow of the Knut \& Alice Wallenberg Foundation. 

\end{acknowledgments}




%
\facilities{}

\software{CIGALE \citep{Boquien2019cigale}, pySTARBURST99 \citep{Hawcroft2025pySTARBURST}, BPASS \citep{Elridge2008BPASS, Eldridge2017}, BC03 \citep{Bruzual2003stellar_pops}, TAOIST \citep{Bassett2021IGM_sim} {\tt emcee} \citep{Foreman-Mackey2013} {\tt photutils} \citep{larry_bradley_2025_photutils}.}


\appendix

\section{Redshift confidence} \label{appendix:interlopers}
\begin{figure}
    \centering
    \includegraphics[width=0.97\linewidth]{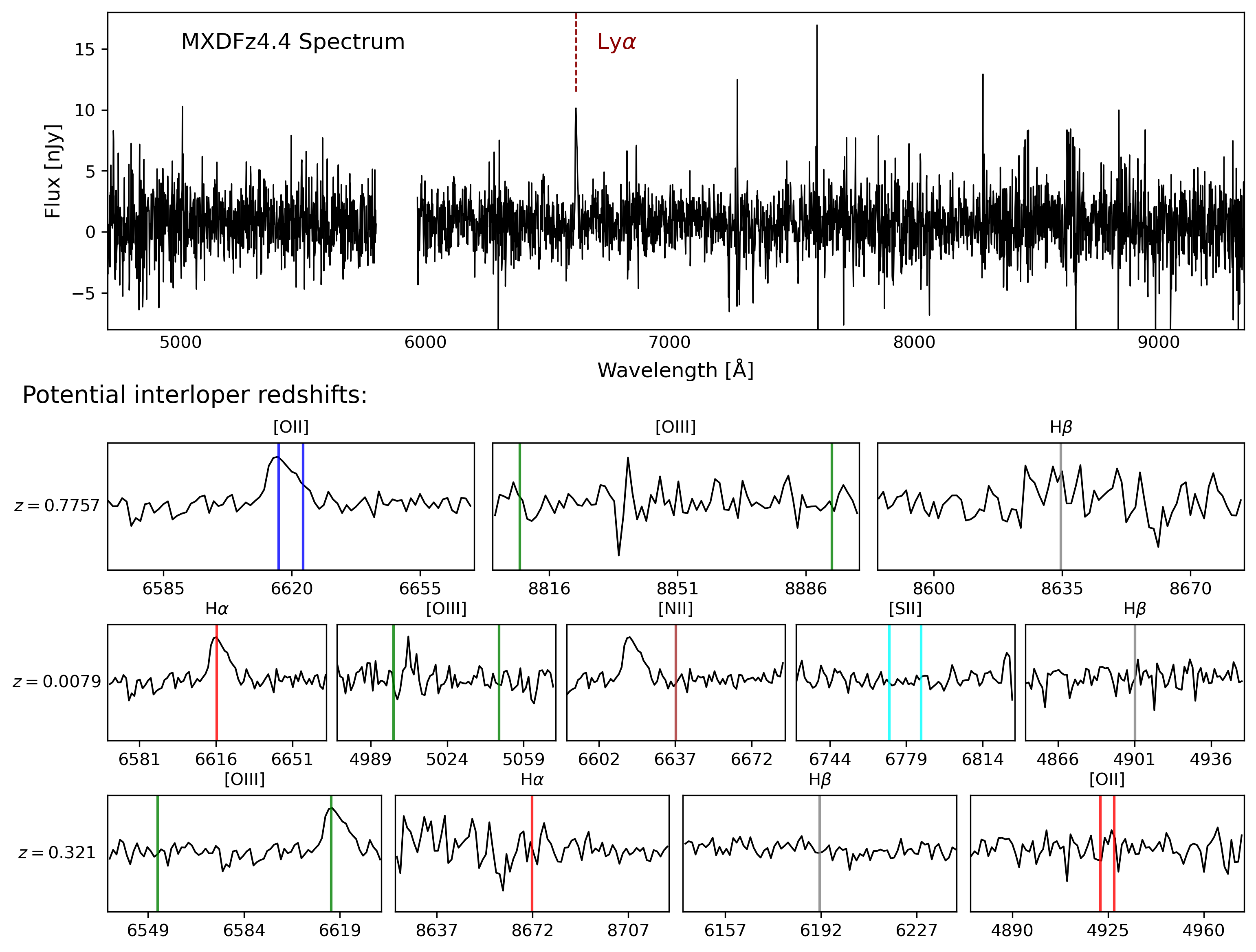}
    \caption{Testing potential interlopers for \objname. We posit that the line at $6615.9\,\mathrm{\AA}$ could be $\mathrm{[O~II]}$, $\mathrm{[O~III]}$ and $\mathrm{H\alpha}$, and check for other lines at the corresponding redshifts. The top spectrum is the full MUSE spectrum of \objname, each row of cutouts below represents a different interloper redshift, noted on the left. The first row plots the case of $\mathrm{[O~II]}$ ($z=0.7757$) and clearly shows the lack of $\mathrm{[O~III]}$ or H$\beta$. The second row plots the case of $\mathrm{H\alpha}$, where we see that $\mathrm{[O~III]}$ and $\mathrm{H\beta}$ are not detected. The third row plots the case of $\mathrm{[O~III]}$, where we show the blue $\mathrm{[O~III]}$ peak is not detected, nor H$\alpha$, H$\beta$, $\mathrm{[O~II]}$. These non-detections, combined with the clear asymmetry in the line at $6615.9\,\mathrm{\AA}$ results in a high redshift confidence for \objname.}
    \label{fig:Interloper plot}
\end{figure}

\begin{figure}
    \centering
    \includegraphics[width=0.97\linewidth]{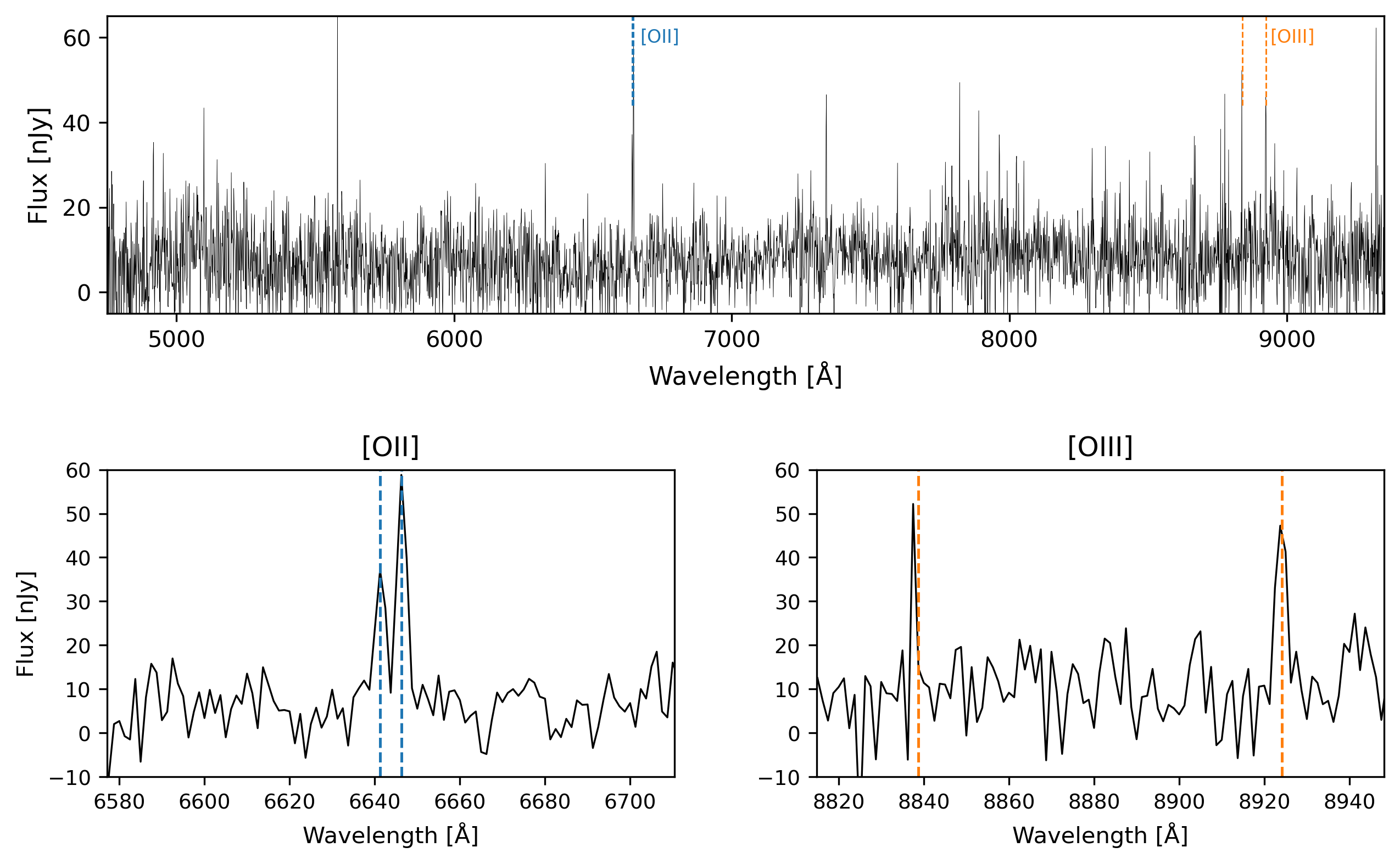}
    \caption{A spectrum of an [O~II] emitter at a similar wavelength as \objname's \lya emission, and with a similar SNR ($\sim10\sigma$; MUSE object ID: 6969). The double peak is clearly resolved. Additionally, the [O~III] doublet is clearly detected, while in \objname's spectrum, only one line is detected.}
    \label{fig:OII_MUSE_spectrum}
\end{figure}

\begin{figure}
    \centering
    \includegraphics[width=0.97\linewidth]{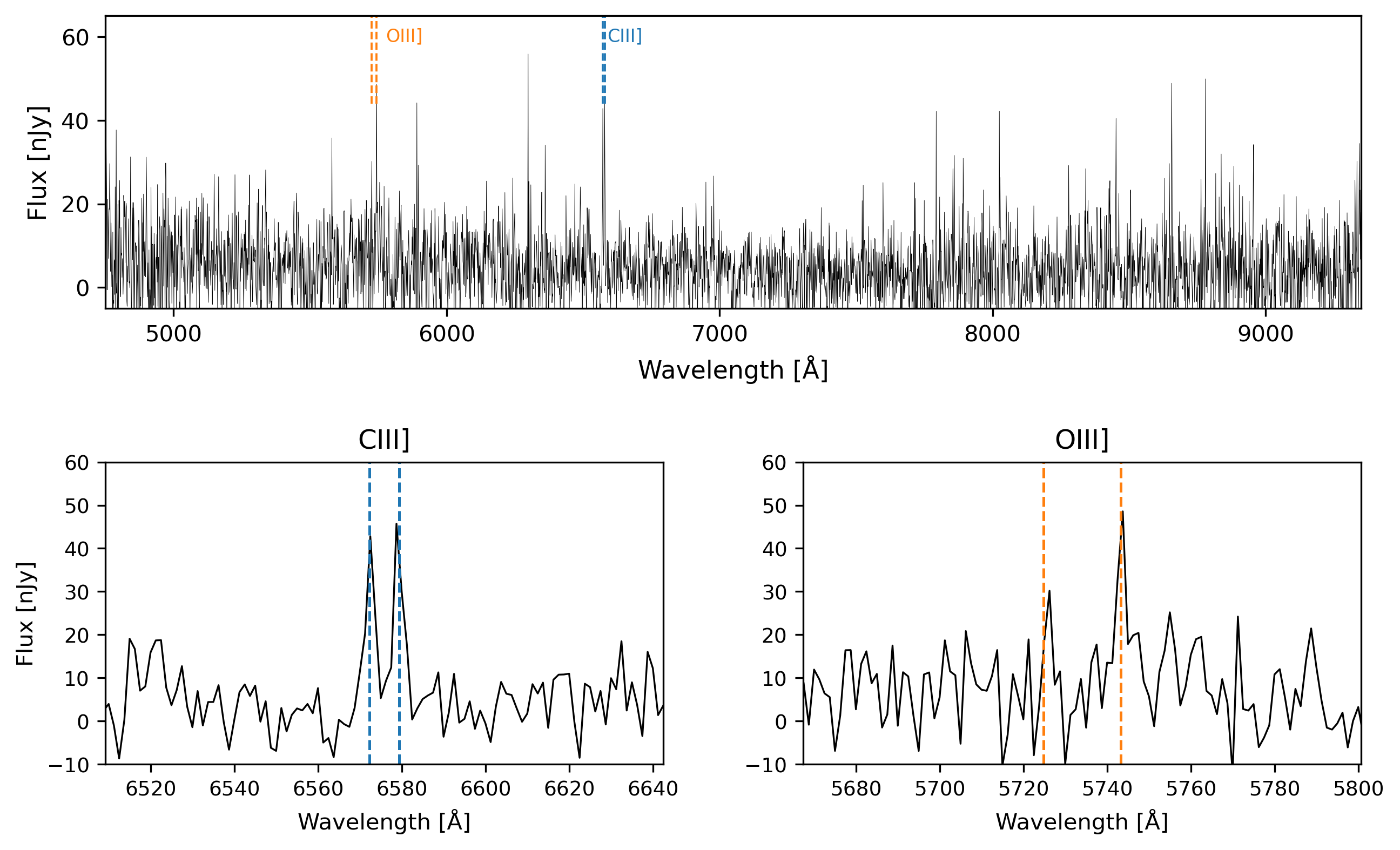}
    \caption{A spectrum of a C~III]$\lambda1907,1909$ emitter detected by MUSE, with C~III] at a similar wavelength as \objname's \lya emission, and with a similar SNR ($\sim7.5\sigma$; MUSE ID: 1784). The double peak of C~III] is clearly resolved. O~III]$\lambda1660,1666$ is also clearly detected in the spectrum.}
    \label{fig:CIII_MUSE_spectrum}
\end{figure}

In order to confirm the redshift of \objname, we carried out a number of tests, positing the alternative possibilities that the emission line seen in the MUSE spectrum at $6615.9\,\mathrm{\AA}$ is not \lya but instead the $\mathrm{[O~II]}~\lambda3727$ doublet, the stronger of the two [O~III] lines ($5007\,\mathrm{\AA}$), H$\alpha$ and the C~III] doublet. The first three cases are shown below in the cutouts in Fig.~\ref{fig:Interloper plot}, clearly showing the lack of the other expected lines in the spectra. In the case of [O~II] and C~III], we show plots of [O~II] and C~III] emitters from the MUSE database in Figs.~\ref{fig:OII_MUSE_spectrum} and \ref{fig:CIII_MUSE_spectrum}, clearly showing that both of these doublets are resolved at $\sim6600\mathrm{\AA}$. It is therefore clear that none of these cases are plausible. 

In the case of H$\alpha$, we do not see either [O~III], [N~II] or [S~II]. Additionally, the line at $6615.9\,\mathrm{\AA}$ is visibly skewed, common for \lya but extremely uncommon for H$\alpha$. Finally, if this line was H$\alpha$, the redshift would be $z=0.008$, and detections would be expected even in the shallower F336W, F275W and F225W images (Fig.~\ref{fig:cutouts}) depending on the SED shape, all of which contain non-detections. Finally, the source's compact morphology is not consistent with a galaxy at $z=0.008$.

If the line were to be the $\mathrm{[O~II]~\lambda3727}$ doublet, we would expect it to be resolved at the spectral resolution of MUSE. We show the spectrum of an $\mathrm{[O~II]}$ emitter at the same SNR ($\sim11\sigma$) and wavelength and show that it is easily resolved (Fig.~\ref{fig:OII_MUSE_spectrum}). Additionally, neither $\mathrm{[O~III]}$ nor H$\beta$ is detected. Even in the case of a very low ionization state, where $\mathrm{[O~II]}$ is expected to be stronger than $\mathrm{[O~III]}$, for an $11\sigma$ $\mathrm{[O~II]}$ detection, we would expect a $>4\sigma$ detection of the stronger $\mathrm{[O~III]}$ line (based on an $\mathrm{[O~III]}/\mathrm{[O~II]}$ ratio of 0.5, typically found in metal-rich low-$z$ galaxies). 

If the line at $6615.9\,\mathrm{\AA}$ were to be the stronger of the two $\mathrm{[O~III]}$ lines, we would expect to detect the weaker line at $3-4\sigma$ as this line ratio is $\sim3$ (independent of astrophysical conditions in the galaxy). In addition to this, not one of H$\alpha$, H$\beta$ or $\mathrm{[O~II]}$ is detected.

Finally, Fig.~\ref{fig:CIII_MUSE_spectrum} clearly shows that the C~III]$\lambda1907,1909$ doublet is resolved at the wavelength that \lya is detected in \objname's spectrum. 

Together with the clearly asymmetric line-profile, these cases show that there is high confidence in \objname's \lya line and therefore its redshift of $z=\redshift$.

\subsection{Chance alignment with a low-redshift interloper or supernova}
We also discuss the possibility of a chance alignment between a low redshift galaxy and a background LAE, such that the continuum that we see, including the detection in F435W, does not actually come from the line emitter we observe in the MUSE spectrum. In order to assess this probability, we start with the number density of quiescent galaxies from \cite{Brammer2011quiescent_num_density} (lowest redshift and mass bin, as the SED is inconsistent with a very high mass galaxy): $\approx7.3\times10^{-4}\,\mathrm{Mpc^{-3}}$. The foreground galaxy must be quiescent, as it is not detected in the filters bluewards of F435W, and no emission lines are detected in 140h-depth MUSE data. From this we calculate the number of such galaxies that should exist in the MXDF in the $1\sigma$ interval of the JADES-derived photometric redshift: $0.391<z<0.476$, arriving at 0.039. As the alignment must be exact, we then compute the number of $0.2\arcsec$ circles (approximate centroid uncertainty between the two emissions, \lya and continuum) which fit in the MXDF. This number, using Thue's circle packing theorem \citep{Chang2010Thue_theorem_proof}, is 20405. Combining this number with the $3.9\,\%$ chance of the necessary galaxy existing in the MXDF field of view, gives a chance of an alignment with any single LAE of 1 in 523205. We multiply this by the number of LAEs in a feasible redshift range to detect LyC in F435W: $4.391<z<4.6$, 40. This gives a final probability of a chance alignment producing the F435W flux we observe, of 1 in 13080, or $0.0076\,\%$.

We also rule out the possibility of the F435W flux that we detect being created by a supernova in a faint host galaxy which aligns with \objname. This is done by showing the F435W is not a point source, rather it is extended and elongated. The F435W PSF FWHM is $3.4\ \mathrm{pixels}$ and the FWHM of \objname's F435W flux is $5.3\ \mathrm{pixels}$. The semi-major to semi-minor axis ratio of \objname in F435W is 1.21 whereas for the PSF it is 1.01 (circular).


\section{JWST/NIRISS Spectrum of \objname from the NGDEEP program} \label{appendix:NGDEEP}
There is JWST/NIRISS Wide-Field Slitless spectroscopy available in the MXDF, from the NGDEEP program \citep{Bagley2024NGDEEP}. There are no lines detected in \objname's spectrum, despite NGDEEP's long exposure time in F200W of $>$31,000 sec. However, considering the $5\sigma$ flux limit of the spectra $1.2\times10^{-18}\,\mathrm{erg/s/cm^2/\AA}$, it is unsurprising that no lines are detected. The strongest line that falls in NIRISS' wavelength coverage is $[\mathrm{O~II}]\,\lambda3727$, which would fall at $\sim2.03\,\mu\mathrm{m}$. The $5\sigma$ limit of the spectrum is $\sim30\%$ more than the \lya flux observed by MUSE, therefore it is not surprising that $[\mathrm{O~II}]$ is not detected. There are even scenarios of Lyman photon escape, the density-bounded case (see, e.g. Fig.~1 of \citealt{Jaskot2025LyCreview}), where $[\mathrm{O~II}]$ is suppressed, due to a lack of low-ionization gas in the ISM.
\begin{figure}
    \centering
    \includegraphics[width=0.9\linewidth]{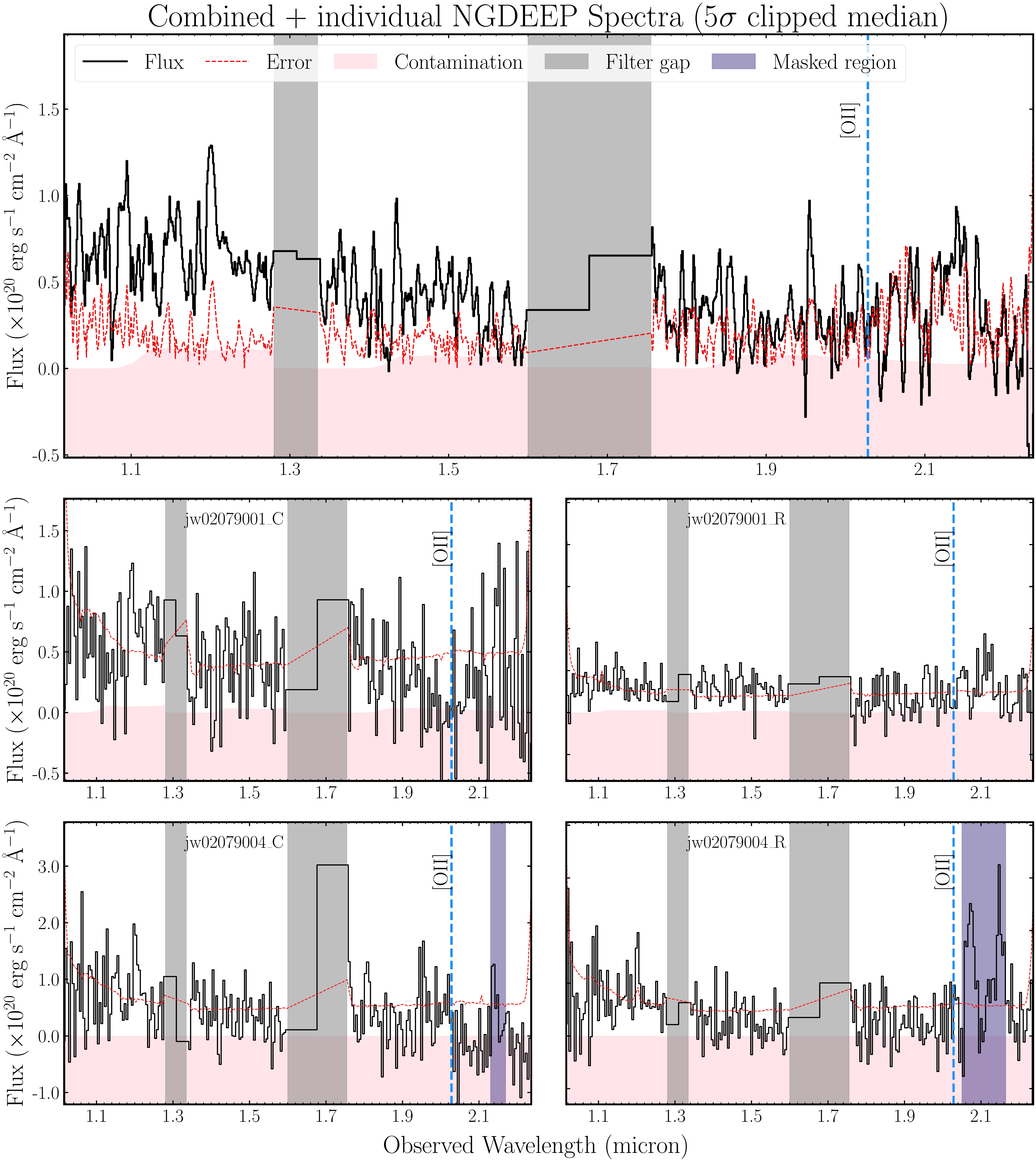}
    \caption{The JWST/NIRISS NGDEEP spectrum of \objname (F115W, F150W and F200W filters). The four individual orientations are shown in the lower panels and the combined ($5\sigma$ - clipped median) spectrum is shown above. Filter gaps are shaded in grey and contaminants based on the 2D spectra are masked in blue. No lines are detected. The position of $\mathrm{[O~II]}$ in F200W is shown with a dashed blue line.}
    \label{fig:placeholder}
\end{figure}


\bibliography{MXDFLCE}{}

@ARTICLE{Rafelski2015,
       author = {{Rafelski}, Marc and {Teplitz}, Harry I. and {Gardner}, Jonathan P. and {Coe}, Dan and {Bond}, Nicholas A. and {Koekemoer}, Anton M. and {Grogin}, Norman and {Kurczynski}, Peter and {McGrath}, Elizabeth J. and {Bourque}, Matthew and {Atek}, Hakim and {Brown}, Thomas M. and {Colbert}, James W. and {Codoreanu}, Alex and {Ferguson}, Henry C. and {Finkelstein}, Steven L. and {Gawiser}, Eric and {Giavalisco}, Mauro and {Gronwall}, Caryl and {Hanish}, Daniel J. and {Lee}, Kyoung-Soo and {Mehta}, Vihang and {de Mello}, Duilia F. and {Ravindranath}, Swara and {Ryan}, Russell E. and {Scarlata}, Claudia and {Siana}, Brian and {Soto}, Emmaris and {Voyer}, Elysse N.},
        title = "{UVUDF: Ultraviolet Through Near-infrared Catalog and Photometric Redshifts of Galaxies in the Hubble Ultra Deep Field}",
      journal = {\aj},
         year = 2015,
        month = jul,
       volume = {150},
       number = {1},
          eid = {31},
        pages = {31},
          doi = {10.1088/0004-6256/150/1/31},
archivePrefix = {arXiv},
       eprint = {1505.01160},
 primaryClass = {astro-ph.GA},
       adsurl = {https://ui.adsabs.harvard.edu/abs/2015AJ....150...31R}
}

@ARTICLE{Sun2024,
       author = {{Sun}, Lei and {Wang}, Xin and {Teplitz}, Harry I. and {Mehta}, Vihang and {Alavi}, Anahita and {Rafelski}, Marc and {Windhorst}, Rogier A. and {Scarlata}, Claudia and {Gardner}, Jonathan P. and {Smith}, Brent M. and {Sunnquist}, Ben and {Prichard}, Laura and {Cheng}, Yingjie and {Grogin}, Norman and {Hathi}, Nimish P. and {Hayes}, Matthew and {Koekemoer}, Anton M. and {Mobasher}, Bahram and {Nedkova}, Kalina V. and {O'Connell}, Robert and {Robertson}, Brant and {Taamoli}, Sina and {Yung}, L.~Y. Aaron and {Brammer}, Gabriel and {Colbert}, James and {Conselice}, Christopher and {Gawiser}, Eric and {Guo}, Yicheng and {Jansen}, Rolf A. and {Ji}, Zhiyuan and {Lucas}, Ray A. and {Rutkowski}, Michael and {Siana}, Brian and {Vanzella}, Eros and {Ashcraft}, Teresa and {Bagley}, Micaela and {Baronchelli}, Ivano and {Barro}, Guillermo and {Blanche}, Alex and {Broussard}, Adam and {Carleton}, Timothy and {Chartab}, Nima and {Codoreanu}, Alex and {Cohen}, Seth and {Dai}, Y. Sophia and {Darvish}, Behnam and {Dav{\'e}}, Romeel and {Degroot}, Laura and {de Mello}, Duilia and {Dickinson}, Mark and {Emami}, Najmeh and {Ferguson}, Henry and {Ferreira}, Leonardo and {Finkelstein}, Keely and {Finkelstein}, Steven and {Gburek}, Timothy and {Giavalisco}, Mauro and {Grazian}, Andrea and {Gronwall}, Caryl and {Hemmati}, Shoubaneh and {Howell}, Justin and {Iyer}, Kartheik and {Kaviraj}, Sugata and {Kurczynski}, Peter and {Lazar}, Ilin and {MacKenty}, John and {Mantha}, Kameswara Bharadwaj and {Martin}, Alec and {Martin}, Garreth and {McCabe}, Tyler and {Olsen}, Charlotte and {Otteson}, Lillian and {Ravindranath}, Swara and {Redshaw}, Caleb and {Sattari}, Zahra and {Soto}, Emmaris and {Zabelle}, Bonnabelle and {The Uvcandels Team}},
        title = "{The Ultraviolet Luminosity Function at 0.6 < z < 1 from UVCANDELS}",
      journal = {\apj},
         year = 2024,
        month = sep,
       volume = {972},
       number = {1},
          eid = {8},
        pages = {8},
          doi = {10.3847/1538-4357/ad5540},
archivePrefix = {arXiv},
       eprint = {2311.15664},
 primaryClass = {astro-ph.GA},
       adsurl = {https://ui.adsabs.harvard.edu/abs/2024ApJ...972....8S}
}

@ARTICLE{bacon2022musedatareleaseII,
       author = {{Bacon}, Roland and {Brinchmann}, Jarle and {Conseil}, Simon and {Maseda}, Michael and {Nanayakkara}, Themiya and {Wendt}, Martin and {Bacher}, Raphael and {Mary}, David and {Weilbacher}, Peter M. and {Krajnovi{\'c}}, Davor and {Boogaard}, Leindert and {Bouch{\'e}}, Nicolas and {Contini}, Thierry and {Epinat}, Beno{\^\i}t and {Feltre}, Anna and {Guo}, Yucheng and {Herenz}, Christian and {Kollatschny}, Wolfram and {Kusakabe}, Haruka and {Leclercq}, Floriane and {Michel-Dansac}, L{\'e}o and {Pello}, Roser and {Richard}, Johan and {Roth}, Martin and {Salvignol}, Gregory and {Schaye}, Joop and {Steinmetz}, Matthias and {Tresse}, Laurence and {Urrutia}, Tanya and {Verhamme}, Anne and {Vitte}, Eloise and {Wisotzki}, Lutz and {Zoutendijk}, Sebastiaan L.},
        title = "{The MUSE Hubble Ultra Deep Field surveys: Data release II}",
      journal = {\aap},
         year = 2023,
        month = feb,
       volume = {670},
          eid = {A4},
        pages = {A4},
          doi = {10.1051/0004-6361/202244187},
archivePrefix = {arXiv},
       eprint = {2211.08493},
 primaryClass = {astro-ph.GA},
       adsurl = {https://ui.adsabs.harvard.edu/abs/2023A&A...670A...4B}
}

@ARTICLE{Flury2022LyC_lowz_survey,
       author = {{Flury}, Sophia R. and {Jaskot}, Anne E. and {Ferguson}, Harry C. and {Worseck}, G{\'a}bor and {Makan}, Kirill and {Chisholm}, John and {Saldana-Lopez}, Alberto and {Schaerer}, Daniel and {McCandliss}, Stephan and {Wang}, Bingjie and {Ford}, N.~M. and {Heckman}, Timothy and {Ji}, Zhiyuan and {Giavalisco}, Mauro and {Amorin}, Ricardo and {Atek}, Hakim and {Blaizot}, Jeremy and {Borthakur}, Sanchayeeta and {Carr}, Cody and {Castellano}, Marco and {Cristiani}, Stefano and {De Barros}, Stephane and {Dickinson}, Mark and {Finkelstein}, Steven L. and {Fleming}, Brian and {Fontanot}, Fabio and {Garel}, Thibault and {Grazian}, Andrea and {Hayes}, Matthew and {Henry}, Alaina and {Mauerhofer}, Valentin and {Micheva}, Genoveva and {Oey}, M.~S. and {Ostlin}, Goran and {Papovich}, Casey and {Pentericci}, Laura and {Ravindranath}, Swara and {Rosdahl}, Joakim and {Rutkowski}, Michael and {Santini}, Paola and {Scarlata}, Claudia and {Teplitz}, Harry and {Thuan}, Trinh and {Trebitsch}, Maxime and {Vanzella}, Eros and {Verhamme}, Anne and {Xu}, Xinfeng},
        title = "{The Low-redshift Lyman Continuum Survey. I. New, Diverse Local Lyman Continuum Emitters}",
      journal = {\apjs},
         year = 2022,
        month = may,
       volume = {260},
       number = {1},
          eid = {1},
        pages = {1},
          doi = {10.3847/1538-4365/ac5331},
archivePrefix = {arXiv},
       eprint = {2201.11716},
 primaryClass = {astro-ph.GA},
       adsurl = {https://ui.adsabs.harvard.edu/abs/2022ApJS..260....1F}
}

@article{Simmonds2023xi_ion_LAEs_z6,
   title={The ionizing photon production efficiency at z ∼ 6 for Lyman-alpha emitters using JEMS and MUSE},
   volume={523},
   ISSN={1365-2966},
   url={http://dx.doi.org/10.1093/mnras/stad1749},
   DOI={10.1093/mnras/stad1749},
   number={4},
   journal={Monthly Notices of the Royal Astronomical Society},
   publisher={Oxford University Press (OUP)},
   author={Simmonds, C and Tacchella, S and Maseda, M and Williams, C C and Baker, W M and Witten, C E C and Johnson, B D and Robertson, B and Saxena, A and Sun, F and Witstok, J and Bhatawdekar, R and Boyett, K and Bunker, A J and Charlot, S and Curtis-Lake, E and Egami, E and Eisenstein, D J and Ji, Z and Maiolino, R and Sandles, L and Smit, R and Übler, H and Willott, C J},
   year={2023},
   month=jun, pages={5468–5486} }

@ARTICLE{Izotov2024lya_lowz_Zpoor,
       author = {{Izotov}, Y.~I. and {Thuan}, T.~X. and {Guseva}, N.~G. and {Schaerer}, D. and {Worseck}, G. and {Verhamme}, A.},
        title = "{Ly {\ensuremath{\alpha}} emission in low-redshift most metal-deficient compact star-forming galaxies}",
      journal = {\mnras},
         year = 2024,
        month = jan,
       volume = {527},
       number = {1},
        pages = {281-297},
          doi = {10.1093/mnras/stad3151},
archivePrefix = {arXiv},
       eprint = {2310.08441},
 primaryClass = {astro-ph.GA},
       adsurl = {https://ui.adsabs.harvard.edu/abs/2024MNRAS.527..281I}
}

@ARTICLE{Izotov2022Lya_LyC_MgIISFGs,
       author = {{Izotov}, Y.~I. and {Chisholm}, J. and {Worseck}, G. and {Guseva}, N.~G. and {Schaerer}, D. and {Prochaska}, J.~X.},
        title = "{Lyman alpha and Lyman continuum emission of Mg II-selected star-forming galaxies}",
      journal = {\mnras},
         year = 2022,
        month = sep,
       volume = {515},
       number = {2},
        pages = {2864-2881},
          doi = {10.1093/mnras/stac1899},
archivePrefix = {arXiv},
       eprint = {2207.04483},
 primaryClass = {astro-ph.GA},
       adsurl = {https://ui.adsabs.harvard.edu/abs/2022MNRAS.515.2864I}
}

@ARTICLE{Izotov2020lya_lowz_SFGsOIII/OII,
       author = {{Izotov}, Y.~I. and {Schaerer}, D. and {Worseck}, G. and {Verhamme}, A. and {Guseva}, N.~G. and {Thuan}, T.~X. and {Orlitov{\'a}}, I. and {Fricke}, K.~J.},
        title = "{Diverse properties of Ly {\ensuremath{\alpha}} emission in low-redshift compact star-forming galaxies with extremely high [O III]/[O II] ratios}",
      journal = {\mnras},
         year = 2020,
        month = jan,
       volume = {491},
       number = {1},
        pages = {468-482},
          doi = {10.1093/mnras/stz3041},
archivePrefix = {arXiv},
       eprint = {1910.12773},
 primaryClass = {astro-ph.GA},
       adsurl = {https://ui.adsabs.harvard.edu/abs/2020MNRAS.491..468I}
}

@article{begley2024lya_LyC_z45,
  title={Connecting the escape fraction of Lyman-alpha and Lyman-continuum photons in star-forming galaxies at z≃ 4--5},
  author={Begley, R and Cullen, F and McLure, RJ and Shapley, AE and Dunlop, JS and Carnall, AC and McLeod, DJ and Donnan, CT and Hamadouche, ML and Stanton, TM},
  journal={Monthly Notices of the Royal Astronomical Society},
  volume={527},
  number={2},
  pages={4040--4051},
  year={2024},
  publisher={Oxford University Press}
}

@article{Rosdahl2022SPHINX_LyC_EoR,
   title={LyC escape from <scp>sphinx</scp> galaxies in the Epoch of Reionization},
   volume={515},
   ISSN={1365-2966},
   url={http://dx.doi.org/10.1093/mnras/stac1942},
   DOI={10.1093/mnras/stac1942},
   number={2},
   journal={Monthly Notices of the Royal Astronomical Society},
   publisher={Oxford University Press (OUP)},
   author={Rosdahl, Joakim and Blaizot, Jérémy and Katz, Harley and Kimm, Taysun and Garel, Thibault and Haehnelt, Martin and Keating, Laura C and Martin-Alvarez, Sergio and Michel-Dansac, Léo and Ocvirk, Pierre},
   year={2022},
   month=jul, pages={2386–2414} }

@ARTICLE{Jaskot2025LyCreview,
       author = {{Jaskot}, Anne E.},
        title = "{Ionizing Radiation Escape from Low-Redshift Galaxies and Its Connection to Cosmic Reionization}",
      journal = {\araa},
         year = 2025,
        month = aug,
       volume = {63},
       number = {1},
        pages = {45-82},
          doi = {10.1146/annurev-astro-111324-074935},
archivePrefix = {arXiv},
       eprint = {2508.18411},
 primaryClass = {astro-ph.GA},
       adsurl = {https://ui.adsabs.harvard.edu/abs/2025ARA&A..63...45J}
}

@ARTICLE{Jaskot2024multivariateLyC_lowz,
       author = {{Jaskot}, Anne E. and {Silveyra}, Anneliese C. and {Plantinga}, Anna and {Flury}, Sophia R. and {Hayes}, Matthew and {Chisholm}, John and {Heckman}, Timothy and {Pentericci}, Laura and {Schaerer}, Daniel and {Trebitsch}, Maxime and {Verhamme}, Anne and {Carr}, Cody and {Ferguson}, Henry C. and {Ji}, Zhiyuan and {Giavalisco}, Mauro and {Henry}, Alaina and {Marques-Chaves}, Rui and {{\"O}stlin}, G{\"o}ran and {Saldana-Lopez}, Alberto and {Scarlata}, Claudia and {Worseck}, G{\'a}bor and {Xu}, Xinfeng},
        title = "{Multivariate Predictors of Lyman Continuum Escape. I. A Survival Analysis of the Low-redshift Lyman Continuum Survey}",
      journal = {\apj},
         year = 2024,
        month = sep,
       volume = {972},
       number = {1},
          eid = {92},
        pages = {92},
          doi = {10.3847/1538-4357/ad58b9},
archivePrefix = {arXiv},
       eprint = {2406.10171},
 primaryClass = {astro-ph.GA},
       adsurl = {https://ui.adsabs.harvard.edu/abs/2024ApJ...972...92J}
}

@ARTICLE{Kerutt2024LAE_LyC,
       author = {{Kerutt}, J. and {Oesch}, P.~A. and {Wisotzki}, L. and {Verhamme}, A. and {Atek}, H. and {Herenz}, E.~C. and {Illingworth}, G.~D. and {Kusakabe}, H. and {Matthee}, J. and {Mauerhofer}, V. and {Montes}, M. and {Naidu}, R.~P. and {Nelson}, E. and {Reddy}, N. and {Schaye}, J. and {Simmonds}, C. and {Urrutia}, T. and {Vitte}, E.},
        title = "{Lyman continuum leaker candidates at z {\ensuremath{\sim}} 3-4 in the HDUV based on a spectroscopic sample of MUSE LAEs}",
      journal = {\aap},
         year = 2024,
        month = apr,
       volume = {684},
          eid = {A42},
        pages = {A42},
          doi = {10.1051/0004-6361/202346656},
archivePrefix = {arXiv},
       eprint = {2312.08791},
 primaryClass = {astro-ph.GA},
       adsurl = {https://ui.adsabs.harvard.edu/abs/2024A&A...684A..42K}
}

@ARTICLE{Citro2025LAEs_noLyC,
       author = {{Citro}, Annalisa and {Scarlata}, Claudia M. and {Mantha}, Kameswara B. and {Williams}, Liliya R. and {Rafelski}, Marc and {Revalski}, Mitchell and {Hayes}, Matthew J. and {Henry}, Alaina and {Rutkowski}, Michael J. and {Teplitz}, Harry I. and {Grazian}, Andrea and {Alavi}, Anahita},
        title = "{Challenging the LyC{\textendash}Ly{\ensuremath{\alpha}} Relation: Strong Ly{\ensuremath{\alpha}} Emitters without LyC Leakage at z {\ensuremath{\sim}} 2.3}",
      journal = {\apj},
         year = 2025,
        month = jun,
       volume = {986},
       number = {2},
          eid = {184},
        pages = {184},
          doi = {10.3847/1538-4357/add5e6},
archivePrefix = {arXiv},
       eprint = {2406.07618},
 primaryClass = {astro-ph.GA},
       adsurl = {https://ui.adsabs.harvard.edu/abs/2025ApJ...986..184C}
}

@ARTICLE{Rieke2023JADES_DR1,
       author = {{Rieke}, Marcia J. and {Robertson}, Brant and {Tacchella}, Sandro and {Hainline}, Kevin and {Johnson}, Benjamin D. and {Hausen}, Ryan and {Ji}, Zhiyuan and {Willmer}, Christopher N.~A. and {Eisenstein}, Daniel J. and {Pusk{\'a}s}, D{\'a}vid and {Alberts}, Stacey and {Arribas}, Santiago and {Baker}, William M. and {Baum}, Stefi and {Bhatawdekar}, Rachana and {Bonaventura}, Nina and {Boyett}, Kristan and {Bunker}, Andrew J. and {Cameron}, Alex J. and {Carniani}, Stefano and {Charlot}, Stephane and {Chevallard}, Jacopo and {Chen}, Zuyi and {Curti}, Mirko and {Curtis-Lake}, Emma and {Danhaive}, A. Lola and {DeCoursey}, Christa and {Dressler}, Alan and {Egami}, Eiichi and {Endsley}, Ryan and {Helton}, Jakob M. and {Hviding}, Raphael E. and {Kumari}, Nimisha and {Looser}, Tobias J. and {Lyu}, Jianwei and {Maiolino}, Roberto and {Maseda}, Michael V. and {Nelson}, Erica J. and {Rieke}, George and {Rix}, Hans-Walter and {Sandles}, Lester and {Saxena}, Aayush and {Sharpe}, Katherine and {Shivaei}, Irene and {Skarbinski}, Maya and {Smit}, Renske and {Stark}, Daniel P. and {Stone}, Meredith and {Suess}, Katherine A. and {Sun}, Fengwu and {Topping}, Michael and {{\"U}bler}, Hannah and {Villanueva}, Natalia C. and {Wallace}, Imaan E.~B. and {Williams}, Christina C. and {Willott}, Chris and {Whitler}, Lily and {Witstok}, Joris and {Woodrum}, Charity},
        title = "{JADES Initial Data Release for the Hubble Ultra Deep Field: Revealing the Faint Infrared Sky with Deep JWST NIRCam Imaging}",
      journal = {\apjs},
         year = 2023,
        month = nov,
       volume = {269},
       number = {1},
          eid = {16},
        pages = {16},
          doi = {10.3847/1538-4365/acf44d},
archivePrefix = {arXiv},
       eprint = {2306.02466},
 primaryClass = {astro-ph.GA},
       adsurl = {https://ui.adsabs.harvard.edu/abs/2023ApJS..269...16R}
}

@ARTICLE{Eisenstein2023JADES,
       author = {{Eisenstein}, Daniel J. and {Willott}, Chris and {Alberts}, Stacey and {Arribas}, Santiago and {Bonaventura}, Nina and {Bunker}, Andrew J. and {Cameron}, Alex J. and {Carniani}, Stefano and {Charlot}, Stephane and {Curtis-Lake}, Emma and {D'Eugenio}, Francesco and {Endsley}, Ryan and {Ferruit}, Pierre and {Giardino}, Giovanna and {Hainline}, Kevin and {Hausen}, Ryan and {Jakobsen}, Peter and {Johnson}, Benjamin D. and {Maiolino}, Roberto and {Rieke}, Marcia and {Rieke}, George and {Rix}, Hans-Walter and {Robertson}, Brant and {Stark}, Daniel P. and {Tacchella}, Sandro and {Williams}, Christina C. and {Willmer}, Christopher N.~A. and {Baker}, William M. and {Baum}, Stefi and {Bhatawdekar}, Rachana and {Boyett}, Kristan and {Chen}, Zuyi and {Chevallard}, Jacopo and {Circosta}, Chiara and {Curti}, Mirko and {Danhaive}, A. Lola and {DeCoursey}, Christa and {de Graaff}, Anna and {Dressler}, Alan and {Egami}, Eiichi and {Helton}, Jakob M. and {Hviding}, Raphael E. and {Ji}, Zhiyuan and {Jones}, Gareth C. and {Kumari}, Nimisha and {L{\"u}tzgendorf}, Nora and {Laseter}, Isaac and {Looser}, Tobias J. and {Lyu}, Jianwei and {Maseda}, Michael V. and {Nelson}, Erica and {Parlanti}, Eleonora and {Perna}, Michele and {Pusk{\'a}s}, D{\'a}vid and {Rawle}, Tim and {Rodr{\'\i}guez Del Pino}, Bruno and {Sandles}, Lester and {Saxena}, Aayush and {Scholtz}, Jan and {Sharpe}, Katherine and {Shivaei}, Irene and {Silcock}, Maddie S. and {Simmonds}, Charlotte and {Skarbinski}, Maya and {Smit}, Renske and {Stone}, Meredith and {Suess}, Katherine A. and {Sun}, Fengwu and {Tang}, Mengtao and {Topping}, Michael W. and {{\"U}bler}, Hannah and {Villanueva}, Natalia C. and {Wallace}, Imaan E.~B. and {Whitler}, Lily and {Witstok}, Joris and {Woodrum}, Charity},
        title = "{Overview of the JWST Advanced Deep Extragalactic Survey (JADES)}",
      journal = {arXiv e-prints},
         year = 2023,
        month = jun,
          eid = {arXiv:2306.02465},
        pages = {arXiv:2306.02465},
          doi = {10.48550/arXiv.2306.02465},
archivePrefix = {arXiv},
       eprint = {2306.02465},
 primaryClass = {astro-ph.GA},
       adsurl = {https://ui.adsabs.harvard.edu/abs/2023arXiv230602465E}
}

@ARTICLE{Chisholm2022LyC_beta,
       author = {{Chisholm}, J. and {Saldana-Lopez}, A. and {Flury}, S. and {Schaerer}, D. and {Jaskot}, A. and {Amor{\'\i}n}, R. and {Atek}, H. and {Finkelstein}, S.~L. and {Fleming}, B. and {Ferguson}, H. and {Fern{\'a}ndez}, V. and {Giavalisco}, M. and {Hayes}, M. and {Heckman}, T. and {Henry}, A. and {Ji}, Z. and {Marques-Chaves}, R. and {Mauerhofer}, V. and {McCandliss}, S. and {Oey}, M.~S. and {{\"O}stlin}, G. and {Rutkowski}, M. and {Scarlata}, C. and {Thuan}, T. and {Trebitsch}, M. and {Wang}, B. and {Worseck}, G. and {Xu}, X.},
        title = "{The far-ultraviolet continuum slope as a Lyman Continuum escape estimator at high redshift}",
      journal = {\mnras},
         year = 2022,
        month = dec,
       volume = {517},
       number = {4},
        pages = {5104-5120},
          doi = {10.1093/mnras/stac2874},
archivePrefix = {arXiv},
       eprint = {2207.05771},
 primaryClass = {astro-ph.GA},
       adsurl = {https://ui.adsabs.harvard.edu/abs/2022MNRAS.517.5104C}
}

@ARTICLE{Izotov2018LyC_OIII/OII,
       author = {{Izotov}, Y.~I. and {Worseck}, G. and {Schaerer}, D. and {Guseva}, N.~G. and {Thuan}, T.~X. and {Fricke}, A., Verhamme and {Orlitov{\'a}}, I.},
        title = "{Low-redshift Lyman continuum leaking galaxies with high [O III]/[O II] ratios}",
      journal = {\mnras},
         year = 2018,
        month = aug,
       volume = {478},
       number = {4},
        pages = {4851-4865},
          doi = {10.1093/mnras/sty1378},
archivePrefix = {arXiv},
       eprint = {1805.09865},
 primaryClass = {astro-ph.GA},
       adsurl = {https://ui.adsabs.harvard.edu/abs/2018MNRAS.478.4851I}
}

@ARTICLE{Choustikov2024LyC_tracers,
       author = {{Choustikov}, Nicholas and {Katz}, Harley and {Saxena}, Aayush and {Cameron}, Alex J. and {Devriendt}, Julien and {Slyz}, Adrianne and {Rosdahl}, Joki and {Blaizot}, Jeremy and {Michel-Dansac}, Leo},
        title = "{The Physics of Indirect Estimators of Lyman Continuum Escape and their Application to High-Redshift JWST Galaxies}",
      journal = {\mnras},
         year = 2024,
        month = apr,
       volume = {529},
       number = {4},
        pages = {3751-3767},
          doi = {10.1093/mnras/stae776},
archivePrefix = {arXiv},
       eprint = {2304.08526},
 primaryClass = {astro-ph.GA},
       adsurl = {https://ui.adsabs.harvard.edu/abs/2024MNRAS.529.3751C}
}

@ARTICLE{Maji2022pred_LyC_fromLya,
       author = {{Maji}, Moupiya and {Verhamme}, Anne and {Rosdahl}, Joakim and {Garel}, Thibault and {Blaizot}, J{\'e}r{\'e}my and {Mauerhofer}, Valentin and {Pittavino}, Marta and {Victoria Feser}, Maria-Pia and {Chuniaud}, Mathieu and {Kimm}, Taysun and {Katz}, Harley and {Haehnelt}, Martin},
        title = "{Predicting Lyman-continuum emission of galaxies using their physical and Lyman-alpha emission properties}",
      journal = {\aap},
         year = 2022,
        month = jul,
       volume = {663},
          eid = {A66},
        pages = {A66},
          doi = {10.1051/0004-6361/202142740},
archivePrefix = {arXiv},
       eprint = {2204.02440},
 primaryClass = {astro-ph.GA},
       adsurl = {https://ui.adsabs.harvard.edu/abs/2022A&A...663A..66M}
}

@ARTICLE{Izotov2016LyC_compactgals,
       author = {{Izotov}, Y.~I. and {Schaerer}, D. and {Thuan}, T.~X. and {Worseck}, G. and {Guseva}, N.~G. and {Orlitov{\'a}}, I. and {Verhamme}, A.},
        title = "{Detection of high Lyman continuum leakage from four low-redshift compact star-forming galaxies}",
      journal = {\mnras},
         year = 2016,
        month = oct,
       volume = {461},
       number = {4},
        pages = {3683-3701},
          doi = {10.1093/mnras/stw1205},
archivePrefix = {arXiv},
       eprint = {1605.05160},
 primaryClass = {astro-ph.GA},
       adsurl = {https://ui.adsabs.harvard.edu/abs/2016MNRAS.461.3683I}
}

@ARTICLE{Naidu2020UVbright_high_fesc,
       author = {{Naidu}, Rohan P. and {Tacchella}, Sandro and {Mason}, Charlotte A. and {Bose}, Sownak and {Oesch}, Pascal A. and {Conroy}, Charlie},
        title = "{Rapid Reionization by the Oligarchs: The Case for Massive, UV-bright, Star-forming Galaxies with High Escape Fractions}",
      journal = {\apj},
         year = 2020,
        month = apr,
       volume = {892},
       number = {2},
          eid = {109},
        pages = {109},
          doi = {10.3847/1538-4357/ab7cc9},
archivePrefix = {arXiv},
       eprint = {1907.13130},
 primaryClass = {astro-ph.GA},
       adsurl = {https://ui.adsabs.harvard.edu/abs/2020ApJ...892..109N}
}

@ARTICLE{Verhamme2017LyC_strongLAEs_lowz,
       author = {{Verhamme}, A. and {Orlitov{\'a}}, I. and {Schaerer}, D. and {Izotov}, Y. and {Worseck}, G. and {Thuan}, T.~X. and {Guseva}, N.},
        title = "{Lyman-{\ensuremath{\alpha}} spectral properties of five newly discovered Lyman continuum emitters}",
      journal = {\aap},
         year = 2017,
        month = jan,
       volume = {597},
          eid = {A13},
        pages = {A13},
          doi = {10.1051/0004-6361/201629264},
archivePrefix = {arXiv},
       eprint = {1609.03477},
 primaryClass = {astro-ph.GA},
       adsurl = {https://ui.adsabs.harvard.edu/abs/2017A&A...597A..13V}
}

@ARTICLE{Verhamme2015Lya_to_get_LyC,
       author = {{Verhamme}, Anne and {Orlitov{\'a}}, Ivana and {Schaerer}, Daniel and {Hayes}, Matthew},
        title = "{Using Lyman-{\ensuremath{\alpha}} to detect galaxies that leak Lyman continuum}",
      journal = {\aap},
         year = 2015,
        month = jun,
       volume = {578},
          eid = {A7},
        pages = {A7},
          doi = {10.1051/0004-6361/201423978},
archivePrefix = {arXiv},
       eprint = {1404.2958},
 primaryClass = {astro-ph.GA},
       adsurl = {https://ui.adsabs.harvard.edu/abs/2015A&A...578A...7V}
}

@ARTICLE{Prichard2022LCEs,
       author = {{Prichard}, Laura J. and {Rafelski}, Marc and {Cooke}, Jeff and {Me{\v{s}}tri{\'c}}, Uros and {Bassett}, Robert and {Ryan-Weber}, Emma V. and {Sunnquist}, Ben and {Alavi}, Anahita and {Hathi}, Nimish and {Wang}, Xin and {Revalski}, Mitchell and {Bajaj}, Varun and {O'Meara}, John M. and {Spitler}, Lee},
        title = "{Lyman Continuum Galaxy Candidates in COSMOS}",
      journal = {\apj},
         year = 2022,
        month = jan,
       volume = {924},
       number = {1},
          eid = {14},
        pages = {14},
          doi = {10.3847/1538-4357/ac3004},
archivePrefix = {arXiv},
       eprint = {2110.06945},
 primaryClass = {astro-ph.GA},
       adsurl = {https://ui.adsabs.harvard.edu/abs/2022ApJ...924...14P}
}

@ARTICLE{Steidel2001LyC_z3,
       author = {{Steidel}, Charles C. and {Pettini}, Max and {Adelberger}, Kurt L.},
        title = "{Lyman-Continuum Emission from Galaxies at Z \raisebox{-0.5ex}\textasciitilde= 3.4}",
      journal = {\apj},
         year = 2001,
        month = jan,
       volume = {546},
       number = {2},
        pages = {665-671},
          doi = {10.1086/318323},
archivePrefix = {arXiv},
       eprint = {astro-ph/0008283},
 primaryClass = {astro-ph},
       adsurl = {https://ui.adsabs.harvard.edu/abs/2001ApJ...546..665S}
}

@ARTICLE{Kennicutt1998SFR,
       author = {{Kennicutt}, Jr., Robert C.},
        title = "{Star Formation in Galaxies Along the Hubble Sequence}",
      journal = {\araa},
         year = 1998,
        month = jan,
       volume = {36},
        pages = {189-232},
          doi = {10.1146/annurev.astro.36.1.189},
archivePrefix = {arXiv},
       eprint = {astro-ph/9807187},
 primaryClass = {astro-ph},
       adsurl = {https://ui.adsabs.harvard.edu/abs/1998ARA&A..36..189K}
}

@ARTICLE{ForemanMackey2019emcee,
       author = {{Foreman-Mackey}, Daniel and {Farr}, Will and {Sinha}, Manodeep and {Archibald}, Anne and {Hogg}, David and {Sanders}, Jeremy and {Zuntz}, Joe and {Williams}, Peter and {Nelson}, Andrew and {de Val-Borro}, Miguel and {Erhardt}, Tobias and {Pashchenko}, Ilya and {Pla}, Oriol},
        title = "{emcee v3: A Python ensemble sampling toolkit for affine-invariant MCMC}",
      journal = {The Journal of Open Source Software},
         year = 2019,
        month = nov,
       volume = {4},
       number = {43},
          eid = {1864},
        pages = {1864},
          doi = {10.21105/joss.01864},
archivePrefix = {arXiv},
       eprint = {1911.07688},
 primaryClass = {astro-ph.IM},
       adsurl = {https://ui.adsabs.harvard.edu/abs/2019JOSS....4.1864F}
}

@ARTICLE{SaldanaLopez2025HF,
       author = {{Saldana-Lopez}, A. and {Hayes}, M.~J. and {Le Reste}, A. and {Scarlata}, C. and {Melinder}, J. and {Henry}, A. and {Leclercq}, F. and {Garel}, T. and {Amor{\'\i}n}, R. and {Atek}, H. and {Bait}, O. and {Carr}, C.~A. and {Chisholm}, J. and {Flury}, S.~R. and {Heckman}, T.~M. and {Jaskot}, A.~E. and {Jung}, I. and {Ji}, Z. and {Komarova}, L. and {Lin}, Y.-H. and {Oey}, M.~S. and {{\"O}stlin}, G. and {Pentericci}, L. and {Runnholm}, A. and {Schaerer}, D. and {Thuan}, T.~X. and {Xu}, X.},
        title = "{The Ly{\ensuremath{\alpha}} and Continuum Origins Survey. II. The Connection between the Escape of Ionizing Radiation and Ly{\ensuremath{\alpha}} Halos in Star-forming Galaxies}",
      journal = {\apj},
         year = 2026,
        month = mar,
       volume = {999},
       number = {1},
          eid = {71},
        pages = {71},
          doi = {10.3847/1538-4357/ae36a8},
archivePrefix = {arXiv},
       eprint = {2504.07074},
 primaryClass = {astro-ph.GA},
       adsurl = {https://ui.adsabs.harvard.edu/abs/2026ApJ...999...71S}
}

@ARTICLE{Mestric2025ion3,
       author = {{Me{\v{s}}tri{\'c}}, U. and {Vanzella}, E. and {Beckett}, A. and {Rafelski}, M. and {Grillo}, C. and {Giavalisco}, M. and {Messa}, M. and {Castellano}, M. and {Calura}, F. and {Cupani}, G. and {Zanella}, A. and {Bergamini}, P. and {Meneghetti}, M. and {Mercurio}, A. and {Rosati}, P. and {Nonino}, M. and {Caputi}, K. and {Comastri}, A.},
        title = "{Unraveling the Lyman continuum emission of Ion3: Insights from HST multiband imaging and X-Shooter spectroscopy}",
      journal = {\aap},
         year = 2025,
        month = jun,
       volume = {698},
          eid = {A203},
        pages = {A203},
          doi = {10.1051/0004-6361/202451959},
archivePrefix = {arXiv},
       eprint = {2504.18711},
 primaryClass = {astro-ph.GA},
       adsurl = {https://ui.adsabs.harvard.edu/abs/2025A&A...698A.203M}
}

@ARTICLE{Vanzella2018Ion3,
       author = {{Vanzella}, E. and {Nonino}, M. and {Cupani}, G. and {Castellano}, M. and {Sani}, E. and {Mignoli}, M. and {Calura}, F. and {Meneghetti}, M. and {Gilli}, R. and {Comastri}, A. and {Mercurio}, A. and {Caminha}, G.~B. and {Caputi}, K. and {Rosati}, P. and {Grillo}, C. and {Cristiani}, S. and {Balestra}, I. and {Fontana}, A. and {Giavalisco}, M.},
        title = "{Direct Lyman continuum and Ly {\ensuremath{\alpha}} escape observed at redshift 4}",
      journal = {\mnras},
         year = 2018,
        month = may,
       volume = {476},
       number = {1},
        pages = {L15-L19},
          doi = {10.1093/mnrasl/sly023},
archivePrefix = {arXiv},
       eprint = {1712.07661},
 primaryClass = {astro-ph.GA},
       adsurl = {https://ui.adsabs.harvard.edu/abs/2018MNRAS.476L..15V}
}

@ARTICLE{FLury2022LzLCS_results,
       author = {{Flury}, Sophia R. and {Jaskot}, Anne E. and {Ferguson}, Harry C. and {Worseck}, G{\'a}bor and {Makan}, Kirill and {Chisholm}, John and {Saldana-Lopez}, Alberto and {Schaerer}, Daniel and {McCandliss}, Stephan R. and {Xu}, Xinfeng and {Wang}, Bingjie and {Oey}, M.~S. and {Ford}, N.~M. and {Heckman}, Timothy and {Ji}, Zhiyuan and {Giavalisco}, Mauro and {Amor{\'\i}n}, Ricardo and {Atek}, Hakim and {Blaizot}, Jeremy and {Borthakur}, Sanchayeeta and {Carr}, Cody and {Castellano}, Marco and {De Barros}, Stephane and {Dickinson}, Mark and {Finkelstein}, Steven L. and {Fleming}, Brian and {Fontanot}, Fabio and {Garel}, Thibault and {Grazian}, Andrea and {Hayes}, Matthew and {Henry}, Alaina and {Mauerhofer}, Valentin and {Micheva}, Genoveva and {Ostlin}, Goran and {Papovich}, Casey and {Pentericci}, Laura and {Ravindranath}, Swara and {Rosdahl}, Joakim and {Rutkowski}, Michael and {Santini}, Paola and {Scarlata}, Claudia and {Teplitz}, Harry and {Thuan}, Trinh and {Trebitsch}, Maxime and {Vanzella}, Eros and {Verhamme}, Anne},
        title = "{The Low-redshift Lyman Continuum Survey. II. New Insights into LyC Diagnostics}",
      journal = {\apj},
         year = 2022,
        month = may,
       volume = {930},
       number = {2},
          eid = {126},
        pages = {126},
          doi = {10.3847/1538-4357/ac61e4},
archivePrefix = {arXiv},
       eprint = {2203.15649},
 primaryClass = {astro-ph.GA},
       adsurl = {https://ui.adsabs.harvard.edu/abs/2022ApJ...930..126F}
}

@ARTICLE{Siana2007LyC_z1,
       author = {{Siana}, Brian and {Teplitz}, Harry I. and {Colbert}, James and {Ferguson}, Henry C. and {Dickinson}, Mark and {Brown}, Thomas M. and {Conselice}, Christopher J. and {de Mello}, Duilia F. and {Gardner}, Jonathan P. and {Giavalisco}, Mauro and {Menanteau}, Felipe},
        title = "{New Constraints on the Lyman Continuum Escape Fraction at z\raisebox{-0.5ex}\textasciitilde1.3}",
      journal = {\apj},
         year = 2007,
        month = oct,
       volume = {668},
       number = {1},
        pages = {62-73},
          doi = {10.1086/521185},
archivePrefix = {arXiv},
       eprint = {0706.4093},
 primaryClass = {astro-ph},
       adsurl = {https://ui.adsabs.harvard.edu/abs/2007ApJ...668...62S}
}

@ARTICLE{Munoz2026burstiness,
       author = {{Mu{\~n}oz}, Julian B. and {Chisholm}, John and {Sun}, Guochao and {Samuel}, Jenna and {Mirocha}, Jordan and {Bregou}, Emily and {Venditti}, Alessandra and {Qezlou}, Mahdi and {Simmonds}, Charlotte and {Endsley}, Ryan},
        title = "{Relatively Fast and Reasonably Furious: Evidence for Increased Burstiness in Smaller Halos at Cosmic Dawn}",
      journal = {\mnras},
         year = 2026,
        month = mar,
          doi = {10.1093/mnras/stag415},
       adsurl = {https://ui.adsabs.harvard.edu/abs/2026MNRAS.tmp..405M}
}

@ARTICLE{Matthee2024MXDF_lya_transmission,
       author = {{Matthee}, Jorryt and {Golling}, Christopher and {Mackenzie}, Ruari and {Pezzulli}, Gabriele and {Lilly}, Simon and {Schaye}, Joop and {Bacon}, Roland and {Kusakabe}, Haruka and {Urrutia}, Tanya and {Boogaard}, Leindert and {Brinchmann}, Jarle and {Maseda}, Michael V. and {Garel}, Thibault and {Bouch{\'e}}, Nicolas F. and {Wisotzki}, Lutz},
        title = "{Large-scale excess H I absorption around z {\ensuremath{\approx}} 4 galaxies detected in a background galaxy spectrum in the MUSE eXtremely deep field}",
      journal = {\mnras},
         year = 2024,
        month = apr,
       volume = {529},
       number = {3},
        pages = {2794-2806},
          doi = {10.1093/mnras/stae673},
archivePrefix = {arXiv},
       eprint = {2305.15346},
 primaryClass = {astro-ph.GA},
       adsurl = {https://ui.adsabs.harvard.edu/abs/2024MNRAS.529.2794M}
}

@ARTICLE{Williams2018JAGUAR,
       author = {{Williams}, Christina C. and {Curtis-Lake}, Emma and {Hainline}, Kevin N. and {Chevallard}, Jacopo and {Robertson}, Brant E. and {Charlot}, Stephane and {Endsley}, Ryan and {Stark}, Daniel P. and {Willmer}, Christopher N.~A. and {Alberts}, Stacey and {Amorin}, Ricardo and {Arribas}, Santiago and {Baum}, Stefi and {Bunker}, Andrew and {Carniani}, Stefano and {Crandall}, Sara and {Egami}, Eiichi and {Eisenstein}, Daniel J. and {Ferruit}, Pierre and {Husemann}, Bernd and {Maseda}, Michael V. and {Maiolino}, Roberto and {Rawle}, Timothy D. and {Rieke}, Marcia and {Smit}, Renske and {Tacchella}, Sandro and {Willott}, Chris J.},
        title = "{The JWST Extragalactic Mock Catalog: Modeling Galaxy Populations from the UV through the Near-IR over 13 Billion Years of Cosmic History}",
      journal = {\apjs},
         year = 2018,
        month = jun,
       volume = {236},
       number = {2},
          eid = {33},
        pages = {33},
          doi = {10.3847/1538-4365/aabcbb},
archivePrefix = {arXiv},
       eprint = {1802.05272},
 primaryClass = {astro-ph.GA},
       adsurl = {https://ui.adsabs.harvard.edu/abs/2018ApJS..236...33W}
}

@ARTICLE{Chang2010Thue_theorem_proof,
       author = {{Chang}, Hai-Chau and {Wang}, Lih-Chung},
        title = "{A Simple Proof of Thue's Theorem on Circle Packing}",
      journal = {arXiv e-prints},
         year = 2010,
        month = sep,
          eid = {arXiv:1009.4322},
        pages = {arXiv:1009.4322},
          doi = {10.48550/arXiv.1009.4322},
archivePrefix = {arXiv},
       eprint = {1009.4322},
 primaryClass = {math.MG},
       adsurl = {https://ui.adsabs.harvard.edu/abs/2010arXiv1009.4322C}
}

@ARTICLE{Brammer2011quiescent_num_density,
       author = {{Brammer}, Gabriel B. and {Whitaker}, K.~E. and {van Dokkum}, P.~G. and {Marchesini}, D. and {Franx}, M. and {Kriek}, M. and {Labb{\'e}}, I. and {Lee}, K.-S. and {Muzzin}, A. and {Quadri}, R.~F. and {Rudnick}, G. and {Williams}, R.},
        title = "{The Number Density and Mass Density of Star-forming and Quiescent Galaxies at 0.4 <= z <= 2.2}",
      journal = {\apj},
         year = 2011,
        month = sep,
       volume = {739},
       number = {1},
          eid = {24},
        pages = {24},
          doi = {10.1088/0004-637X/739/1/24},
archivePrefix = {arXiv},
       eprint = {1104.2595},
 primaryClass = {astro-ph.CO},
       adsurl = {https://ui.adsabs.harvard.edu/abs/2011ApJ...739...24B}
}

@ARTICLE{Schaerer2022LyC_CIV,
       author = {{Schaerer}, D. and {Izotov}, Y.~I. and {Worseck}, G. and {Berg}, D. and {Chisholm}, J. and {Jaskot}, A. and {Nakajima}, K. and {Ravindranath}, S. and {Thuan}, T.~X. and {Verhamme}, A.},
        title = "{Strong Lyman continuum emitting galaxies show intense C IV {\ensuremath{\lambda}}1550 emission}",
      journal = {\aap},
         year = 2022,
        month = feb,
       volume = {658},
          eid = {L11},
        pages = {L11},
          doi = {10.1051/0004-6361/202243149},
archivePrefix = {arXiv},
       eprint = {2202.07768},
 primaryClass = {astro-ph.GA},
       adsurl = {https://ui.adsabs.harvard.edu/abs/2022A&A...658L..11S}
}

@ARTICLE{Saxena2022LyC_CIV,
       author = {{Saxena}, A. and {Cryer}, E. and {Ellis}, R.~S. and {Pentericci}, L. and {Calabr{\`o}}, A. and {Mascia}, S. and {Saldana-Lopez}, A. and {Schaerer}, D. and {Katz}, H. and {Llerena}, M. and {Amor{\'\i}n}, R.},
        title = "{Strong C IV emission from star-forming galaxies: a case for high Lyman continuum photon escape}",
      journal = {\mnras},
         year = 2022,
        month = nov,
       volume = {517},
       number = {1},
        pages = {1098-1111},
          doi = {10.1093/mnras/stac2742},
archivePrefix = {arXiv},
       eprint = {2206.06161},
 primaryClass = {astro-ph.GA},
       adsurl = {https://ui.adsabs.harvard.edu/abs/2022MNRAS.517.1098S}
}

@ARTICLE{Izotov2024_CIV_low_metal,
       author = {{Izotov}, Y.~I. and {Schaerer}, D. and {Guseva}, N.~G. and {Thuan}, T.~X. and {Worseck}, G.},
        title = "{Extremely strong C IV {\ensuremath{\lambda}}1550 nebular emission in the extremely low-metallicity star-forming galaxy J2229+2725}",
      journal = {\mnras},
         year = 2024,
        month = feb,
       volume = {528},
       number = {1},
        pages = {L10-L14},
          doi = {10.1093/mnrasl/slad166},
archivePrefix = {arXiv},
       eprint = {2311.02015},
 primaryClass = {astro-ph.GA},
       adsurl = {https://ui.adsabs.harvard.edu/abs/2024MNRAS.528L..10I}
}

@ARTICLE{Madau1996SF_history,
       author = {{Madau}, Piero and {Ferguson}, Henry C. and {Dickinson}, Mark E. and {Giavalisco}, Mauro and {Steidel}, Charles C. and {Fruchter}, Andrew},
        title = "{High-redshift galaxies in the Hubble Deep Field: colour selection and star formation history to z\raisebox{-0.5ex}\textasciitilde4}",
      journal = {\mnras},
         year = 1996,
        month = dec,
       volume = {283},
       number = {4},
        pages = {1388-1404},
          doi = {10.1093/mnras/283.4.1388},
archivePrefix = {arXiv},
       eprint = {astro-ph/9607172},
 primaryClass = {astro-ph},
       adsurl = {https://ui.adsabs.harvard.edu/abs/1996MNRAS.283.1388M}
}

@ARTICLE{Fan2006UVBackground_EoR,
       author = {{Fan}, Xiaohui and {Strauss}, Michael A. and {Becker}, Robert H. and {White}, Richard L. and {Gunn}, James E. and {Knapp}, Gillian R. and {Richards}, Gordon T. and {Schneider}, Donald P. and {Brinkmann}, J. and {Fukugita}, Masataka},
        title = "{Constraining the Evolution of the Ionizing Background and the Epoch of Reionization with z\raisebox{-0.5ex}\textasciitilde6 Quasars. II. A Sample of 19 Quasars}",
      journal = {\aj},
         year = 2006,
        month = jul,
       volume = {132},
       number = {1},
        pages = {117-136},
          doi = {10.1086/504836},
archivePrefix = {arXiv},
       eprint = {astro-ph/0512082},
 primaryClass = {astro-ph},
       adsurl = {https://ui.adsabs.harvard.edu/abs/2006AJ....132..117F}
}

@ARTICLE{Rodriguez-Gomez2015Illustris_mergers,
       author = {{Rodriguez-Gomez}, Vicente and {Genel}, Shy and {Vogelsberger}, Mark and {Sijacki}, Debora and {Pillepich}, Annalisa and {Sales}, Laura V. and {Torrey}, Paul and {Snyder}, Greg and {Nelson}, Dylan and {Springel}, Volker and {Ma}, Chung-Pei and {Hernquist}, Lars},
        title = "{The merger rate of galaxies in the Illustris simulation: a comparison with observations and semi-empirical models}",
      journal = {\mnras},
         year = 2015,
        month = may,
       volume = {449},
       number = {1},
        pages = {49-64},
          doi = {10.1093/mnras/stv264},
archivePrefix = {arXiv},
       eprint = {1502.01339},
 primaryClass = {astro-ph.GA},
       adsurl = {https://ui.adsabs.harvard.edu/abs/2015MNRAS.449...49R}
}

@ARTICLE{Pahl2025xi_ion,
       author = {{Pahl}, Anthony and {Topping}, Michael W. and {Shapley}, Alice and {Sanders}, Ryan and {Reddy}, Naveen A. and {Clarke}, Leonardo and {Kehoe}, Emily and {Bento}, Trinity and {Brammer}, Gabe},
        title = "{A Spectroscopic Analysis of the Ionizing Photon Production Efficiency in JADES and CEERS: Implications for the Ionizing Photon Budget}",
      journal = {\apj},
         year = 2025,
        month = mar,
       volume = {981},
       number = {2},
          eid = {134},
        pages = {134},
          doi = {10.3847/1538-4357/adb1ab},
archivePrefix = {arXiv},
       eprint = {2407.03399},
 primaryClass = {astro-ph.GA},
       adsurl = {https://ui.adsabs.harvard.edu/abs/2025ApJ...981..134P}
}

@ARTICLE{Llerena2025_xi_ion,
       author = {{Llerena}, M. and {Pentericci}, L. and {Napolitano}, L. and {Mascia}, S. and {Amor{\'\i}n}, R. and {Calabr{\`o}}, A. and {Castellano}, M. and {Cleri}, N.~J. and {Giavalisco}, M. and {Grogin}, N.~A. and {Hathi}, N.~P. and {Hirschmann}, M. and {Koekemoer}, A.~M. and {Nanayakkara}, T. and {Pacucci}, F. and {Shen}, L. and {Wilkins}, S.~M. and {Yoon}, I. and {Yung}, L.~Y.~A. and {Bhatawdekar}, R. and {Lucas}, R.~A. and {Wang}, X. and {Arrabal Haro}, P. and {Bagley}, M.~B. and {Finkelstein}, S.~L. and {Kartaltepe}, J.~S. and {Merlin}, E. and {Papovich}, C. and {Pirzkal}, N. and {Santini}, P.},
        title = "{The ionizing photon production efficiency of star-forming galaxies at z {\ensuremath{\sim}} 4{\textendash}10}",
      journal = {\aap},
         year = 2025,
        month = jun,
       volume = {698},
          eid = {A302},
        pages = {A302},
          doi = {10.1051/0004-6361/202453251},
archivePrefix = {arXiv},
       eprint = {2412.01358},
 primaryClass = {astro-ph.GA},
       adsurl = {https://ui.adsabs.harvard.edu/abs/2025A&A...698A.302L}
}

@ARTICLE{Donnan2024UVLF,
       author = {{Donnan}, C.~T. and {McLure}, R.~J. and {Dunlop}, J.~S. and {McLeod}, D.~J. and {Magee}, D. and {Arellano-C{\'o}rdova}, K.~Z. and {Barrufet}, L. and {Begley}, R. and {Bowler}, R.~A.~A. and {Carnall}, A.~C. and {Cullen}, F. and {Ellis}, R.~S. and {Fontana}, A. and {Illingworth}, G.~D. and {Grogin}, N.~A. and {Hamadouche}, M.~L. and {Koekemoer}, A.~M. and {Liu}, F.-Y. and {Mason}, C. and {Santini}, P. and {Stanton}, T.~M.},
        title = "{JWST PRIMER: a new multifield determination of the evolving galaxy UV luminosity function at redshifts z ≃ 9 - 15}",
      journal = {\mnras},
         year = 2024,
        month = sep,
       volume = {533},
       number = {3},
        pages = {3222-3237},
          doi = {10.1093/mnras/stae2037},
archivePrefix = {arXiv},
       eprint = {2403.03171},
 primaryClass = {astro-ph.GA},
       adsurl = {https://ui.adsabs.harvard.edu/abs/2024MNRAS.533.3222D}
}

@ARTICLE{Finkelstein2024UVLF,
       author = {{Finkelstein}, Steven L. and {Leung}, Gene C.~K. and {Bagley}, Micaela B. and {Dickinson}, Mark and {Ferguson}, Henry C. and {Papovich}, Casey and {Akins}, Hollis B. and {Arrabal Haro}, Pablo and {Dav{\'e}}, Romeel and {Dekel}, Avishai and {Kartaltepe}, Jeyhan S. and {Kocevski}, Dale D. and {Koekemoer}, Anton M. and {Pirzkal}, Nor and {Somerville}, Rachel S. and {Yung}, L.~Y. Aaron and {Amor{\'\i}n}, Ricardo O. and {Backhaus}, Bren E. and {Behroozi}, Peter and {Bisigello}, Laura and {Bromm}, Volker and {Casey}, Caitlin M. and {Ch{\'a}vez Ortiz}, {\'O}scar A. and {Cheng}, Yingjie and {Chworowsky}, Katherine and {Cleri}, Nikko J. and {Cooper}, M.~C. and {Davis}, Kelcey and {de la Vega}, Alexander and {Elbaz}, David and {Franco}, Maximilien and {Fontana}, Adriano and {Fujimoto}, Seiji and {Giavalisco}, Mauro and {Grogin}, Norman A. and {Holwerda}, Benne W. and {Huertas-Company}, Marc and {Hirschmann}, Michaela and {Iyer}, Kartheik G. and {Jogee}, Shardha and {Jung}, Intae and {Larson}, Rebecca L. and {Lucas}, Ray A. and {Mobasher}, Bahram and {Morales}, Alexa M. and {Morley}, Caroline V. and {Mukherjee}, Sagnick and {P{\'e}rez-Gonz{\'a}lez}, Pablo G. and {Ravindranath}, Swara and {Rodighiero}, Giulia and {Rowland}, Melanie J. and {Tacchella}, Sandro and {Taylor}, Anthony J. and {Trump}, Jonathan R. and {Wilkins}, Stephen M.},
        title = "{The Complete CEERS Early Universe Galaxy Sample: A Surprisingly Slow Evolution of the Space Density of Bright Galaxies at z {\ensuremath{\sim}} 8.5{\textendash}14.5}",
      journal = {\apjl},
         year = 2024,
        month = jul,
       volume = {969},
       number = {1},
          eid = {L2},
        pages = {L2},
          doi = {10.3847/2041-8213/ad4495},
archivePrefix = {arXiv},
       eprint = {2311.04279},
 primaryClass = {astro-ph.GA},
       adsurl = {https://ui.adsabs.harvard.edu/abs/2024ApJ...969L...2F}
}

@ARTICLE{Asada2025UVLF,
       author = {{Asada}, Yoshihisa and {Willott}, Chris and {Muzzin}, Adam and {Brada{\v{c}}}, Maru{\v{s}}a and {Brammer}, Gabriel and {Desprez}, Guillaume and {Iyer}, Kartheik and {Marchesini}, Danilo and {Martis}, Nicholas and {Noirot}, Ga{\"e}l and {Sarrouh}, Ghassan and {Sawicki}, Marcin and {Withers}, Sunna and {Fujimoto}, Seiji and {Felicioni}, Giordano and {Goovaerts}, Ilias and {Jude{\v{z}}}, Jon and {Jagga}, Naadiyah and {Merchant}, Maya and {M{\'e}rida}, Rosa and {Robbins}, Luke},
        title = "{Earliest Galaxy Evolution in the CANUCS+Technicolor fields: Galaxy Properties at $z\sim10-16$ seen with the Full NIRCam Medium and Broad Band Filters}",
      journal = {arXiv e-prints},
         year = 2025,
        month = jul,
          eid = {arXiv:2507.03124},
        pages = {arXiv:2507.03124},
          doi = {10.48550/arXiv.2507.03124},
archivePrefix = {arXiv},
       eprint = {2507.03124},
 primaryClass = {astro-ph.GA},
       adsurl = {https://ui.adsabs.harvard.edu/abs/2025arXiv250703124A}
}

@ARTICLE{Harikane2025UVLF_spec,
       author = {{Harikane}, Yuichi and {Nakajima}, Kimihiko and {Ouchi}, Masami and {Umeda}, Hiroya and {Isobe}, Yuki and {Ono}, Yoshiaki and {Xu}, Yi and {Zhang}, Yechi},
        title = "{Pure Spectroscopic Constraints on UV Luminosity Functions and Cosmic Star Formation History from 25 Galaxies at z $_{spec}$ = 8.61-13.20 Confirmed with JWST/NIRSpec}",
      journal = {\apj},
         year = 2024,
        month = jan,
       volume = {960},
       number = {1},
          eid = {56},
        pages = {56},
          doi = {10.3847/1538-4357/ad0b7e},
archivePrefix = {arXiv},
       eprint = {2304.06658},
 primaryClass = {astro-ph.GA},
       adsurl = {https://ui.adsabs.harvard.edu/abs/2024ApJ...960...56H}
}

@ARTICLE{Pahl2021LyCfesc_z3,
       author = {{Pahl}, Anthony J. and {Shapley}, Alice and {Steidel}, Charles C. and {Chen}, Yuguang and {Reddy}, Naveen A.},
        title = "{An uncontaminated measurement of the escaping Lyman continuum at z   3}",
      journal = {\mnras},
         year = 2021,
        month = aug,
       volume = {505},
       number = {2},
        pages = {2447-2467},
          doi = {10.1093/mnras/stab1374},
archivePrefix = {arXiv},
       eprint = {2104.02081},
 primaryClass = {astro-ph.GA},
       adsurl = {https://ui.adsabs.harvard.edu/abs/2021MNRAS.505.2447P}
}

@ARTICLE{Steidel2018KLCS,
       author = {{Steidel}, Charles C. and {Bogosavljevi{\'c}}, Milan and {Shapley}, Alice E. and {Reddy}, Naveen A. and {Rudie}, Gwen C. and {Pettini}, Max and {Trainor}, Ryan F. and {Strom}, Allison L.},
        title = "{The Keck Lyman Continuum Spectroscopic Survey (KLCS): The Emergent Ionizing Spectrum of Galaxies at z {\ensuremath{\sim}} 3}",
      journal = {\apj},
         year = 2018,
        month = dec,
       volume = {869},
       number = {2},
          eid = {123},
        pages = {123},
          doi = {10.3847/1538-4357/aaed28},
archivePrefix = {arXiv},
       eprint = {1805.06071},
 primaryClass = {astro-ph.GA},
       adsurl = {https://ui.adsabs.harvard.edu/abs/2018ApJ...869..123S}
}

@article{Kakiichi2021LyC_Lyaspectra,
   title={Radiation Hydrodynamics of Turbulent H ii Regions in Molecular Clouds: A Physical Origin of LyC Leakage and the Associated Lyα Spectra},
   volume={908},
   ISSN={1538-4357},
   url={http://dx.doi.org/10.3847/1538-4357/abc2d9},
   DOI={10.3847/1538-4357/abc2d9},
   number={1},
   journal={The Astrophysical Journal},
   publisher={American Astronomical Society},
   author={Kakiichi, Koki and Gronke, Max},
   year={2021},
   month=feb, pages={30} }

@ARTICLE{Gazagnes2020LyaLyCescape,
       author = {{Gazagnes}, S. and {Chisholm}, J. and {Schaerer}, D. and {Verhamme}, A. and {Izotov}, Y.},
        title = "{The origin of the escape of Lyman {\ensuremath{\alpha}} and ionizing photons in Lyman continuum emitters}",
      journal = {\aap},
         year = 2020,
        month = jul,
       volume = {639},
          eid = {A85},
        pages = {A85},
          doi = {10.1051/0004-6361/202038096},
archivePrefix = {arXiv},
       eprint = {2005.07215},
 primaryClass = {astro-ph.GA},
       adsurl = {https://ui.adsabs.harvard.edu/abs/2020A&A...639A..85G}
}

@ARTICLE{Mary2020ORIGIN,
       author = {{Mary}, David and {Bacon}, Roland and {Conseil}, Simon and {Piqueras}, Laure and {Schutz}, Antony},
        title = "{ORIGIN: Blind detection of faint emission line galaxies in MUSE datacubes}",
      journal = {\aap},
         year = 2020,
        month = mar,
       volume = {635},
          eid = {A194},
        pages = {A194},
          doi = {10.1051/0004-6361/201937001},
archivePrefix = {arXiv},
       eprint = {2002.00214},
 primaryClass = {astro-ph.IM},
       adsurl = {https://ui.adsabs.harvard.edu/abs/2020A&A...635A.194M}
}

@ARTICLE{Horne1986CCDspectroscopy,
       author = {{Horne}, K.},
        title = "{An optimal extraction algorithm for CCD spectroscopy.}",
      journal = {\pasp},
         year = 1986,
        month = jun,
       volume = {98},
        pages = {609-617},
          doi = {10.1086/131801},
       adsurl = {https://ui.adsabs.harvard.edu/abs/1986PASP...98..609H}
}

@ARTICLE{Sobral2019Lya_esc_from_EW,
       author = {{Sobral}, David and {Matthee}, Jorryt},
        title = "{Predicting Ly{\ensuremath{\alpha}} escape fractions with a simple observable. Ly{\ensuremath{\alpha}} in emission as an empirically calibrated star formation rate indicator}",
      journal = {\aap},
         year = 2019,
        month = mar,
       volume = {623},
          eid = {A157},
        pages = {A157},
          doi = {10.1051/0004-6361/201833075},
archivePrefix = {arXiv},
       eprint = {1803.08923},
 primaryClass = {astro-ph.GA},
       adsurl = {https://ui.adsabs.harvard.edu/abs/2019A&A...623A.157S}
}

@ARTICLE{Goovaerts2024Lya_esc_fraction,
       author = {{Goovaerts}, I. and {Thai}, T.~T. and {Pello}, R. and {Tuan-Anh}, P. and {Laporte}, N. and {Matthee}, J. and {Nanayakkara}, T. and {Pharo}, J.},
        title = "{Charting the Lyman-{\ensuremath{\alpha}} escape fraction in the range 2.9 < z < 6.7 and consequences for the LAE reionisation contribution}",
      journal = {\aap},
         year = 2024,
        month = oct,
       volume = {690},
          eid = {A302},
        pages = {A302},
          doi = {10.1051/0004-6361/202451432},
archivePrefix = {arXiv},
       eprint = {2408.00517},
 primaryClass = {astro-ph.GA},
       adsurl = {https://ui.adsabs.harvard.edu/abs/2024A&A...690A.302G}
}

@ARTICLE{Bassett2021IGM_sim,
       author = {{Bassett}, R. and {Ryan-Weber}, E.~V. and {Cooke}, J. and {Me{\v{s}}tri{\'c}}, U. and {Kakiichi}, K. and {Prichard}, L. and {Rafelski}, M.},
        title = "{IGM transmission bias for z {\ensuremath{\geq}} 2.9 Lyman continuum detected galaxies}",
      journal = {\mnras},
         year = 2021,
        month = mar,
       volume = {502},
       number = {1},
        pages = {108-126},
          doi = {10.1093/mnras/stab070},
archivePrefix = {arXiv},
       eprint = {2101.00727},
 primaryClass = {astro-ph.GA},
       adsurl = {https://ui.adsabs.harvard.edu/abs/2021MNRAS.502..108B}
}

@ARTICLE{Hawcroft2025pySTARBURST,
       author = {{Hawcroft}, Calum and {Leitherer}, Claus and {Arangur{\'e}}, Oskar and {Chisholm}, John and {Ekstr{\"o}m}, Sylvia and {Martinet}, S{\'e}bastien and {Martins}, Lucimara P. and {Meynet}, Georges and {Morisset}, Christophe and {Sander}, Andreas A.~C. and {Wofford}, Aida},
        title = "{pySTARBURST99: The Next Generation of STARBURST99}",
      journal = {\apjs},
         year = 2025,
        month = sep,
       volume = {280},
       number = {1},
          eid = {5},
        pages = {5},
          doi = {10.3847/1538-4365/adddb6},
archivePrefix = {arXiv},
       eprint = {2505.24841},
 primaryClass = {astro-ph.GA},
       adsurl = {https://ui.adsabs.harvard.edu/abs/2025ApJS..280....5H}
}

@ARTICLE{Elridge2008BPASS,
       author = {{Eldridge}, John J. and {Izzard}, Robert G. and {Tout}, Christopher A.},
        title = "{The effect of massive binaries on stellar populations and supernova progenitors}",
      journal = {\mnras},
         year = 2008,
        month = mar,
       volume = {384},
       number = {3},
        pages = {1109-1118},
          doi = {10.1111/j.1365-2966.2007.12738.x},
archivePrefix = {arXiv},
       eprint = {0711.3079},
 primaryClass = {astro-ph},
       adsurl = {https://ui.adsabs.harvard.edu/abs/2008MNRAS.384.1109E}
}

@ARTICLE{Henry2015LyC_Lya_GPs,
       author = {{Henry}, Alaina and {Scarlata}, Claudia and {Martin}, Crystal L. and {Erb}, Dawn},
        title = "{Ly{\ensuremath{\alpha}} Emission from Green Peas: The Role of Circumgalactic Gas Density, Covering, and Kinematics}",
      journal = {\apj},
         year = 2015,
        month = aug,
       volume = {809},
       number = {1},
          eid = {19},
        pages = {19},
          doi = {10.1088/0004-637X/809/1/19},
archivePrefix = {arXiv},
       eprint = {1505.05149},
 primaryClass = {astro-ph.GA},
       adsurl = {https://ui.adsabs.harvard.edu/abs/2015ApJ...809...19H}
}

@ARTICLE{Kron1980photometry,
       author = {{Kron}, R.~G.},
        title = "{Photometry of a complete sample of faint galaxies.}",
      journal = {\apjs},
         year = 1980,
        month = jun,
       volume = {43},
        pages = {305-325},
          doi = {10.1086/190669},
       adsurl = {https://ui.adsabs.harvard.edu/abs/1980ApJS...43..305K}
}

@ARTICLE{Beckwith2006HUDF,
       author = {{Beckwith}, Steven V.~W. and {Stiavelli}, Massimo and {Koekemoer}, Anton M. and {Caldwell}, John A.~R. and {Ferguson}, Henry C. and {Hook}, Richard and {Lucas}, Ray A. and {Bergeron}, Louis E. and {Corbin}, Michael and {Jogee}, Shardha and {Panagia}, Nino and {Robberto}, Massimo and {Royle}, Patricia and {Somerville}, Rachel S. and {Sosey}, Megan},
        title = "{The Hubble Ultra Deep Field}",
      journal = {\aj},
         year = 2006,
        month = nov,
       volume = {132},
       number = {5},
        pages = {1729-1755},
          doi = {10.1086/507302},
archivePrefix = {arXiv},
       eprint = {astro-ph/0607632},
 primaryClass = {astro-ph},
       adsurl = {https://ui.adsabs.harvard.edu/abs/2006AJ....132.1729B}
}

@ARTICLE{Bagley2024NGDEEP,
       author = {{Bagley}, Micaela B. and {Pirzkal}, Nor and {Finkelstein}, Steven L. and {Papovich}, Casey and {Berg}, Danielle A. and {Lotz}, Jennifer M. and {Leung}, Gene C.~K. and {Ferguson}, Henry C. and {Koekemoer}, Anton M. and {Dickinson}, Mark and {Kartaltepe}, Jeyhan S. and {Kocevski}, Dale D. and {Somerville}, Rachel S. and {Yung}, L.~Y. Aaron and {Backhaus}, Bren E. and {Casey}, Caitlin M. and {Castellano}, Marco and {Ch{\'a}vez Ortiz}, {\'O}scar A. and {Chworowsky}, Katherine and {Cox}, Isabella G. and {Dav{\'e}}, Romeel and {Davis}, Kelcey and {Estrada-Carpenter}, Vicente and {Fontana}, Adriano and {Fujimoto}, Seiji and {Gardner}, Jonathan P. and {Giavalisco}, Mauro and {Grazian}, Andrea and {Grogin}, Norman A. and {Hathi}, Nimish P. and {Hutchison}, Taylor A. and {Jaskot}, Anne E. and {Jung}, Intae and {Kewley}, Lisa J. and {Kirkpatrick}, Allison and {Larson}, Rebecca L. and {Matharu}, Jasleen and {Natarajan}, Priyamvada and {Pentericci}, Laura and {P{\'e}rez-Gonz{\'a}lez}, Pablo G. and {Ravindranath}, Swara and {Rothberg}, Barry and {Ryan}, Russell and {Shen}, Lu and {Simons}, Raymond C. and {Snyder}, Gregory F. and {Trump}, Jonathan R. and {Wilkins}, Stephen M.},
        title = "{The Next Generation Deep Extragalactic Exploratory Public (NGDEEP) Survey}",
      journal = {\apjl},
         year = 2024,
        month = apr,
       volume = {965},
       number = {1},
          eid = {L6},
        pages = {L6},
          doi = {10.3847/2041-8213/ad2f31},
archivePrefix = {arXiv},
       eprint = {2302.05466},
 primaryClass = {astro-ph.GA},
       adsurl = {https://ui.adsabs.harvard.edu/abs/2024ApJ...965L...6B}
}

@ARTICLE{LeReste2025LaCOS,
       author = {{Le Reste}, Alexandra and {Scarlata}, Claudia and {Hayes}, Matthew J. and {Melinder}, Jens and {Saldana-Lopez}, Alberto and {Smith}, Aaron and {Runnholm}, Axel and {Lin}, Yu-Heng and {Amor{\'\i}n}, Ricardo O. and {Atek}, Hakim and {Borthakur}, Sanchayeeta and {Carr}, Cody A. and {Fleming}, Brian and {Flury}, Sophia R. and {Giavalisco}, Mauro and {Henry}, Alaina and {Jaskot}, Anne E. and {Ji}, Zhiyuan and {Jung}, Intae and {Leclercq}, Floriane and {Marques-Chaves}, Rui and {McCandliss}, Stephan R. and {Oey}, M.~S. and {{\"O}stlin}, G{\"o}ran and {Ravindranath}, Swara and {Schaerer}, Daniel and {Thuan}, Trinh X. and {Xu}, Xinfeng},
        title = "{The Ly{\ensuremath{\alpha}} and Continuum Origins Survey. I. Survey Description and Ly{\ensuremath{\alpha}} Imaging}",
      journal = {\apjs},
         year = 2025,
        month = sep,
       volume = {280},
       number = {1},
          eid = {27},
        pages = {27},
          doi = {10.3847/1538-4365/adf227},
archivePrefix = {arXiv},
       eprint = {2504.07056},
 primaryClass = {astro-ph.GA},
       adsurl = {https://ui.adsabs.harvard.edu/abs/2025ApJS..280...27L}
}

@ARTICLE{Giavalisco2004GOODS,
       author = {{Giavalisco}, M. and {Ferguson}, H.~C. and {Koekemoer}, A.~M. and {Dickinson}, M. and {Alexander}, D.~M. and {Bauer}, F.~E. and {Bergeron}, J. and {Biagetti}, C. and {Brandt}, W.~N. and {Casertano}, S. and {Cesarsky}, C. and {Chatzichristou}, E. and {Conselice}, C. and {Cristiani}, S. and {Da Costa}, L. and {Dahlen}, T. and {de Mello}, D. and {Eisenhardt}, P. and {Erben}, T. and {Fall}, S.~M. and {Fassnacht}, C. and {Fosbury}, R. and {Fruchter}, A. and {Gardner}, J.~P. and {Grogin}, N. and {Hook}, R.~N. and {Hornschemeier}, A.~E. and {Idzi}, R. and {Jogee}, S. and {Kretchmer}, C. and {Laidler}, V. and {Lee}, K.~S. and {Livio}, M. and {Lucas}, R. and {Madau}, P. and {Mobasher}, B. and {Moustakas}, L.~A. and {Nonino}, M. and {Padovani}, P. and {Papovich}, C. and {Park}, Y. and {Ravindranath}, S. and {Renzini}, A. and {Richardson}, M. and {Riess}, A. and {Rosati}, P. and {Schirmer}, M. and {Schreier}, E. and {Somerville}, R.~S. and {Spinrad}, H. and {Stern}, D. and {Stiavelli}, M. and {Strolger}, L. and {Urry}, C.~M. and {Vandame}, B. and {Williams}, R. and {Wolf}, C.},
        title = "{The Great Observatories Origins Deep Survey: Initial Results from Optical and Near-Infrared Imaging}",
      journal = {\apjl},
         year = 2004,
        month = jan,
       volume = {600},
       number = {2},
        pages = {L93-L98},
          doi = {10.1086/379232},
archivePrefix = {arXiv},
       eprint = {astro-ph/0309105},
 primaryClass = {astro-ph},
       adsurl = {https://ui.adsabs.harvard.edu/abs/2004ApJ...600L..93G}
}

@ARTICLE{Grogin2011,
       author = {{Grogin}, Norman A. and {Kocevski}, Dale D. and {Faber}, S.~M. and {Ferguson}, Henry C. and {Koekemoer}, Anton M. and {Riess}, Adam G. and {Acquaviva}, Viviana and {Alexander}, David M. and {Almaini}, Omar and {Ashby}, Matthew L.~N. and {Barden}, Marco and {Bell}, Eric F. and {Bournaud}, Fr{\'e}d{\'e}ric and {Brown}, Thomas M. and {Caputi}, Karina I. and {Casertano}, Stefano and {Cassata}, Paolo and {Castellano}, Marco and {Challis}, Peter and {Chary}, Ranga-Ram and {Cheung}, Edmond and {Cirasuolo}, Michele and {Conselice}, Christopher J. and {Roshan Cooray}, Asantha and {Croton}, Darren J. and {Daddi}, Emanuele and {Dahlen}, Tomas and {Dav{\'e}}, Romeel and {de Mello}, Du{\'\i}lia F. and {Dekel}, Avishai and {Dickinson}, Mark and {Dolch}, Timothy and {Donley}, Jennifer L. and {Dunlop}, James S. and {Dutton}, Aaron A. and {Elbaz}, David and {Fazio}, Giovanni G. and {Filippenko}, Alexei V. and {Finkelstein}, Steven L. and {Fontana}, Adriano and {Gardner}, Jonathan P. and {Garnavich}, Peter M. and {Gawiser}, Eric and {Giavalisco}, Mauro and {Grazian}, Andrea and {Guo}, Yicheng and {Hathi}, Nimish P. and {H{\"a}ussler}, Boris and {Hopkins}, Philip F. and {Huang}, Jia-Sheng and {Huang}, Kuang-Han and {Jha}, Saurabh W. and {Kartaltepe}, Jeyhan S. and {Kirshner}, Robert P. and {Koo}, David C. and {Lai}, Kamson and {Lee}, Kyoung-Soo and {Li}, Weidong and {Lotz}, Jennifer M. and {Lucas}, Ray A. and {Madau}, Piero and {McCarthy}, Patrick J. and {McGrath}, Elizabeth J. and {McIntosh}, Daniel H. and {McLure}, Ross J. and {Mobasher}, Bahram and {Moustakas}, Leonidas A. and {Mozena}, Mark and {Nandra}, Kirpal and {Newman}, Jeffrey A. and {Niemi}, Sami-Matias and {Noeske}, Kai G. and {Papovich}, Casey J. and {Pentericci}, Laura and {Pope}, Alexandra and {Primack}, Joel R. and {Rajan}, Abhijith and {Ravindranath}, Swara and {Reddy}, Naveen A. and {Renzini}, Alvio and {Rix}, Hans-Walter and {Robaina}, Aday R. and {Rodney}, Steven A. and {Rosario}, David J. and {Rosati}, Piero and {Salimbeni}, Sara and {Scarlata}, Claudia and {Siana}, Brian and {Simard}, Luc and {Smidt}, Joseph and {Somerville}, Rachel S. and {Spinrad}, Hyron and {Straughn}, Amber N. and {Strolger}, Louis-Gregory and {Telford}, Olivia and {Teplitz}, Harry I. and {Trump}, Jonathan R. and {van der Wel}, Arjen and {Villforth}, Carolin and {Wechsler}, Risa H. and {Weiner}, Benjamin J. and {Wiklind}, Tommy and {Wild}, Vivienne and {Wilson}, Grant and {Wuyts}, Stijn and {Yan}, Hao-Jing and {Yun}, Min S.},
        title = "{CANDELS: The Cosmic Assembly Near-infrared Deep Extragalactic Legacy Survey}",
      journal = {\apjs},
         year = 2011,
        month = dec,
       volume = {197},
       number = {2},
          eid = {35},
        pages = {35},
          doi = {10.1088/0067-0049/197/2/35},
archivePrefix = {arXiv},
       eprint = {1105.3753},
 primaryClass = {astro-ph.CO},
       adsurl = {https://ui.adsabs.harvard.edu/abs/2011ApJS..197...35G}
}

@ARTICLE{Koekemoer2011CANDELS,
       author = {{Koekemoer}, Anton M. and {Faber}, S.~M. and {Ferguson}, Henry C. and {Grogin}, Norman A. and {Kocevski}, Dale D. and {Koo}, David C. and {Lai}, Kamson and {Lotz}, Jennifer M. and {Lucas}, Ray A. and {McGrath}, Elizabeth J. and {Ogaz}, Sara and {Rajan}, Abhijith and {Riess}, Adam G. and {Rodney}, Steve A. and {Strolger}, Louis and {Casertano}, Stefano and {Castellano}, Marco and {Dahlen}, Tomas and {Dickinson}, Mark and {Dolch}, Timothy and {Fontana}, Adriano and {Giavalisco}, Mauro and {Grazian}, Andrea and {Guo}, Yicheng and {Hathi}, Nimish P. and {Huang}, Kuang-Han and {van der Wel}, Arjen and {Yan}, Hao-Jing and {Acquaviva}, Viviana and {Alexander}, David M. and {Almaini}, Omar and {Ashby}, Matthew L.~N. and {Barden}, Marco and {Bell}, Eric F. and {Bournaud}, Fr{\'e}d{\'e}ric and {Brown}, Thomas M. and {Caputi}, Karina I. and {Cassata}, Paolo and {Challis}, Peter J. and {Chary}, Ranga-Ram and {Cheung}, Edmond and {Cirasuolo}, Michele and {Conselice}, Christopher J. and {Roshan Cooray}, Asantha and {Croton}, Darren J. and {Daddi}, Emanuele and {Dav{\'e}}, Romeel and {de Mello}, Duilia F. and {de Ravel}, Loic and {Dekel}, Avishai and {Donley}, Jennifer L. and {Dunlop}, James S. and {Dutton}, Aaron A. and {Elbaz}, David and {Fazio}, Giovanni G. and {Filippenko}, Alexei V. and {Finkelstein}, Steven L. and {Frazer}, Chris and {Gardner}, Jonathan P. and {Garnavich}, Peter M. and {Gawiser}, Eric and {Gruetzbauch}, Ruth and {Hartley}, Will G. and {H{\"a}ussler}, Boris and {Herrington}, Jessica and {Hopkins}, Philip F. and {Huang}, Jia-Sheng and {Jha}, Saurabh W. and {Johnson}, Andrew and {Kartaltepe}, Jeyhan S. and {Khostovan}, Ali A. and {Kirshner}, Robert P. and {Lani}, Caterina and {Lee}, Kyoung-Soo and {Li}, Weidong and {Madau}, Piero and {McCarthy}, Patrick J. and {McIntosh}, Daniel H. and {McLure}, Ross J. and {McPartland}, Conor and {Mobasher}, Bahram and {Moreira}, Heidi and {Mortlock}, Alice and {Moustakas}, Leonidas A. and {Mozena}, Mark and {Nandra}, Kirpal and {Newman}, Jeffrey A. and {Nielsen}, Jennifer L. and {Niemi}, Sami and {Noeske}, Kai G. and {Papovich}, Casey J. and {Pentericci}, Laura and {Pope}, Alexandra and {Primack}, Joel R. and {Ravindranath}, Swara and {Reddy}, Naveen A. and {Renzini}, Alvio and {Rix}, Hans-Walter and {Robaina}, Aday R. and {Rosario}, David J. and {Rosati}, Piero and {Salimbeni}, Sara and {Scarlata}, Claudia and {Siana}, Brian and {Simard}, Luc and {Smidt}, Joseph and {Snyder}, Diana and {Somerville}, Rachel S. and {Spinrad}, Hyron and {Straughn}, Amber N. and {Telford}, Olivia and {Teplitz}, Harry I. and {Trump}, Jonathan R. and {Vargas}, Carlos and {Villforth}, Carolin and {Wagner}, Cory R. and {Wandro}, Pat and {Wechsler}, Risa H. and {Weiner}, Benjamin J. and {Wiklind}, Tommy and {Wild}, Vivienne and {Wilson}, Grant and {Wuyts}, Stijn and {Yun}, Min S.},
        title = "{CANDELS: The Cosmic Assembly Near-infrared Deep Extragalactic Legacy Survey{\textemdash}The Hubble Space Telescope Observations, Imaging Data Products, and Mosaics}",
      journal = {\apjs},
         year = 2011,
        month = dec,
       volume = {197},
       number = {2},
          eid = {36},
        pages = {36},
          doi = {10.1088/0067-0049/197/2/36},
archivePrefix = {arXiv},
       eprint = {1105.3754},
 primaryClass = {astro-ph.CO},
       adsurl = {https://ui.adsabs.harvard.edu/abs/2011ApJS..197...36K}
}

@ARTICLE{Brammer2012.3DHST,
       author = {{Brammer}, Gabriel B. and {van Dokkum}, Pieter G. and {Franx}, Marijn and {Fumagalli}, Mattia and {Patel}, Shannon and {Rix}, Hans-Walter and {Skelton}, Rosalind E. and {Kriek}, Mariska and {Nelson}, Erica and {Schmidt}, Kasper B. and {Bezanson}, Rachel and {da Cunha}, Elisabete and {Erb}, Dawn K. and {Fan}, Xiaohui and {F{\"o}rster Schreiber}, Natascha and {Illingworth}, Garth D. and {Labb{\'e}}, Ivo and {Leja}, Joel and {Lundgren}, Britt and {Magee}, Dan and {Marchesini}, Danilo and {McCarthy}, Patrick and {Momcheva}, Ivelina and {Muzzin}, Adam and {Quadri}, Ryan and {Steidel}, Charles C. and {Tal}, Tomer and {Wake}, David and {Whitaker}, Katherine E. and {Williams}, Anna},
        title = "{3D-HST: A Wide-field Grism Spectroscopic Survey with the Hubble Space Telescope}",
      journal = {\apjs},
         year = 2012,
        month = jun,
       volume = {200},
       number = {2},
          eid = {13},
        pages = {13},
          doi = {10.1088/0067-0049/200/2/13},
archivePrefix = {arXiv},
       eprint = {1204.2829},
 primaryClass = {astro-ph.CO},
       adsurl = {https://ui.adsabs.harvard.edu/abs/2012ApJS..200...13B}
}

@ARTICLE{Teplitz2013UVUDF,
       author = {{Teplitz}, Harry I. and {Rafelski}, Marc and {Kurczynski}, Peter and {Bond}, Nicholas A. and {Grogin}, Norman and {Koekemoer}, Anton M. and {Atek}, Hakim and {Brown}, Thomas M. and {Coe}, Dan and {Colbert}, James W. and {Ferguson}, Henry C. and {Finkelstein}, Steven L. and {Gardner}, Jonathan P. and {Gawiser}, Eric and {Giavalisco}, Mauro and {Gronwall}, Caryl and {Hanish}, Daniel J. and {Lee}, Kyoung-Soo and {de Mello}, Duilia F. and {Ravindranath}, Swara and {Ryan}, Russell E. and {Siana}, Brian D. and {Scarlata}, Claudia and {Soto}, Emmaris and {Voyer}, Elysse N. and {Wolfe}, Arthur M.},
        title = "{UVUDF: Ultraviolet Imaging of the Hubble Ultra Deep Field with Wide-Field Camera 3}",
      journal = {\aj},
         year = 2013,
        month = dec,
       volume = {146},
       number = {6},
          eid = {159},
        pages = {159},
          doi = {10.1088/0004-6256/146/6/159},
archivePrefix = {arXiv},
       eprint = {1305.1357},
 primaryClass = {astro-ph.CO},
       adsurl = {https://ui.adsabs.harvard.edu/abs/2013AJ....146..159T}
}

@ARTICLE{Oesch2018HDUV,
       author = {{Oesch}, P.~A. and {Montes}, M. and {Reddy}, N. and {Bouwens}, R.~J. and {Illingworth}, G.~D. and {Magee}, D. and {Atek}, H. and {Carollo}, C.~M. and {Cibinel}, A. and {Franx}, M. and {Holden}, B. and {Labb{\'e}}, I. and {Nelson}, E.~J. and {Steidel}, C.~C. and {van Dokkum}, P.~G. and {Morselli}, L. and {Naidu}, R.~P. and {Wilkins}, S.},
        title = "{HDUV: The Hubble Deep UV Legacy Survey}",
      journal = {\apjs},
         year = 2018,
        month = jul,
       volume = {237},
       number = {1},
          eid = {12},
        pages = {12},
          doi = {10.3847/1538-4365/aacb30},
archivePrefix = {arXiv},
       eprint = {1806.01853},
 primaryClass = {astro-ph.GA},
       adsurl = {https://ui.adsabs.harvard.edu/abs/2018ApJS..237...12O}
}

@ARTICLE{Wang2025UVCANDELS,
       author = {{Wang}, Xin and {Teplitz}, Harry I. and {Smith}, Brent M. and {Windhorst}, Rogier A. and {Rafelski}, Marc and {Mehta}, Vihang and {Alavi}, Anahita and {Ji}, Zhiyuan and {Brammer}, Gabriel and {Colbert}, James and {Grogin}, Norman and {Hathi}, Nimish P. and {Koekemoer}, Anton M. and {Prichard}, Laura and {Scarlata}, Claudia and {Sunnquist}, Ben and {Arrabal Haro}, Pablo and {Conselice}, Christopher and {Gawiser}, Eric and {Guo}, Yicheng and {Hayes}, Matthew and {Jansen}, Rolf A. and {Lucas}, Ray A. and {O'Connell}, Robert and {Robertson}, Brant and {Rutkowski}, Michael and {Siana}, Brian and {Vanzella}, Eros and {Ashcraft}, Teresa and {Bagley}, Micaela and {Baronchelli}, Ivano and {Barro}, Guillermo and {Blanche}, Alex and {Broussard}, Adam and {Carleton}, Timothy and {Chartab}, Nima and {Cheng}, Yingjie and {Codoreanu}, Alex and {Cohen}, Seth and {Dai}, Y. Sophia and {Darvish}, Behnam and {Dav{\'e}}, Romeel and {Degroot}, Laura and {de Mello}, Duilia and {Dickinson}, Mark and {Emami}, Najmeh and {Ferguson}, Henry and {Ferreira}, Leonardo and {Finkelstein}, Keely and {Finkelstein}, Steven and {Gardner}, Jonathan P. and {Gburek}, Timothy and {Giavalisco}, Mauro and {Grazian}, Andrea and {Gronwall}, Caryl and {Hemmati}, Shoubaneh and {Howell}, Justin and {Iyer}, Kartheik and {Kaviraj}, Sugata and {Kurczynski}, Peter and {Lazar}, Ilin and {MacKenty}, John and {Mantha}, Kameswara Bharadwaj and {Martin}, Alec and {Martin}, Garreth and {McCabe}, Tyler and {Mobasher}, Bahram and {Nedkova}, Kalina and {Olsen}, Charlotte and {Otteson}, Lillian and {Ravindranath}, Swara and {Redshaw}, Caleb and {Sattari}, Zahra and {Soto}, Emmaris and {Yung}, L.~Y. Aaron and {Zabelle}, Bonnabelle and {UVCANDELS Team}},
        title = "{The Lyman Continuum Escape Fraction of Star-forming Galaxies at 2.4 {\ensuremath{\lesssim}} z {\ensuremath{\lesssim}} 3.0 from UVCANDELS}",
      journal = {\apj},
         year = 2025,
        month = feb,
       volume = {980},
       number = {1},
          eid = {74},
        pages = {74},
          doi = {10.3847/1538-4357/ada4ab},
archivePrefix = {arXiv},
       eprint = {2308.09064},
 primaryClass = {astro-ph.GA},
       adsurl = {https://ui.adsabs.harvard.edu/abs/2025ApJ...980...74W}
}

@ARTICLE{Nedkova2024ApJ,
       author = {{Nedkova}, Kalina V. and {Rafelski}, Marc and {Teplitz}, Harry I. and {Mehta}, Vihang and {Degroot}, Laura and {Ravindranath}, Swara and {Alavi}, Anahita and {Beckett}, Alexander and {Grogin}, Norman A. and {H{\"a}u{\ss}ler}, Boris and {Koekemoer}, Anton M. and {Oyarz{\'u}n}, Grecco A. and {Prichard}, Laura and {Revalski}, Mitchell and {Snyder}, Gregory F. and {Sunnquist}, Ben and {Wang}, Xin and {Windhorst}, Rogier A. and {Chartab}, Nima and {Conselice}, Christopher J. and {Guo}, Yicheng and {Hathi}, Nimish and {Hayes}, Matthew J. and {Ji}, Zhiyuan and {Kim}, Keunho J. and {Lucas}, Ray A. and {Mobasher}, Bahram and {O'Connell}, Robert W. and {Sattari}, Zahra and {Smith}, Brent M. and {Taamoli}, Sina and {Yung}, L.~Y. Aaron and {The Uvcandels Team}},
        title = "{UVCANDELS: The Role of Dust on the Stellar Mass{\textendash}Size Relation of Disk Galaxies at 0.5 {\ensuremath{\leq}} z {\ensuremath{\leq}} 3.0}",
      journal = {\apj},
         year = 2024,
        month = aug,
       volume = {970},
       number = {2},
          eid = {188},
        pages = {188},
          doi = {10.3847/1538-4357/ad4ede},
archivePrefix = {arXiv},
       eprint = {2405.10908},
 primaryClass = {astro-ph.GA},
       adsurl = {https://ui.adsabs.harvard.edu/abs/2024ApJ...970..188N}
}

@ARTICLE{Haussler2013MNRAS,
       author = {{H{\"a}u{\ss}ler}, Boris and {Bamford}, Steven P. and {Vika}, Marina and {Rojas}, Alex L. and {Barden}, Marco and {Kelvin}, Lee S. and {Alpaslan}, Mehmet and {Robotham}, Aaron S.~G. and {Driver}, Simon P. and {Baldry}, I.~K. and {Brough}, Sarah and {Hopkins}, Andrew M. and {Liske}, Jochen and {Nichol}, Robert C. and {Popescu}, Cristina C. and {Tuffs}, Richard J.},
        title = "{MegaMorph - multiwavelength measurement of galaxy structure: complete S{\'e}rsic profile information from modern surveys}",
      journal = {\mnras},
         year = 2013,
        month = mar,
       volume = {430},
       number = {1},
        pages = {330-369},
          doi = {10.1093/mnras/sts633},
archivePrefix = {arXiv},
       eprint = {1212.3332},
 primaryClass = {astro-ph.CO},
       adsurl = {https://ui.adsabs.harvard.edu/abs/2013MNRAS.430..330H}
}

@ARTICLE{Peng2002Galfit,
       author = {{Peng}, Chien Y. and {Ho}, Luis C. and {Impey}, Chris D. and {Rix}, Hans-Walter},
        title = "{Detailed Structural Decomposition of Galaxy Images}",
      journal = {\aj},
         year = 2002,
        month = jul,
       volume = {124},
       number = {1},
        pages = {266-293},
          doi = {10.1086/340952},
archivePrefix = {arXiv},
       eprint = {astro-ph/0204182},
 primaryClass = {astro-ph},
       adsurl = {https://ui.adsabs.harvard.edu/abs/2002AJ....124..266P}
}

@ARTICLE{Peng2010Galfit,
       author = {{Peng}, Chien Y. and {Ho}, Luis C. and {Impey}, Chris D. and {Rix}, Hans-Walter},
        title = "{Detailed Decomposition of Galaxy Images. II. Beyond Axisymmetric Models}",
      journal = {\aj},
         year = 2010,
        month = jun,
       volume = {139},
       number = {6},
        pages = {2097-2129},
          doi = {10.1088/0004-6256/139/6/2097},
archivePrefix = {arXiv},
       eprint = {0912.0731},
 primaryClass = {astro-ph.CO},
       adsurl = {https://ui.adsabs.harvard.edu/abs/2010AJ....139.2097P}
}

@ARTICLE{Shibuya2019morphologies_lya,
       author = {{Shibuya}, Takatoshi and {Ouchi}, Masami and {Harikane}, Yuichi and {Nakajima}, Kimihiko},
        title = "{Morphologies of {\ensuremath{\sim}}190,000 Galaxies at z = 0-10 Revealed with HST Legacy Data. III. Continuum Profile and Size Evolution of Ly{\ensuremath{\alpha}} Emitters}",
      journal = {\apj},
         year = 2019,
        month = feb,
       volume = {871},
       number = {2},
          eid = {164},
        pages = {164},
          doi = {10.3847/1538-4357/aaf64b},
archivePrefix = {arXiv},
       eprint = {1809.00765},
 primaryClass = {astro-ph.GA},
       adsurl = {https://ui.adsabs.harvard.edu/abs/2019ApJ...871..164S}
}

@ARTICLE{Mas-Hesse2003lya_escape,
       author = {{Mas-Hesse}, J.~M. and {Kunth}, D. and {Tenorio-Tagle}, G. and {Leitherer}, C. and {Terlevich}, R.~J. and {Terlevich}, E.},
        title = "{Ly{\ensuremath{\alpha}} Emission in Starbursts: Implications for Galaxies at High Redshift}",
      journal = {\apj},
         year = 2003,
        month = dec,
       volume = {598},
       number = {2},
        pages = {858-877},
          doi = {10.1086/379116},
archivePrefix = {arXiv},
       eprint = {astro-ph/0309396},
 primaryClass = {astro-ph},
       adsurl = {https://ui.adsabs.harvard.edu/abs/2003ApJ...598..858M}
}

@ARTICLE{Hayes2007Lya_escape_starburst,
       author = {{Hayes}, Matthew and {{\"O}stlin}, G{\"o}ran and {Atek}, Hakim and {Kunth}, Daniel and {Mas-Hesse}, J. Miguel and {Leitherer}, Claus and {Jim{\'e}nez-Bail{\'o}n}, Elena and {Adamo}, Angela},
        title = "{The escape of Lyman photons from a young starburst: the case of Haro11{\textdagger}}",
      journal = {\mnras},
         year = 2007,
        month = dec,
       volume = {382},
       number = {4},
        pages = {1465-1480},
          doi = {10.1111/j.1365-2966.2007.12482.x},
archivePrefix = {arXiv},
       eprint = {0710.2622},
 primaryClass = {astro-ph},
       adsurl = {https://ui.adsabs.harvard.edu/abs/2007MNRAS.382.1465H}
}

@ARTICLE{Austin2025UVslopes,
       author = {{Austin}, Duncan and {Conselice}, Christopher J. and {Adams}, Nathan J. and {Harvey}, Thomas and {Duan}, Qiao and {Trussler}, James and {Li}, Qiong and {Juod{\v{z}}balis}, Ignas and {Ormerod}, Katherine and {Ferreira}, Leonardo and {Westcott}, Lewi and {Harris}, Honor and {Wilkins}, Stephen M. and {Bhatawdekar}, Rachana and {Caruana}, Joseph and {Coe}, Dan and {Cohen}, Seth H. and {Driver}, Simon P. and {D'Silva}, Jordan C.~J. and {Frye}, Brenda and {Furtak}, Lukas J. and {Grogin}, Norman A. and {Hathi}, Nimish P. and {Holwerda}, Benne W. and {Jansen}, Rolf A. and {Koekemoer}, Anton M. and {Marshall}, Madeline A. and {Nonino}, Mario and {Ortiz}, III, Rafael and {Pirzkal}, Nor and {Robotham}, Aaron and {Ryan}, Jr., Russell E. and {Summers}, Jake and {Willmer}, Christopher N.~A. and {Windhorst}, Rogier A. and {Yan}, Haojing and {Zackrisson}, Erik},
        title = "{EPOCHS. III. Unbiased UV Continuum Slopes at 6.5 < z < 13 from Combined PEARLS GTO and Public JWST/NIRCam Imaging}",
      journal = {\apj},
         year = 2025,
        month = dec,
       volume = {995},
       number = {1},
          eid = {43},
        pages = {43},
          doi = {10.3847/1538-4357/ae07db},
archivePrefix = {arXiv},
       eprint = {2404.10751},
 primaryClass = {astro-ph.GA},
       adsurl = {https://ui.adsabs.harvard.edu/abs/2025ApJ...995...43A}
}

@ARTICLE{Rahner2017stellar_winds_fesc,
       author = {{Rahner}, Daniel and {Pellegrini}, Eric W. and {Glover}, Simon C.~O. and {Klessen}, Ralf S.},
        title = "{Winds and radiation in unison: a new semi-analytic feedback model for cloud dissolution}",
      journal = {\mnras},
         year = 2017,
        month = oct,
       volume = {470},
       number = {4},
        pages = {4453-4472},
          doi = {10.1093/mnras/stx1532},
archivePrefix = {arXiv},
       eprint = {1704.04240},
 primaryClass = {astro-ph.GA},
       adsurl = {https://ui.adsabs.harvard.edu/abs/2017MNRAS.470.4453R}
}

@ARTICLE{Komarova2025winds_LCEs,
       author = {{Komarova}, Lena and {Oey}, M.~S. and {Marques-Chaves}, Rui and {Amor{\'\i}n}, Ricardo and {Henry}, Alaina and {Schaerer}, Daniel and {Saldana-Lopez}, Alberto and {Le Reste}, Alexandra and {Scarlata}, Claudia and {Hayes}, Matthew J. and {Bait}, Omkar and {Borthakur}, Sanchayeeta and {Carr}, Cody and {Chisholm}, John and {Ferguson}, Harry C. and {Gutierrez Fernandez}, Vital and {Fleming}, Brian and {Flury}, Sophia R. and {Giavalisco}, Mauro and {Grazian}, Andrea and {Heckman}, Timothy and {Jaskot}, Anne E. and {Ji}, Zhiyuan and {{\"O}stlin}, G{\"o}ran and {Pentericci}, Laura and {Ravindranath}, Swara and {Thuan}, Trinh and {V{\'\i}lchez}, Jose M. and {Worseck}, Gabor and {Xu}, Xinfeng},
        title = "{Power-law Emission-line Wings and Radiation-driven Superwinds in Local Lyman Continuum Emitters}",
      journal = {\apj},
         year = 2025,
        month = dec,
       volume = {994},
       number = {2},
          eid = {192},
        pages = {192},
          doi = {10.3847/1538-4357/ae0e0a},
archivePrefix = {arXiv},
       eprint = {2506.19623},
 primaryClass = {astro-ph.GA},
       adsurl = {https://ui.adsabs.harvard.edu/abs/2025ApJ...994..192K}
}

@ARTICLE{Goovaerts2023LAEs_LBGs,
       author = {{Goovaerts}, I. and {Pello}, R. and {Thai}, T.~T. and {Tuan-Anh}, P. and {Richard}, J. and {Claeyssens}, A. and {Carinos}, E. and {de la Vieuville}, G. and {Matthee}, J.},
        title = "{Evolution of the Lyman-{\ensuremath{\alpha}}-emitting fraction and UV properties of lensed star-forming galaxies in the range 2.9 < z < 6.7}",
      journal = {\aap},
         year = 2023,
        month = oct,
       volume = {678},
          eid = {A174},
        pages = {A174},
          doi = {10.1051/0004-6361/202347110},
archivePrefix = {arXiv},
       eprint = {2307.15559},
 primaryClass = {astro-ph.GA},
       adsurl = {https://ui.adsabs.harvard.edu/abs/2023A&A...678A.174G}
}

@ARTICLE{Endsley2023UVfaintgals_reion,
       author = {{Endsley}, Ryan and {Stark}, Daniel P. and {Whitler}, Lily and {Topping}, Michael W. and {Chen}, Zuyi and {Plat}, Ad{\`e}le and {Chisholm}, John and {Charlot}, St{\'e}phane},
        title = "{A JWST/NIRCam study of key contributors to reionization: the star-forming and ionizing properties of UV-faint z   7-8 galaxies}",
      journal = {\mnras},
         year = 2023,
        month = sep,
       volume = {524},
       number = {2},
        pages = {2312-2330},
          doi = {10.1093/mnras/stad1919},
archivePrefix = {arXiv},
       eprint = {2208.14999},
 primaryClass = {astro-ph.GA},
       adsurl = {https://ui.adsabs.harvard.edu/abs/2023MNRAS.524.2312E}
}

@ARTICLE{Sun2023burstiness_highz,
       author = {{Sun}, Guochao and {Faucher-Gigu{\`e}re}, Claude-Andr{\'e} and {Hayward}, Christopher C. and {Shen}, Xuejian and {Wetzel}, Andrew and {Cochrane}, Rachel K.},
        title = "{Bursty Star Formation Naturally Explains the Abundance of Bright Galaxies at Cosmic Dawn}",
      journal = {\apjl},
         year = 2023,
        month = oct,
       volume = {955},
       number = {2},
          eid = {L35},
        pages = {L35},
          doi = {10.3847/2041-8213/acf85a},
archivePrefix = {arXiv},
       eprint = {2307.15305},
 primaryClass = {astro-ph.GA},
       adsurl = {https://ui.adsabs.harvard.edu/abs/2023ApJ...955L..35S}
}

@ARTICLE{Shen2023UVvariability_highz,
       author = {{Shen}, Xuejian and {Vogelsberger}, Mark and {Boylan-Kolchin}, Michael and {Tacchella}, Sandro and {Kannan}, Rahul},
        title = "{The impact of UV variability on the abundance of bright galaxies at z {\ensuremath{\geq}} 9}",
      journal = {\mnras},
         year = 2023,
        month = nov,
       volume = {525},
       number = {3},
        pages = {3254-3261},
          doi = {10.1093/mnras/stad2508},
archivePrefix = {arXiv},
       eprint = {2305.05679},
 primaryClass = {astro-ph.GA},
       adsurl = {https://ui.adsabs.harvard.edu/abs/2023MNRAS.525.3254S}
}

@ARTICLE{Simmonds2024bursty_gals_ionizing,
       author = {{Simmonds}, C. and {Tacchella}, S. and {Hainline}, K. and {Johnson}, B.~D. and {McClymont}, W. and {Robertson}, B. and {Saxena}, A. and {Sun}, F. and {Witten}, C. and {Baker}, W.~M. and {Bhatawdekar}, R. and {Boyett}, K. and {Bunker}, A.~J. and {Charlot}, S. and {Curtis-Lake}, E. and {Egami}, E. and {Eisenstein}, D.~J. and {Hausen}, R. and {Maiolino}, R. and {Maseda}, M.~V. and {Scholtz}, J. and {Williams}, C.~C. and {Willott}, C. and {Witstok}, J.},
        title = "{Low-mass bursty galaxies in JADES efficiently produce ionizing photons and could represent the main drivers of reionization}",
      journal = {\mnras},
         year = 2024,
        month = jan,
       volume = {527},
       number = {3},
        pages = {6139-6157},
          doi = {10.1093/mnras/stad3605},
archivePrefix = {arXiv},
       eprint = {2310.01112},
 primaryClass = {astro-ph.GA},
       adsurl = {https://ui.adsabs.harvard.edu/abs/2024MNRAS.527.6139S}
}

@ARTICLE{Burgarella2005SED,
       author = {{Burgarella}, D. and {Buat}, V. and {Iglesias-P{\'a}ramo}, J.},
        title = "{Star formation and dust attenuation properties in galaxies from a statistical ultraviolet-to-far-infrared analysis}",
      journal = {\mnras},
         year = 2005,
        month = jul,
       volume = {360},
       number = {4},
        pages = {1413-1425},
          doi = {10.1111/j.1365-2966.2005.09131.x},
archivePrefix = {arXiv},
       eprint = {astro-ph/0504434},
 primaryClass = {astro-ph},
       adsurl = {https://ui.adsabs.harvard.edu/abs/2005MNRAS.360.1413B}
}

@ARTICLE{Boquien2019cigale,
       author = {{Boquien}, M. and {Burgarella}, D. and {Roehlly}, Y. and {Buat}, V. and {Ciesla}, L. and {Corre}, D. and {Inoue}, A.~K. and {Salas}, H.},
        title = "{CIGALE: a python Code Investigating GALaxy Emission}",
      journal = {\aap},
         year = 2019,
        month = feb,
       volume = {622},
          eid = {A103},
        pages = {A103},
          doi = {10.1051/0004-6361/201834156},
archivePrefix = {arXiv},
       eprint = {1811.03094},
 primaryClass = {astro-ph.GA},
       adsurl = {https://ui.adsabs.harvard.edu/abs/2019A&A...622A.103B}
}

@ARTICLE{Calzetti2000dust_attenuation,
       author = {{Calzetti}, Daniela and {Armus}, Lee and {Bohlin}, Ralph C. and {Kinney}, Anne L. and {Koornneef}, Jan and {Storchi-Bergmann}, Thaisa},
        title = "{The Dust Content and Opacity of Actively Star-forming Galaxies}",
      journal = {\apj},
         year = 2000,
        month = apr,
       volume = {533},
       number = {2},
        pages = {682-695},
          doi = {10.1086/308692},
archivePrefix = {arXiv},
       eprint = {astro-ph/9911459},
 primaryClass = {astro-ph},
       adsurl = {https://ui.adsabs.harvard.edu/abs/2000ApJ...533..682C}
}

@ARTICLE{Bruzual2003stellar_pops,
       author = {{Bruzual}, G. and {Charlot}, S.},
        title = "{Stellar population synthesis at the resolution of 2003}",
      journal = {\mnras},
         year = 2003,
        month = oct,
       volume = {344},
       number = {4},
        pages = {1000-1028},
          doi = {10.1046/j.1365-8711.2003.06897.x},
archivePrefix = {arXiv},
       eprint = {astro-ph/0309134},
 primaryClass = {astro-ph},
       adsurl = {https://ui.adsabs.harvard.edu/abs/2003MNRAS.344.1000B}
}

@ARTICLE{Wisotzki2016lya_halos,
       author = {{Wisotzki}, L. and {Bacon}, R. and {Blaizot}, J. and {Brinchmann}, J. and {Herenz}, E.~C. and {Schaye}, J. and {Bouch{\'e}}, N. and {Cantalupo}, S. and {Contini}, T. and {Carollo}, C.~M. and {Caruana}, J. and {Courbot}, J.-B. and {Emsellem}, E. and {Kamann}, S. and {Kerutt}, J. and {Leclercq}, F. and {Lilly}, S.~J. and {Patr{\'\i}cio}, V. and {Sandin}, C. and {Steinmetz}, M. and {Straka}, L.~A. and {Urrutia}, T. and {Verhamme}, A. and {Weilbacher}, P.~M. and {Wendt}, M.},
        title = "{Extended Lyman {\ensuremath{\alpha}} haloes around individual high-redshift galaxies revealed by MUSE}",
      journal = {\aap},
         year = 2016,
        month = mar,
       volume = {587},
          eid = {A98},
        pages = {A98},
          doi = {10.1051/0004-6361/201527384},
archivePrefix = {arXiv},
       eprint = {1509.05143},
 primaryClass = {astro-ph.GA},
       adsurl = {https://ui.adsabs.harvard.edu/abs/2016A&A...587A..98W}
}

@software{larry_bradley_2025_photutils,
  author       = {Larry Bradley and
                  Brigitta Sip{\H o}cz and
                  Thomas Robitaille and
                  Erik Tollerud and
                  Z\`e Vin{\'{\i}}cius and
                  Christoph Deil and
                  Kyle Barbary and
                  Tom J Wilson and
                  Ivo Busko and
                  Axel Donath and
                  Hans Moritz G{\"u}nther and
                  Mihai Cara and
                  P. L. Lim and
                  Sebastian Me{\ss}linger and
                  Simon Conseil and
                  Michael Droettboom and
                  Azalee Bostroem and
                  E. M. Bray and
                  Lars Andersen Bratholm and
                  Zach Burnett and
                  William Jamieson and
                  Adam Ginsburg and
                  Dan Taranu and
                  Geert Barentsen and
                  Matt Craig and
                  Brett M. Morris and
                  Marshall Perrin and
                  Shivangee Rathi},
  title        = {astropy/photutils: 2.3.0},
  month        = sep,
  year         = 2025,
  publisher    = {Zenodo},
  version      = {2.3.0},
  doi          = {10.5281/zenodo.17129028},
  url          = {https://doi.org/10.5281/zenodo.17129028},
  swhid        = {swh:1:dir:dd51869167d76d722ba87e3f80f9f4199ec08c3f
                   ;origin=https://doi.org/10.5281/zenodo.596036;visi
                   t=swh:1:snp:30a5f50b0586911dc674668853d9abc352a2bc
                   22;anchor=swh:1:rel:e97861da904cf010c499a4211cd8a6
                   12373e912a;path=astropy-photutils-2294e35
                  },
}

@ARTICLE{Lofthouse2020MAGG,
       author = {{Lofthouse}, Emma K. and {Fumagalli}, Michele and {Fossati}, Matteo and {O'Meara}, John M. and {Murphy}, Michael T. and {Christensen}, Lise and {Prochaska}, J. Xavier and {Cantalupo}, Sebastiano and {Bielby}, Richard M. and {Cooke}, Ryan J. and {Lusso}, Elisabeta and {Morris}, Simon L.},
        title = "{MUSE Analysis of Gas around Galaxies (MAGG) - I: Survey design and the environment of a near pristine gas cloud at z ≍ 3.5}",
      journal = {\mnras},
         year = 2020,
        month = jan,
       volume = {491},
       number = {2},
        pages = {2057-2074},
          doi = {10.1093/mnras/stz3066},
archivePrefix = {arXiv},
       eprint = {1910.13458},
 primaryClass = {astro-ph.GA},
       adsurl = {https://ui.adsabs.harvard.edu/abs/2020MNRAS.491.2057L}
}

@ARTICLE{Becker2015quasar_abs_lines,
       author = {{Becker}, George D. and {Bolton}, James S. and {Lidz}, Adam},
        title = "{Reionisation and High-Redshift Galaxies: The View from Quasar Absorption Lines}",
      journal = {\pasa},
         year = 2015,
        month = dec,
       volume = {32},
          eid = {e045},
        pages = {e045},
          doi = {10.1017/pasa.2015.45},
archivePrefix = {arXiv},
       eprint = {1510.03368},
 primaryClass = {astro-ph.CO},
       adsurl = {https://ui.adsabs.harvard.edu/abs/2015PASA...32...45B}
}

@ARTICLE{Eldridge2017,
       author = {{Eldridge}, J.~J. and {Stanway}, E.~R. and {Xiao}, L. and {McClelland}, L.~A.~S. and {Taylor}, G. and {Ng}, M. and {Greis}, S.~M.~L. and {Bray}, J.~C.},
        title = "{Binary Population and Spectral Synthesis Version 2.1: Construction, Observational Verification, and New Results}",
      journal = {\pasa},
         year = 2017,
        month = nov,
       volume = {34},
          eid = {e058},
        pages = {e058},
          doi = {10.1017/pasa.2017.51},
archivePrefix = {arXiv},
       eprint = {1710.02154},
 primaryClass = {astro-ph.SR},
       adsurl = {https://ui.adsabs.harvard.edu/abs/2017PASA...34...58E}
}

@ARTICLE{Foreman-Mackey2013,
       author = {{Foreman-Mackey}, Daniel and {Hogg}, David W. and {Lang}, Dustin and {Goodman}, Jonathan},
        title = "{emcee: The MCMC Hammer}",
      journal = {\pasp},
         year = 2013,
        month = mar,
       volume = {125},
       number = {925},
        pages = {306},
          doi = {10.1086/670067},
archivePrefix = {arXiv},
       eprint = {1202.3665},
 primaryClass = {astro-ph.IM},
       adsurl = {https://ui.adsabs.harvard.edu/abs/2013PASP..125..306F}
}

@BOOK{Osterbrock1974,
       author = {{Osterbrock}, Donald E.},
        title = "{Astrophysics of gaseous nebulae}",
         year = 1974,
       adsurl = {https://ui.adsabs.harvard.edu/abs/1974agn..book.....O}
}

@ARTICLE{Hui1999lya_profiles,
       author = {{Hui}, Lam and {Stebbins}, Albert and {Burles}, Scott},
        title = "{A Geometrical Test of the Cosmological Energy Contents Using the Ly{\ensuremath{\alpha}} Forest}",
      journal = {\apjl},
         year = 1999,
        month = jan,
       volume = {511},
       number = {1},
        pages = {L5-L8},
          doi = {10.1086/311826},
archivePrefix = {arXiv},
       eprint = {astro-ph/9807190},
 primaryClass = {astro-ph},
       adsurl = {https://ui.adsabs.harvard.edu/abs/1999ApJ...511L...5H}
}

@ARTICLE{Inoue2014IGM,
       author = {{Inoue}, Akio K. and {Shimizu}, Ikkoh and {Iwata}, Ikuru and {Tanaka}, Masayuki},
        title = "{An updated analytic model for attenuation by the intergalactic medium}",
      journal = {\mnras},
         year = 2014,
        month = aug,
       volume = {442},
       number = {2},
        pages = {1805-1820},
          doi = {10.1093/mnras/stu936},
archivePrefix = {arXiv},
       eprint = {1402.0677},
 primaryClass = {astro-ph.CO},
       adsurl = {https://ui.adsabs.harvard.edu/abs/2014MNRAS.442.1805I}
}

@ARTICLE{Inoue2011IGM_UVspectra,
       author = {{Inoue}, Akio K.},
        title = "{Rest-frame ultraviolet-to-optical spectral characteristics of extremely metal-poor and metal-free galaxies}",
      journal = {\mnras},
         year = 2011,
        month = aug,
       volume = {415},
       number = {3},
        pages = {2920-2931},
          doi = {10.1111/j.1365-2966.2011.18906.x},
archivePrefix = {arXiv},
       eprint = {1102.5150},
 primaryClass = {astro-ph.CO},
       adsurl = {https://ui.adsabs.harvard.edu/abs/2011MNRAS.415.2920I}
}

@ARTICLE{Rudie2013,
       author = {{Rudie}, Gwen C. and {Steidel}, Charles C. and {Shapley}, Alice E. and {Pettini}, Max},
        title = "{The Column Density Distribution and Continuum Opacity of the Intergalactic and Circumgalactic Medium at Redshift langzrang = 2.4}",
      journal = {\apj},
         year = 2013,
        month = jun,
       volume = {769},
       number = {2},
          eid = {146},
        pages = {146},
          doi = {10.1088/0004-637X/769/2/146},
archivePrefix = {arXiv},
       eprint = {1304.6719},
 primaryClass = {astro-ph.CO},
       adsurl = {https://ui.adsabs.harvard.edu/abs/2013ApJ...769..146R}
}

@ARTICLE{Bielby2020lya_CGM,
       author = {{Bielby}, Richard M. and {Fumagalli}, Michele and {Fossati}, Matteo and {Rafelski}, Marc and {Oppenheimer}, Benjamin and {Cantalupo}, Sebastiano and {Christensen}, Lise and {Fynbo}, J.~P.~U. and {Lopez}, Sebastian and {Morris}, Simon L. and {D'Odorico}, Valentina and {Peroux}, Celine},
        title = "{Into the Ly {\ensuremath{\alpha}} jungle: exploring the circumgalactic medium of galaxies at z {\ensuremath{\sim}} 4-5 with MUSE}",
      journal = {\mnras},
         year = 2020,
        month = apr,
       volume = {493},
       number = {4},
        pages = {5336-5356},
          doi = {10.1093/mnras/staa546},
archivePrefix = {arXiv},
       eprint = {2001.09058},
 primaryClass = {astro-ph.GA},
       adsurl = {https://ui.adsabs.harvard.edu/abs/2020MNRAS.493.5336B}
}

@ARTICLE{Ejdetjarn2026Haro11LyC_lya,
       author = {{Ejdetj{\"a}rn}, Timmy and {{\"O}stlin}, G{\"o}ran and {Rosdahl}, Joakim and {Blaizot}, J{\'e}r{\'e}my and {Agertz}, Oscar},
        title = "{Radiation hydrodynamic simulation of the Haro 11 galaxy: the escape of LyC and Ly$α$ in a dwarf galaxy merger}",
      journal = {arXiv e-prints},
         year = 2026,
        month = jan,
          eid = {arXiv:2601.20943},
        pages = {arXiv:2601.20943},
archivePrefix = {arXiv},
       eprint = {2601.20943},
 primaryClass = {astro-ph.GA},
       adsurl = {https://ui.adsabs.harvard.edu/abs/2026arXiv260120943E}
}

@ARTICLE{Markov2025dustyLAE_lensed,
       author = {{Markov}, V. and {Brada{\v{c}}}, M. and {Estrada-Carpenter}, V. and {Desprez}, G. and {Rihtar{\v{s}}i{\v{c}}}, G. and {Jude{\v{z}}}, J. and {Tripodi}, R. and {Sawicki}, M. and {Noirot}, G. and {Martis}, N. and {Willott}, C. and {Sarrouh}, G.~T.~E. and {Withers}, S. and {Muzzin}, A. and {Asada}, Y. and {Gallerani}, S. and {Ferrara}, A. and {Goovaerts}, I. and {Harshan}, A. and {Fujimoto}, S.},
        title = "{Resolving dust and Ly{\ensuremath{\alpha}} emission in a lensed galaxy at the epoch of reionization with JWST/CANUCS}",
      journal = {arXiv e-prints},
         year = 2025,
        month = dec,
          eid = {arXiv:2512.13778},
        pages = {arXiv:2512.13778},
          doi = {10.48550/arXiv.2512.13778},
archivePrefix = {arXiv},
       eprint = {2512.13778},
 primaryClass = {astro-ph.GA},
       adsurl = {https://ui.adsabs.harvard.edu/abs/2025arXiv251213778M}
}

@ARTICLE{Endsley2024SF_ionizing_gals_z6-9,
       author = {{Endsley}, Ryan and {Stark}, Daniel P. and {Whitler}, Lily and {Topping}, Michael W. and {Johnson}, Benjamin D. and {Robertson}, Brant and {Tacchella}, Sandro and {Alberts}, Stacey and {Baker}, William M. and {Bhatawdekar}, Rachana and {Boyett}, Kristan and {Bunker}, Andrew J. and {Cameron}, Alex J. and {Carniani}, Stefano and {Charlot}, Stephane and {Chen}, Zuyi and {Chevallard}, Jacopo and {Curtis-Lake}, Emma and {Danhaive}, A. Lola and {Egami}, Eiichi and {Eisenstein}, Daniel J. and {Hainline}, Kevin and {Helton}, Jakob M. and {Ji}, Zhiyuan and {Looser}, Tobias J. and {Maiolino}, Roberto and {Nelson}, Erica and {Pusk{\'a}s}, D{\'a}vid and {Rieke}, George and {Rieke}, Marcia and {Rix}, Hans-Walter and {Sandles}, Lester and {Saxena}, Aayush and {Simmonds}, Charlotte and {Smit}, Renske and {Sun}, Fengwu and {Williams}, Christina C. and {Willmer}, Christopher N.~A. and {Willott}, Chris and {Witstok}, Joris},
        title = "{The star-forming and ionizing properties of dwarf z 6-9 galaxies in JADES: insights on bursty star formation and ionized bubble growth}",
      journal = {\mnras},
         year = 2024,
        month = sep,
       volume = {533},
       number = {1},
        pages = {1111-1142},
          doi = {10.1093/mnras/stae1857},
archivePrefix = {arXiv},
       eprint = {2306.05295},
 primaryClass = {astro-ph.GA},
       adsurl = {https://ui.adsabs.harvard.edu/abs/2024MNRAS.533.1111E}
}

@ARTICLE{Finkelstein2019reionization_fesc,
       author = {{Finkelstein}, Steven L. and {D'Aloisio}, Anson and
         {Paardekooper}, Jan-Pieter and {Ryan}, Russell, Jr. and
         {Behroozi}, Peter and {Finlator}, Kristian and {Livermore}, Rachael and
         {Upton Sanderbeck}, Phoebe R. and {Dalla Vecchia}, Claudio and
         {Khochfar}, Sadegh},
        title = "{Conditions for Reionizing the Universe with a Low Galaxy Ionizing Photon Escape Fraction}",
      journal = {\apj},
         year = 2019,
        month = jul,
       volume = {879},
       number = {1},
          eid = {36},
        pages = {36},
          doi = {10.3847/1538-4357/ab1ea8},
archivePrefix = {arXiv},
       eprint = {1902.02792},
 primaryClass = {astro-ph.CO},
       adsurl = {https://ui.adsabs.harvard.edu/abs/2019ApJ...879...36F}
}

@ARTICLE{OkeGunn1983,
       author = {{Oke}, J.~B. and {Gunn}, J.~E.},
        title = "{Secondary standard stars for absolute spectrophotometry.}",
      journal = {\apj},
         year = 1983,
        month = mar,
       volume = {266},
        pages = {713-717},
          doi = {10.1086/160817},
       adsurl = {https://ui.adsabs.harvard.edu/abs/1983ApJ...266..713O}
}
\bibliographystyle{aasjournalv7}



\end{document}